\documentclass{aa}  

\usepackage{graphicx}
\usepackage{placeins}
\usepackage{txfonts}
\usepackage[colorlinks=true,linkcolor=bluea,allcolors=blue]{hyperref}

\begin{document}

   \title{The diverse cold molecular gas contents, morphologies, and kinematics of type-2 quasars as seen by ALMA}


   \author{C. Ramos Almeida\inst{1,2},
          M. Bischetti\inst{3},
          S. Garc\' ia-Burillo\inst{4}, 
A. Alonso-Herrero\inst{5},
A. Audibert\inst{1,2},
C. Cicone\inst{6},
C. Feruglio\inst{3},
C. N. Tadhunter\inst{7},
J. C. S. Pierce\inst{7,8},
M. Pereira-Santaella\inst{9}
\and P. S. Bessiere\inst{1,2}
          }

   \institute{Instituto de Astrof\' isica de Canarias, Calle V\' ia L\'actea, s/n, E-38205, La Laguna, Tenerife, Spain\\
              \email{cra@iac.es}
              \and Departamento de Astrof\' isica, Universidad de La Laguna, E-38206, La Laguna, Tenerife, Spain
          \and INAF – Osservatorio Astronomico di Trieste, via G.B. Tiepolo 11, 34143 Trieste, Italy 
\and Observatorio Astron\'omico Nacional (OAN-IGN)- Observatorio de Madrid, Alfonso XII, 3, 28014, Madrid, Spain 
\and Centro de Astrobiolog\'{\i}a (CAB, CSIC-INTA), ESAC Campus, E-28692, Villanueva de la Ca\~nada, Madrid, Spain
\and Institute of Theoretical Astrophysics, University of Oslo, PO Box 1029, Blindern 0315, Oslo, Norway
\and Department of Physics \& Astronomy, University of Sheffield, S3 7RH Sheffield, UK
\and Centre for Astrophysics Research, University of Hertfordshire, Hatfield, Hertfordshire, AL10 9AB, UK
\and Centro de Astrobiolog\' ia (CSIC-INTA), Ctra. de Ajalvir, Km 4, 28850, Torrej\'on de Ardoz, Madrid, Spain}

   \date{Received July 29, 2021; accepted December 2, 2021}

 
  \abstract

\abstract{We present CO(2$-$1) and adjacent continuum observations of seven nearby radio-quiet type-2 quasars (QSO2s) obtained with ALMA at $\sim$0.2\arcsec~resolution (370 pc at z$\sim$0.1). These QSO2s are luminous (L$_{\rm [OIII]}>$10$^{8.5}L_{\sun}\sim$M$_B<$-23), and their host galaxies massive (M$_*\sim$10$^{11}$M$_{\rm \sun}$). 
The CO morphologies are diverse, including disks and interacting systems. Two of the QSO2s are red early-type galaxies with no CO(2-1) detected.
In the interacting galaxies, the central kiloparsec contains 18--25\% of the total cold molecular gas, whereas in the spirals it is only $\sim$5--12\%. 
J1010+0612 and J1430+1339 show double-peaked CO flux maps along the major axis of the CO disks that do not have an optical counterpart at the same angular resolution. Based on our analysis of the ionized and molecular gas kinematics and millimeter continuum emission, these CO morphologies are most likely produced by active galactic nucleus (AGN) feedback in the form of outflows, jets, and/or shocks. 
The CO kinematics of the QSO2s with CO(2$-$1) detections are dominated by rotation but also reveal noncircular motions. 
According to our analysis, these noncircular motions correspond to molecular outflows that are mostly coplanar with the CO disks in four of the QSO2s, and either to a coplanar inflow or vertical outflow in the case of J1010+0612. These outflows represent 0.2--0.7\% of the QSO2s' total molecular gas mass and have maximum velocities of 200--350 km~s$^{-1}$, radii from 0.4 to 1.3 kpc, and outflow mass rates of 8--16 M$_{\rm \sun}$~yr$^{-1}$. These outflow properties are intermediate between those of the mild molecular outflows measured for Seyfert galaxies and the fast and energetic outflows shown by ultra-luminous infrared galaxies. This suggests that it is not only AGN luminosity that drives massive molecular outflows. Other factors such as jet power, coupling between winds, jets, and/or ionized outflows and the CO disks, and amount or geometry of dense gas in the nuclear regions might also be relevant. Thus, although we do not find evidence for a significant impact of quasar feedback on the total molecular gas reservoirs and star formation rates, it appears to be modifying the distribution of cold molecular gas in the central kiloparsec of the galaxies.

}


\titlerunning{Cold molecular gas in QSO2s as seen by ALMA}
\authorrunning{C. Ramos Almeida et al.}
   \maketitle
%

\section{Introduction}

Different modes of active galactic nucleus (AGN) feedback are considered to be important processes driving the evolution of massive galaxies, by regulating black hole and galaxy growth \citep{2005Natur.433..604D,2017NatAs...1E.165H}. If AGN feedback is not considered in cosmological simulations, gas can cool efficiently and galaxies keep forming stars, growing too big (e.g., \citealt{2016MNRAS.463.3948D}). Additionally, the predicted galaxy-halo mass relations do not match observations either \citep{1998A&A...331L...1S,2006MNRAS.365...11C,2010ApJ...710..903M}. Currently there is plenty of observational evidence showing that AGN or black hole feedback has an impact on very different scales, which go from the central tens to hundreds of parsecs (e.g., \citealt{2021arXiv210410227G}) to hundreds of kiloparsecs (e.g., \citealt{2019Natur.574..643R,2021Natur.594..187M}). We now need to understand how AGN feedback works in relation to AGN and host galaxy properties and the coupling between the two while considering the short timescales associated with nuclear activity \citep{2004cbhg.symp..169M,2014ApJ...782....9H}.



The most widely studied manifestation of AGN feedback are gas outflows. Radio jets and/or AGN winds can drive multiphase outflows \citep{2018MNRAS.476...80M,2018NatAs...2..181W,2019MNRAS.485.2710J} that have an impact on the star formation efficiency of galaxies, as they can remove, heat up, and/or disrupt the gas available to form stars. 
An important source of uncertainty affecting outflow studies is that the contribution from the different gas phases entrained in the winds has not been determined in representative AGN samples \citep{2018NatAs...2..176C}. If the neutral atomic, ionized, and molecular gas phases are accounted for, AGN-driven outflows might be massive and energetic enough to regulate star formation, at least in the central region of galaxies (e.g., \citealt{2018RMxAA..54..217S}). In the case of low-to-intermediate luminosity AGN (log L$_{\rm bol}\sim$42--46 erg~s$^{-1}$), the molecular gas phase appears to be dominant in terms of mass over the other interstellar medium (ISM) phases (e.g., \citealt{2017A&A...601A.143F,2019MNRAS.483.4586F,2020arXiv200613232F}). However, at log L$_{\rm bol}>$46 erg~s$^{-1}$, ionized outflows might be as massive as their molecular counterparts \citep{2017A&A...601A.143F,2019A&A...628A.118B,2020arXiv200613232F}, although this has to be explored in larger AGN samples with multiphase outflow measurements. 
In \citet{2017MNRAS.470..964R,2019MNRAS.487L..18R} we used near-infrared (NIR) spectroscopy of nearby (z$\sim$0.1) obscured quasars with log L$_{\rm bol}$=45.5--46.5  erg~s$^{-1}$ to study quasar-driven outflows using ionized (Pa$\alpha$ and [Si VI]) and warm molecular emission lines (H$_2$). We found that the outflow properties were different in the two gas phases, as previously claimed by \citet{2013ApJ...775L..15R} for the same type of source. Ionized outflows are faster, but the molecular outflows appear to carry the bulk of the mass (e.g., \citealt{2020arXiv200613232F}). However, estimating total molecular outflow masses from warm molecular gas depends on assuming warm-to-cold gas mass ratios, which are affected by large uncertainties \citep{2019MNRAS.487L..18R}. 

Cold molecular outflows in the nearby universe have been studied in low-to-intermediate luminosity AGN (i.e., Seyfert galaxies with L$_{\rm IR}<$10$^{12}$L$_{\rm \sun}$) and ultra-luminous infrared galaxies (ULIRGs), mainly, but not only, using hydroxyl (OH) and carbon monoxide (CO) transitions (see \citealt{2020A&ARv..28....2V} for a recent review). Focusing on CO-based studies, it has been found that the two types of objects show very different outflow properties, going from maximum velocities of $\sim$100--200 km~s$^{-1}$ and outflow rates of $\la$10 M$_{\sun}$yr$^{-1}$ in Seyfert galaxies (e.g., \citealt{2019A&A...628A..65A,2020arXiv200305663D,2021A&A...645A..21G}), to $>$300 km~s$^{-1}$ and hundreds of M$_{\sun}$yr$^{-1}$ in ULIRGs (e.g., \citealt{2014A&A...562A..21C,2018A&A...616A.171P,2019MNRAS.483.4586F,2020arXiv200613232F,2020A&A...635A..47H}). It is noteworthy that jetted Seyfert galaxies like NGC\,1068 and IC\,5063 \citep{2014A&A...567A.125G,2019A&A...632A..61G,2015A&A...580A...1M} show much higher outflow velocities and masses than those without, indicating the strong influence that jets might have in launching/accelerating molecular outflows \citep{2018NatAs...2..181W,2019MNRAS.485.2710J}. Indeed, even in lower radio-power Seyfert galaxies, small-scale jets can drive nuclear molecular outflows (e.g., \citealt{2016A&A...590A..73A,2019A&A...632A..33A,2020A&A...633A.127F,2021A&A...645A..21G}). Studying molecular outflows in nearby quasars, which have higher AGN luminosities than ULIRGs, and are hosted in galaxies of different morphologies \citep{2012MNRAS.426..276B,2021MNRAS.tmp.2947P}, might help us to identify which factors, including AGN luminosity, are relevant for producing more or less massive molecular outflows. 


In the most powerful quasars and ULIRGs in the local universe, AGN-driven outflows have the potential to deplete the host galaxies of molecular gas on timescales of $\sim$10--50 Myr \citep{2010A&A...518L.155F,2014A&A...562A..21C,2019A&A...628A.118B}. However, as constrained from simulations and observations, only a small percentage of the outflowing gas is capable of escaping the galaxy \citep{2017MNRAS.467.3475N,2019MNRAS.483.4586F}, with the rest raining back down onto it. This result remains valid at high redshifts (z$\sim$6) and at high AGN luminosities (e.g., \citealt{2019MNRAS.489.3939C,2019A&A...630A..59B}). Regardless, it has been shown that AGN feedback is capable of influencing the distribution of molecular gas in the central kpc of nearby galaxies \citep{2019ApJ...875L...8R,2021arXiv210410227G} and/or injecting energy in the haloes, preventing hot gas from cooling \citep{2020MNRAS.491.5406T}. This gentler form of AGN feedback, sometimes referred to as ``maintenance mode,'' could be enough to regulate galaxy growth.

Here we explore for the first time the cold (T$_k<$100 K) molecular gas content traced by the 2--1 line of CO, and adjacent continuum emission, of a sample of nearby (z$\sim$0.1) type-2 quasars (QSO2s) at an angular resolution of $\sim$0.2\arcsec~(370 pc). QSO2s (L$_{\rm [OIII]}>$10$^{8.5}L_{\rm \sun}$; \citealt{2003AJ....126.2125Z,2008AJ....136.2373R}) are optically obscured quasars (i.e., buried AGN) showing narrow emission lines (FWHM$<$2000 km s$^{-1}$) of high equivalent widths.
They have been extensively studied in the optical (e.g., \citealt{2014MNRAS.442..784Z,2014MNRAS.440.3202V,2016ApJ...817..108W}) and to a much lesser extent in the NIR \citep{2013ApJ...775L..15R,2015MNRAS.454..439V,2017MNRAS.470..964R,2019MNRAS.487L..18R}. 
QSO2s are excellent laboratories to search for outflows and study their influence in their host galaxies in the optical and NIR, because the strong AGN continuum and the broad permitted lines produced in the broad-line region are naturally obscured by dust. This makes it easier to detect 1) broad components associated with outflowing gas and 2) stellar absorption features, required to study the stellar populations of the host galaxies. 
In addition, QSO2s might constitute a crucial phase in the coevolution of AGN and their host galaxies in which the AGN is clearing up gas and dust to eventually shine as a type-1 QSO (e.g., \citealt{1988ApJ...325...74S,2008ApJS..175..356H,2009ApJ...696..891H}). However, in contrast with what would be expected from this scenario, \citet{2019ApJ...873...90S} showed that the gas content of nearby QSO2s appears indistinguishable from that of type-1 quasars of the same luminosity. 

From unresolved millimeter observations carried out with the Plateau de Bure Interferometer, IRAM 30 m and APEX, 
gas masses of $\sim$7--25$\times 10^9$ M$_{\sun}$\footnote{Using the Galactic $\alpha_{\rm CO}$=4.35 M$_{\sun}(\rm K~km~s^{-1}~pc^2)^{-1}$ from \citet{2013ARA&A..51..207B}.} have been reported for small samples of QSO2s at z$<$0.4  \citep{2012ApJ...753..135K,2013MNRAS.434..978V,2020MNRAS.498.1560J}. Due to the low sensitivities available in the past, a limited number of AGN have been studied in the high-luminosity regime. Most of them are ULIRGs, and hence, CO-bright \citep{1998ApJ...507..615D,2008ApJS..178..189W,2014A&A...562A..21C}. The 
Atacama Large Millimeter/sub-millimeter Array (ALMA) now enables spatially resolved CO maps
of galaxy samples to be obtained with reasonable integration times. High angular resolution observations are particularly important in the case of luminous quasars and QSO2s in particular. They are commonly found in galaxy groups \citep{2013MNRAS.436..997R} and are often seen to be interacting with other galaxies (\citealt{2012MNRAS.426..276B,2021MNRAS.tmp.2947P}). These companion galaxies might be included in the large apertures of single-dish observations, resulting in larger molecular gas masses/fractions. Furthermore, the high angular resolution is key for identifying molecular outflows, since even in nearby ULIRGs they appear generally compact (r$\sim$1--2 kpc; e.g., \citealt{2010A&A...518L.155F,2014A&A...562A..21C}), although these radii can extend to a few kpc if low surface brightness components are considered \citep{2013A&A...549A..51F,2019ApJ...871...37H,2020A&A...633A.163C}.
In Section \ref{sample} we describe the sample selection and properties of the QSO2s. Section \ref{observations} describes the ALMA observations and data reduction. In Section \ref{methodology} we explain the methodology that we followed to interpret the molecular gas kinematics. Section \ref{results} includes the results found for the individual galaxies (Section \ref{individual}), the millimeter continuum emission of the QSO2s (Section \ref{continuum}) and their molecular gas content (Section \ref{molecular}). In Section \ref{Discussion} we discuss the results on the molecular gas reservoirs and molecular outflows, and in Section \ref{conclusions} we summarize the findings of this work.
We assume a cosmology with H$_0$=70 km~s$^{-1}$ Mpc$^{-1}$, $\Omega_m$=0.3, and $\Omega_{\Lambda}$=0.7. The measurements from other works discussed here have been converted to this cosmology.


\begin{table*}
\caption{Quasar properties.}
\centering
\begin{tabular}{lcccccccccc}
\hline
\hline
SDSS ID  	       &      z  & D$_{\rm L}$ &   Scale   &  log L$_{\rm [OIII]}$  & log L$_{\rm bol}$& log L$_{\rm 1.4GHz}$ & log M$_{\rm BH}$ & log\(\frac{L_{\rm bol}}{L_{\rm Edd}}\) \\
                       &    SDSS	      & (Mpc) &(kpc/\arcsec) & (L$_{\rm \sun}$)	  & (erg/s) & (W/Hz) &(M$_{\rm \sun}$) &  \\
\hline
J023224.24-081140.2   &      0.1001  & 461   &  1.846  &  8.60   &   45.73   & 22.96  & 7.46$\pm$0.33     & -0.33$\pm$0.35  \\
J101043.36+061201.4   &      0.0977  & 449   &  1.807  &  8.68   &   45.81   & 24.37  & 8.36$\pm$0.77     & -0.80$\pm$0.78  \\
J110012.39+084616.3   &      0.1004  & 462   &  1.851  &  9.20   &   46.33   & 24.18  & 7.82$\pm$0.44     &  0.04$\pm$0.45  \\
J115245.66+101623.8   &      0.0699  & 315   &  1.335  &  8.72   &   45.85   & 22.67  & 7.91$\pm$0.33     & -0.72$\pm$0.35  \\
J135646.10+102609.0   &      0.1232  & 576   &  2.213  &  9.21   &   46.34   & 24.36  & 8.58$\pm$0.34     & -1.03$\pm$0.36  \\
J143029.88+133912.0   &      0.0851  & 388   &  1.597  &  9.08   &   46.21   & 23.67  & 8.19$\pm$0.35     & -0.35$\pm$0.37  \\
J150904.22+043441.8   &      0.1115$^*$  & 517   &  2.028  &  8.56   &   45.69   & 23.81  & 8.27$\pm$0.76     & -0.22$\pm$0.77  \\ 
\hline	   					 			    					 			      
\end{tabular}	
\tablefoot{The values of L$_{\rm bol}$ listed here were derived from the non-parametric measurements of the [OIII] luminosity from \citet{2008AJ....136.2373R}, using a bolometric correction factor of 3500 \citep{2004ApJ...613..109H}. Rest-frame radio luminosities are calculated from integrated FIRST fluxes \citep{1995ApJ...450..559B}, assuming a spectral index $\alpha$=-0.7.
Black hole masses and Eddington ratios are from \citet{2018ApJ...859..116K}. * The redshift measured from the NIR spectrum of J1509 is z=0.1118 \citep{2019MNRAS.487L..18R}.}
\label{tab1}
\end{table*}

\begin{table*}
\caption{Galaxy properties.}
\centering
\begin{tabular}{lccccccccccc}
\hline
\hline
ID      & log L$_{\rm IR}$ & SFR & log M$_{\rm *}$   &  \multicolumn{2}{c}{Major axis}   & \multicolumn{2}{c}{Minor axis}  & PA   & i & \multicolumn{2}{c}{Galaxy morphology}	  \\
		& (L$_{\rm \sun}$) & (M$_{\rm \sun}$/yr)& (M$_{\rm \sun}$)    & (\arcsec)  & (kpc)    & (\arcsec)  & (kpc)  & (deg)& (deg) & This work & Galaxy Zoo \\
\hline
J0232 &  10.45   & 3.0     & 10.91$\pm$0.19 &  6.05 & 11.2 & 4.85 & 8.95 & 142 & 37  & Red ETG  & \dots \\
J1010 &  11.44   & 30      & 10.99$\pm$0.20 &  9.17 & 16.6 & 7.55 & 13.6 & 113 & 35  & Interacting ETG & S0a \\
J1100 &  11.50   & 34      & 11.02$\pm$0.22 &  8.51 & 15.7 & 6.72 & 12.4 & 67  & 38  & Barred spiral	& SBb   \\
J1152 &  10.54   & 3.7     & 10.90$\pm$0.16 & 12.39 & 16.5 & 6.75 & 9.01 & 56  & 57  & Red ETG & E  \\
J1356 &  11.80   & 69      & 11.27$\pm$0.19 & 11.45 & 25.3 & 6.56 & 14.5 & 156 & 55  & Merging ETG & E \\
J1430 &  11.06   & 12      & 11.15$\pm$0.11 & 11.80 & 18.8 & 9.27 & 14.8 & 161 & 38  & Post-merger ETG & E  \\
J1509 & 11.49   & 34      & 10.94$\pm$0.31 &  7.40 & 15.0 & 5.35 & 10.8 & 94  & 44  & Barred spiral  & SBa \\
\hline	   					 			    								      
\end{tabular}	
\tablefoot{Rest-frame IR luminosities (8--1000 $\mu$m) and corresponding SFRs were derived from IRAS 60 and PACS/IRAS 100 $\mu$m fluxes. 
The uncertainties of L$_{\rm IR}$ and SFRs are $\sim$0.12 dex and $\sim$0.3 dex, respectively. For J1100 and J1152, the PACS 70 $\mu$m fluxes were used as a proxy for the 60 $\mu$m flux. For J0232 there are no FIR data, but we estimated them by scaling the WISE+PACS spectral energy distribution (SED) of J1152 (also a red ETG) to the WISE SED of J0232.
Stellar masses were calculated from 2MASS XSC K-band magnitudes, as described in the text. J1010 and J1100 show K-band excesses that might be indicative of an
important contribution from AGN-heated dust (see \citealt{2020MNRAS.498.1560J} and \citealt{2019ApJ...873...90S}, respectively) and thus, for them we used J-band magnitudes instead. Indeed, in the case of J1100, we use M$_*$ from \citet{2019ApJ...873...90S}, converted to our cosmology. 
Columns from 5 to 10 list the isophotal major and minor axis at 25 mag~arcsec$^{-2}$, position angle (PA) and inclination (i) from r-band SDSS DR6 photometry (2007). 
Morphological classifications come from visual inspection of the SDSS images shown in Figure \ref{fig1} and from the Galaxy Zoo 2 project \citep{2013MNRAS.435.2835W}.}
\label{tab2}
\end{table*}

\section{Sample selection and properties}
\label{sample}

\begin{table*}
\caption{Ionized and warm molecular outflow properties measured from the broadest component of the [OIII]$\lambda$5007 \AA, Pa$\alpha$ and H$_2$~emission lines.}
\centering
\begin{tabular}{lccccclc}
\hline
\hline
ID    & Emission&   FWHM & V$_{\rm max}$ & r$_{\rm out}$  & Orientation & Data & Reference	  \\
    & line  & (km~s$^{-1}$) & (km~s$^{-1}$) & (kpc) & & &  \\
\hline
J0232 & [OIII] & 770  & -755 & $\le$2.8 & \dots & SDSS & a    \\
J1010 & [OIII] & 1350 & -890 & $\le$1.6 & NW & VLT/VIMOS \& MUSE$^*$ & b \\
J1100 & [OIII] & 1780 & -1240 & 0.46  & SE  & HST/STIS & c    \\
J1152 & [OIII] & 360  & -480 & 0.13  & NE & HST/STIS & c  \\
J1356 & [OIII] & 880  & -990 & $\le$3.1 & SW & VLT/VIMOS   & b \\
J1430 & [OIII] & 955  & -745 & 0.80 & NE & VLT/VIMOS & d  \\
\dots  & Pa$\alpha$ & 1800 & -1100& 0.55 & NE & VLT/SINFONI & e  \\
J1509 & [OIII] & 1500  & -1100 & $\le$3 & \dots & SDSS & f  \\
\dots & Pa$\alpha$ & 1800  & -1200 & 0.65 & \dots & GTC/EMIR & f  \\
\dots & H$_2$ & 1500 & -750& 0.75 & \dots & GTC/EMIR & f \\
\hline	   					 			    								      
\end{tabular}	
\tablefoot{Columns 3, 4, 5, and 6 correspond to the emission line's FWHM, V$_{\rm max}$=-(|V$_{\rm out}$|+FWHM/2), projected radius and spatial orientation. Column 7 lists the details of the corresponding data. * VLT/MUSE (PI: G. Venturi, 0104.B-0476) data retrieved from the ESO Archive Science Portal.} 
\tablebib{(a) \citet{2014MNRAS.440.3202V}; (b) \citet{2014MNRAS.441.3306H}; (c) \citet{2018ApJ...856..102F}; (d) \citet{2015ApJ...800...45H};  (e) \citet{2017MNRAS.470..964R}; (f) \citet{2019MNRAS.487L..18R}.}
\label{tab4a}
\end{table*}

\begin{figure*}
\centering
\includegraphics[width=16cm]{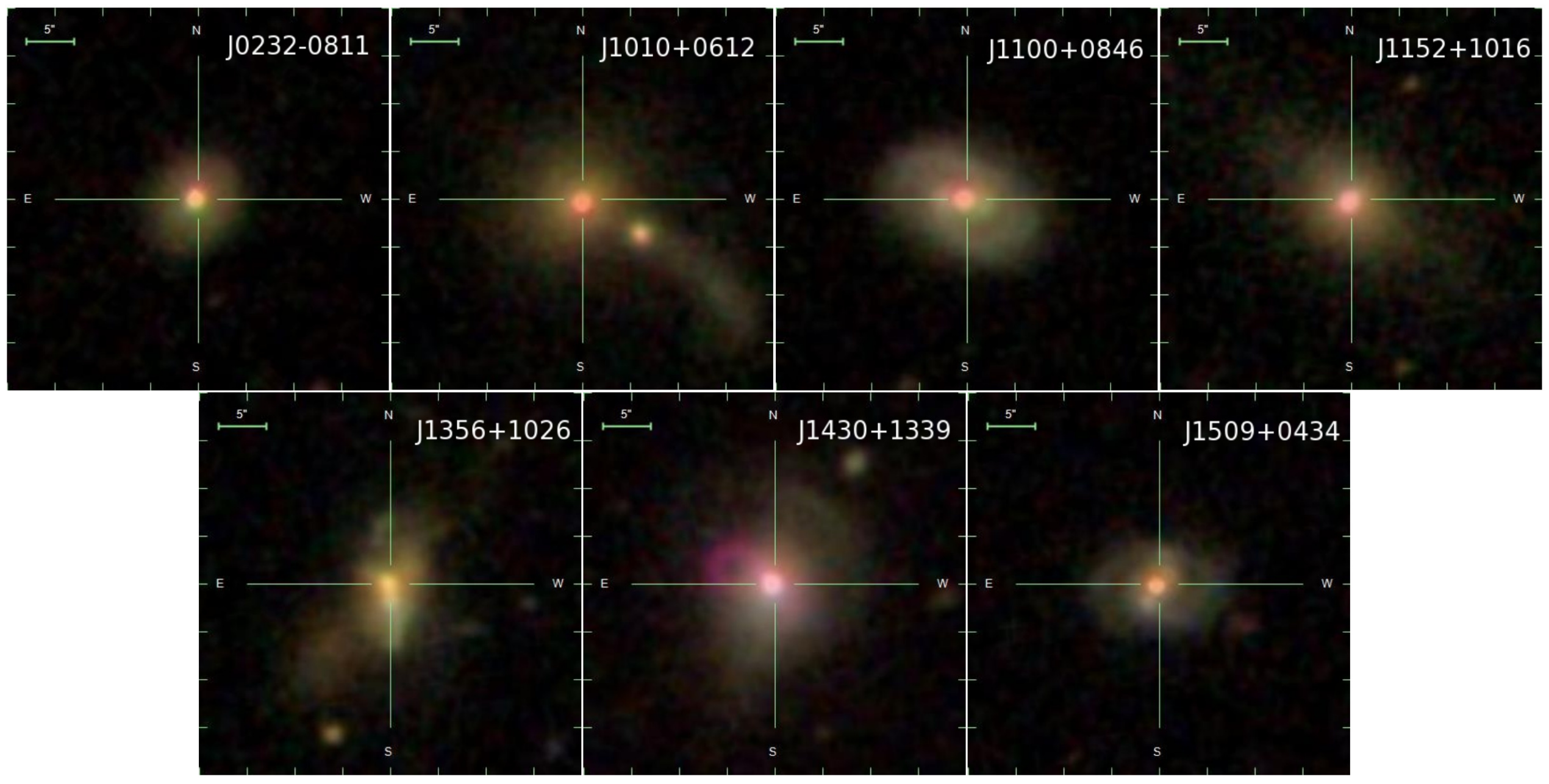}
\caption{SDSS gri color composite images of the seven QSO2s. Their optical morphologies include barred spiral galaxies (J1100 and J1509), red ETGs (J0232 and J1152) and  interacting, merging and post-merger galaxies (J1010, J1356, and J1430). North is up and east to the left. The green horizontal bars at the top left of each panel correspond to 5\arcsec, which, at the average redshift of the targets (z=0.1), corresponds to $\sim$9 kpc. The images are 40\arcsec$\times$40\arcsec~(74$\times$74 kpc$^2$).}  
\label{fig1}
\end{figure*}

Our QSO2 sample was drawn from \citet{2008AJ....136.2373R}, one of the largest compilations of narrow emission line AGN. We selected all the QSO2s with L$_{\rm [OIII]}>10^{8.5}L_{\sun}$ (L$_{\rm bol}>10^{45.6}$ erg~s$^{-1}$ using the bolometric correction of 3500 from \citealt{2004ApJ...613..109H}) and redshifts z$<$0.14. These constraints leave us with a sample of 48 QSO2s, hereinafter referred to as the Quasar Feedback (QSOFEED) sample, with 45.6<log L$_{\rm bol}$<46.5 erg~s$^{-1}$ (average value of 45.9$\pm$0.2 erg~s$^{-1}$). The optical selection of the targets prevents biases in dust and gas content, unlike in the case of infrared-selected samples. QSO2s might have, in principle, higher gas and dust masses than type-1 quasars, but recent results do not support this hypothesis \citep{2019ApJ...873...90S}.  
These luminous QSO2s have stellar masses ranging from 10$^{10.7}$ to $10^{11.6}M_{\sun}$ (average M$_*$=10$^{11.1\pm0.2}M_{\sun}$), calculated from 2MASS Extended Source Catalogue (XSC) K-band magnitudes. We followed the procedure applied in \citet{2021MNRAS.tmp.2947P}, which uses the \citet{2003ApJ...586..794B} equations, extinction and k-corrected K-band magnitudes, but considering a Chabrier IMF and (B-V)=0.95.  Considering that these are type-2 AGN, we do not expect the AGN contamination of the stellar mass estimates to be high. Indeed, the recent study of QSO2 hosts by \citet{2019ApJ...873...90S}, in which they performed SED fitting for a sample of 86 optically selected QSO2s at z$<$0.5, shows that only five objects show strong hot dust emission in the NIR, one of them being one of our targets (J1100; see Tables \ref{tab1} and \ref{tab2}). 
Indeed, the average stellar mass of the QSOFEED sample is similar to the value of M$_*$=10$^{10.9\pm0.2}M_{\sun}$ reported by \citet{2019ApJ...873...90S}.



From the QSOFEED sample we selected a subset of 7 QSO2s with redshifts 0.07$\le$z$\le$0.12. These QSO2s are representative of the whole sample in terms of AGN and radio luminosity, stellar mass, galaxy morphology and ionized outflow properties. These and other properties are listed in Tables \ref{tab1}, \ref{tab2} and \ref{tab4a}.  
The 7 QSO2s show a range of different galaxy morphologies in the optical SDSS images (see Figure \ref{fig1}) including barred spirals, early-type galaxies (ETGs) and interacting, merging and post-merger systems (see Table \ref{tab2}). The two seemingly undisturbed ETGs in the sample, J0232 and J1152, show rest-frame colors M$_u$-M$_g\sim$1.4. These are much redder and closer to typical red-sequence galaxies (M$_u$-M$_g>$1.5; \citealt{2006ApJ...648..268B}) than the other five QSO2s, which show (M$_u$-M$_g$)$\sim$0.95--1.18. In the following, we refer to J0232 and J1152 as red ETGs. J1152 has been revealed as a post-merger system from deep optical imaging (see \citealt{2021MNRAS.tmp.2947P}). Thus, four of the seven QSO2s (57\%) show disturbed morphologies that are indicative of a past interaction. This is consistent with the percentage of disturbance found for the QSOFEED sample based on deep optical imaging taken with the Isaac Newton Telescope (INT), in La Palma, which is 65\% (Pierce et al. in prep.).





\begin{figure}
\includegraphics[width=1.04\columnwidth]{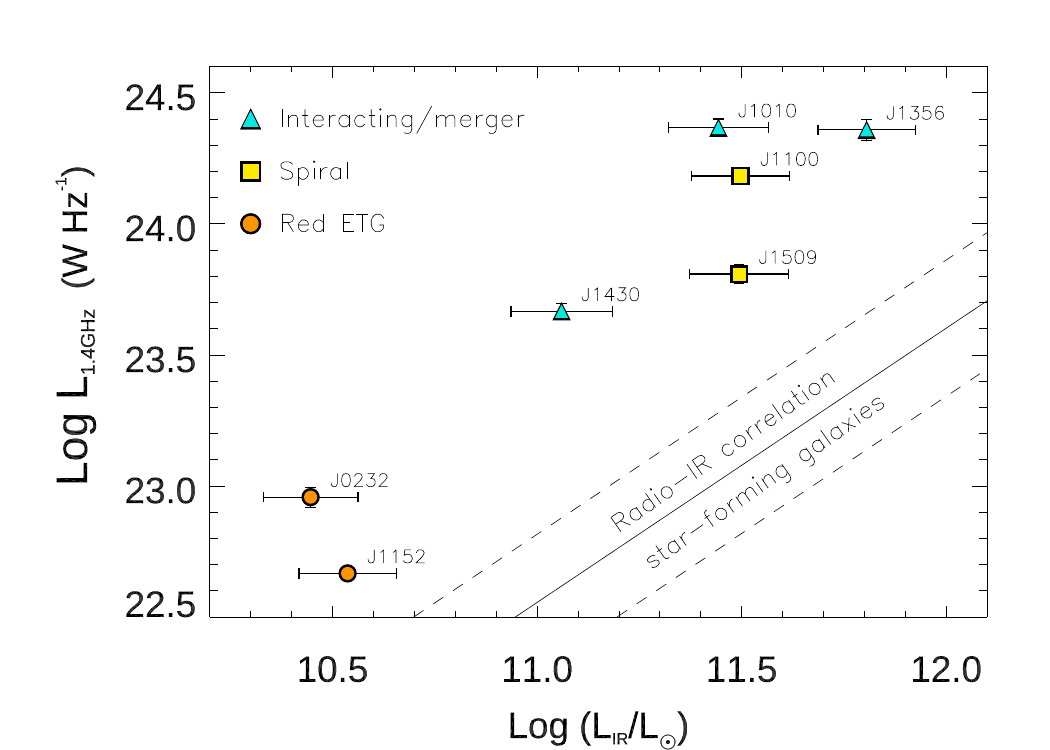}
\caption{Radio (1.4 GHz) versus IR luminosities (8--1000 $\mu$m) for the QSO2s in our sample. The radio-FIR correlation of star-forming galaxies from \citet{2003ApJ...586..794B} is shown as a solid line (slope of 1.05$\pm$0.04), with a correlation scatter of 0.26 dex. All the QSO2s in our sample show a radio excess. Different colors indicate whether the QSO2s are red ETGs, spirals, or interacting galaxies (see Table \ref{tab2}).} 
\label{fig2}
\end{figure}

All the QSO2s are luminous infrared galaxies (LIRGs; see Table \ref{tab2} and Figure \ref{fig2}) except the red ETGs. From their FIR luminosities (L$_{\rm FIR}$), 
calculated from 60 and 100 $\mu$m fluxes \citep{1985ApJ...298L...7H}, we estimated the IR luminosities (L$_{\rm IR}$; 8--1000 $\mu$m) by multiplying L$_{\rm FIR}$ by a factor 1.82$\pm$0.17. We determined this value using the sample of LIRGs at z$<$0.1 presented in \citet{2014ApJ...794..142G}, for which both L$_{\rm FIR}$ and L$_{\rm IR}$ are available. From the total sample of 68 targets, we selected the 43 LIRGs with 10$^{10.2}L_{\sun}\leq$L$_{\rm FIR}\leq$10$^{11.5}L_{\sun}$, which is the FIR luminosity range covered  by the 7 QSO2s studied here.
We then measured the star formation rates (SFRs) of the QSO2s using Eq. 4 in \citet{1998ARA&A..36..189K}, corrected to a Chabrier IMF (i.e., dividing by a factor of 1.59; \citealt{2003PASP..115..763C}). 
SFRs estimated from FIR fluxes are more appropriate for massive galaxies with high dust contents, as is the case for our QSO2s \citep{2003ApJ...586..794B,2020ApJ...899..112S}. In addition, the FIR emission of quasars is less affected by the AGN contribution than the mid-infrared or the radio, and it is not attenuated by dust, as is the case for the ultraviolet and H$\alpha$ fluxes. 
All the QSO2s but the red ETGs have SFRs=$\sim$12--69 M$_{\rm \sun}$yr$^{-1}$, which place them between 0.75 and 1.5 dex above the main sequence (MS) of local SDSS DR7 star-forming galaxies \citep{2016MNRAS.462.1749S}. 
For J1010, J1100, J1356, and J1430, \citet{2020MNRAS.498.1560J} reported SFRs of 35, 34, 84 and 8 M$_{\rm \sun}$~yr$^{-1}$, calculated from the IR luminosity due to star formation, excluding the AGN contribution. These values are in good agreement with ours (30, 34, 69, and 12 M$_{\rm \sun}$~yr$^{-1}$; see Table \ref{tab2}), and thus we are confident that the AGN contribution to the FIR emission of these QSO2s is small. The case of the red ETGs might be different because dust heating from older stars or AGN might be responsible for a non-negligible fraction of the FIR emission \citep{1998ARA&A..36..189K}, which is much lower than in the case of the LIRG QSO2s. For J0232 and J1152 we measure SFRs of 3 and 3.7 M$_{\rm \sun}$yr$^{-1}$, which place them 0.2 and 0.3 dex above the MS. These are high SFRs for red ETGs, and indeed, the fit of the IR SED of J1152 reported in \citet{2019ApJ...873...90S} reveals an important contribution from AGN-heated dust to the FIR, unlike in the cases of J1100, J1356, and J1430.



The black hole masses and Eddington ratios reported in Table \ref{tab1} are from \citet{2018ApJ...859..116K} and were estimated using stellar velocity dispersions measured from SDSS spectra and the M$_{\rm BH}$-$\sigma_*$ relation. The black hole masses range from 10$^{7.5}$ to 10$^{8.6}M_{\sun}$ and the Eddington ratios (f$_{\rm Edd}$) between 0.1 and 1, which are around the median values (10$^{8.2}M_{\sun}$ and 0.2) reported by \citet{2018ApJ...859..116K} from 669 of the QSO2s in \citet{2008AJ....136.2373R}. Thus, our seven QSO2s are near-Eddington to Eddington-limit obscured AGN in the local universe, unlike Seyfert 2 galaxies, which have typical Eddington ratios of f$_{\rm Edd}\sim$0.001--0.1.

\begin{table*}
\caption{Main properties of the ALMA continuum observations.}
\centering
\begin{tabular}{lccccccccccc}
\hline
\hline
ID  	     & $\nu_{\rm obs}$ &  Beamsize    & rms$_{\rm cont}$  &    S$_{\rm cont}$ & Major axis & Minor axis & \multicolumn{2}{c}{200 GHz} & \multicolumn{2}{c}{6 GHz} & $\alpha$ \\
             & 	&  &  &            &  &     & M & PA  & M & PA &  \\
             & (GHz)	& (arcsec$^2$) &  (mJy/beam) &        (mJy)      & (arcsec)  & (arcsec)   & & (deg)  &  & (deg) &  \\
\hline
J0232  & 202.6 & 0.18$\times$0.15  & 0.013   & 0.13$\pm$0.02 &  0.13$\pm$0.05	 &0.09$\pm$0.06	& C &	145$\pm$50  &\dots& \dots & -0.67 \\
J1010  & 203.8 & 0.80$\times$0.69  & 0.037   & 2.62$\pm$0.05 &  0.25$\pm$0.04	 &0.24$\pm$0.06	& C &	41$\pm$69   & C   & 180   & -0.71 \\
J1100  & 203.6 & 0.24$\times$0.20  & 0.017   & 0.57$\pm$0.06 &  0.30$\pm$0.04	 &0.25$\pm$0.04	& C &	72$\pm$76   & C   & 170   & -0.90 \\
J1152 & 221.6 & 0.23$\times$0.15  & 0.023   & 0.15$\pm$0.02 &  0.20$\pm$0.08	 &0.12$\pm$0.08	& A?&	142$\pm$80  &\dots& \dots & -0.65 \\
J1356 & 199.4 & 0.25$\times$0.24  &  0.015  & 1.03$\pm$0.07 &  0.27$\pm$0.03	 &0.20$\pm$0.02	& A &   61$\pm$17   & J   & 20    & -0.79 \\
J1430 & 221.7 & 0.21$\times$0.18  &  0.015  & 0.47$\pm$0.05 &  0.37$\pm$0.05	 &0.23$\pm$0.03	& J &	80$\pm$12   & J   & 60    & -0.55 \\
J1509 & 200.6 & 0.26$\times$0.24  &  0.013  & 0.87$\pm$0.09 &  0.74$\pm$0.10	 &0.55$\pm$0.08	& J &	135$\pm$22  &\dots& \dots & -0.64 \\ 
\hline	   					 			    								      
\end{tabular}
\tablefoot{Columns 2 and 3 correspond to the observed frequency ($\sim$200 GHz for band 5 and $\sim$220 GHz for band 6 observations) and beamsize of the observations. Columns 4 and 5 list the rms of the continuum maps and continuum flux density. Columns from 6 to 9 are the size, morphology (C=compact, A=asymmetric, and J=jet-like; see Sections \ref{individual} and \ref{continuum}) and position angle estimated from a 2D Gaussian fit of the corresponding maps, deconvolved from the beamsize. The last three columns correspond to the morphology and PA derived from VLA data of similar angular resolution at 6 GHz from \citet{2019MNRAS.485.2710J}, and to the spectral index calculated as described in Section \ref{continuum}.}
\label{tab3}
\end{table*}

The influence of radio jets on the gas properties of radio-quiet AGN is another open question that requires further investigation (e.g., \citealt{2014MNRAS.440.3202V,2021arXiv210306805V,2019MNRAS.485.2710J,2021MNRAS.tmp..644J}). Our targets are radio-quiet in terms of their L$_{\rm 1.4 GHz}$/L$_{\rm [O III]}$ values \citep{1999AJ....118.1169X}, but they span a wide range of radio luminosities (log L$_{\rm 1.4 GHz}$=22.7--24.4 W~Hz$^{-1}$; see Figure \ref{fig2} and Table \ref{tab1}). These values are representative of the QSOFEED sample, for which we measure an average luminosity of log L$_{\rm 1.4 GHz}$=23.4$\pm$0.7 W~Hz$^{-1}$. These radio luminosities are intermediate between those of Seyfert galaxies (log L$_{\rm 1.4 GHz}<$22.5 W~Hz$^{-1}$) and luminous high-excitation radio galaxies (HERGs; log L$_{\rm 1.4 GHz}>$25.0 W~Hz$^{-1}$). We note that the red ETGs have the lowest radio luminosities in our sample (log L$_{\rm 1.4 GHz}$=22.7--23.0 W~Hz$^{-1}$), whilst the other QSO2s have log L$_{\rm 1.4 GHz}$=23.7--24.4 W~Hz$^{-1}$ (see Figure \ref{fig2}). 
The 7 QSO2s are well above the radio-FIR correlation of star-forming galaxies \citep{2003ApJ...586..794B}, indicating an excess of radio emission unrelated to star-formation (see Figure \ref{fig2}). In fact, according to \citet{2019MNRAS.485.2710J}, in J1010, J1100, J1356, and J1430, star formation accounts for $\leq$10\% of the radio emission. This, together with the steep radio spectra and the radio morphologies measured from 1--7 GHz VLA data, led them to conclude that jets are responsible for the high radio luminosities derived for these QSO2s.

The 7 QSO2s also cover a wide range of ionized outflow properties, as can be seen from Table \ref{tab4a}. The red ETGs, J0232 and J1152, have slower and less turbulent outflows of ionized gas than the other QSO2s. This is not surprising considering the well-known connection between radio power and ionized gas kinematics \citep{2013MNRAS.433..622M,2014MNRAS.442..784Z}. AGN with log L$_{\rm 1.4 GHz}>$ 23 W~Hz$^{-1}$ are five times more likely to have [O III] line profiles with FWHM$>$1000 km~s$^{-1}$ than AGN with lower L$_{\rm 1.4 GHz}$, with the FWHM peaking at log L$_{\rm 1.4 GHz}\sim$24 W~Hz$^{-1}$ \citep{2013MNRAS.433..622M}. Using the SDSS spectra that are available for all the QSO2s, we measure an average FWHM=1100$\pm$450 km~s$^{-1}$ for the broadest component of the [OIII] lines. We find FWHMs ranging from 370 to 2300 km~s$^{-1}$. By performing a non-parametric analysis, we measure an average W80 (i.e., the width that contains 80\% of the line flux) of 780$\pm$310 km~s$^{-1}$, ranging from 345 to 1800 km~s$^{-1}$. Thus, the emission line kinematics of the 7 QSO2s studied here are representative of the QSOFEED sample.

\section{ALMA observations}
\label{observations}

We targeted the $^{12}$CO(2$-$1) emission line (rest frequency 230.538 GHz) and its underlying continuum emission (rest frequency 220--240 GHz, which corresponds to $\lambda\sim$1.2--1.3 mm) in the 7 QSO2s with ALMA during Cycle 6 (project 2018.1.00870.S; PI: C. Ramos Almeida). 
Observations were performed between October 2018 and August 2019 using the C43-3 and C46-6 antenna configurations, with maximum baseline lengths of 0.5 km and 2.5 km. The typical on-source times were $\sim$0.3 hours and $\sim$0.6 hours per source, respectively.
Band 5 or Band 6 receivers were selected depending on the redshift of each galaxy, providing us with four spectral windows of 1.875 GHz width and a spectral resolution of $\sim$10 km s$^{-1}$. One spectral window was centered on the CO(2$-$1) expected frequency, while the others accounted for continuum emission. We used a single pointing with a field-of-view (FOV) of 26.3\arcsec~in the case of the Band 6 observations, and 28.7--29.3\arcsec~for the Band 5 observations.

We created datacubes without spectral averaging, i.e., keeping the native spectral resolution of $\sim$10 km~s$^{-1}$.
The visibilities from the two configurations, which were calibrated using the CASA 5.4.0 software \citep{2007ASPC..376..127M} in the pipeline mode, were combined for all sources except for J1010, which was observed in the compact configuration only. Moreover, visibilities from all spectral windows were averaged to measure the millimeter continuum emission, by excluding the spectral region associated with the CO(2$-$1) emission line (|v-v$_{\rm sys}$|$<$500 km~s$^{-1}$ for J1100, J1010 and J1509, and |v-v$_{\rm sys}$|$<$600 km~s$^{-1}$ for the other QSO2s). To model the continuum emission next to the line, we fitted a first order polynomial in the uv plane to channels with velocities 500(600)$<$|v-v$_{\rm sys}$|$<$2000 km~s$^{-1}$. The subtraction of this fit provided us with continuum-subtracted visibilities. We thus produced continuum-subtracted data-cubes by using the {\it tclean} CASA task in non-interactive mode, using natural weighting. The Hogbom cleaning algorithm was used in combination with a threshold of three times the rms sensitivity. Similarly, imaging of continuum maps was performed. The resulting rms sensitivity and beamsize of our QSO2 observations are listed in Table \ref{tab3}. In addition to the rms noise, there is a 10\% flux calibration error.  
We note that the combination of the C43-3 and C43-6 configurations allowed us to reach angular resolutions of $\sim$0.18--0.25\arcsec~(300--500 pc at z$\sim$0.1), while recovering extended emission on scales of up to $\sim$6.5--8.6\arcsec~(12--16 kpc). Such a combination was fundamental to recover the total CO emission of the QSO2s. 
As shown in Section \ref{molecular}, four of the QSO2s have APEX CO(2--1) luminosities measured in an aperture of $\sim$28\arcsec~\citep{2020MNRAS.498.1560J}, which we have compared with our total CO fluxes (see Section \ref{integrated}). 

In the case of J0232, the observations in the extended configuration were classified as semi-pass, and the data reduction was manually optimized by the Observatory. Specifically, the spectral index of the phase calibrator was identified from closely paired (within 1-2 days) Band 3 and Band 7 measurements as close as possible in time to the actual observing date ($\sim$20 days). The flux of the calibrator was then extrapolated in time from the two Band 3 measurements made before and after the observations. Using this value, of 0.475 Jy at 97.475 GHz, along with the spectral index (-0.45) we established the flux scaling. These values were used as input for the {\it gfluxscale} calibration step.

\section{Methodology}
\label{methodology}

\begin{figure}
\includegraphics[width=0.48\textwidth]{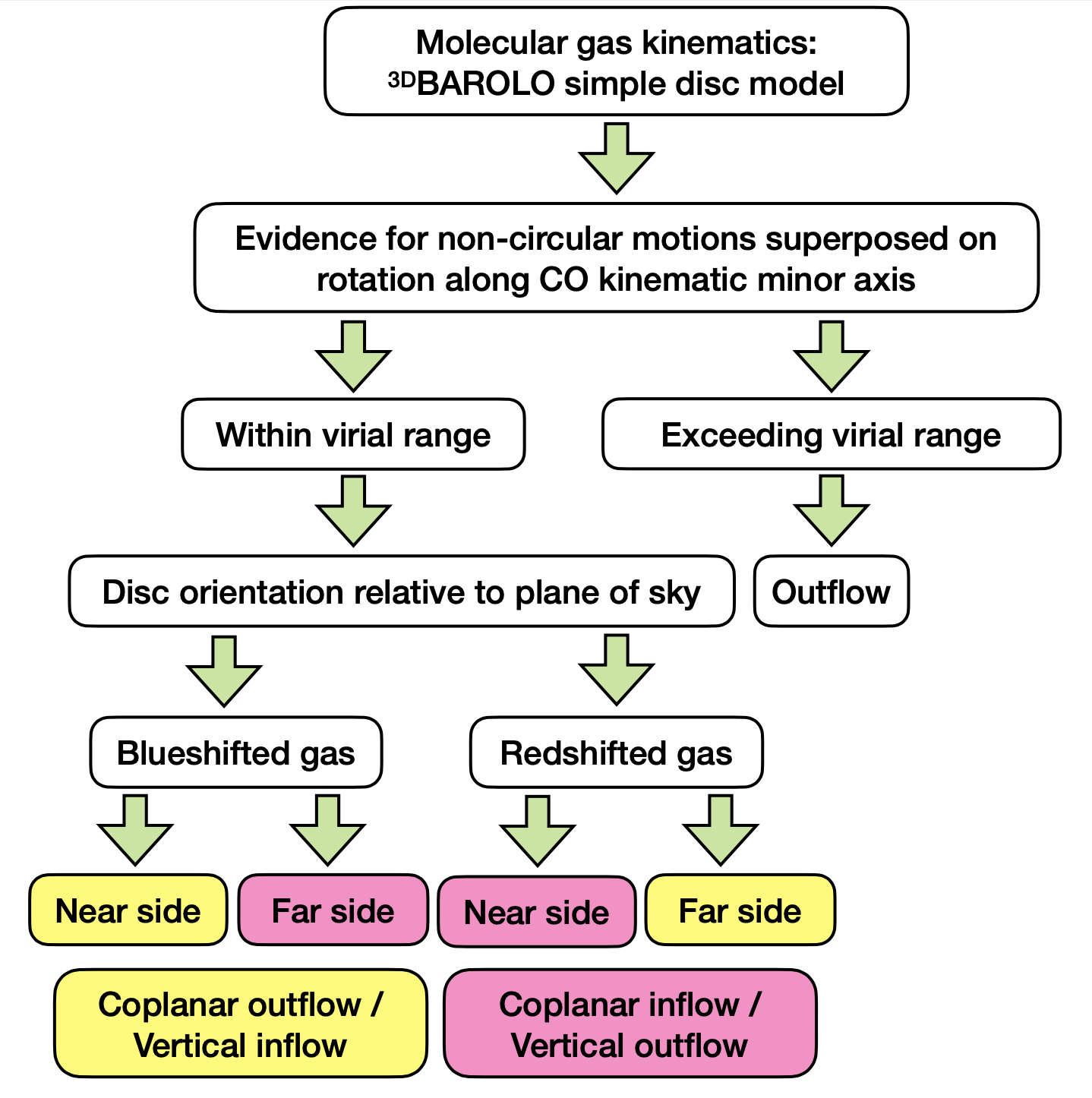}
\caption{Flowchart of the methodology used here for interpreting the molecular gas kinematics of the QSO2s. We note that the coplanar outflow/vertical inflow and coplanar inflow/vertical outflow degeneracy only holds if the vertical motions are spatially extended. Otherwise, line splitting would be detected.}  
\label{kinematics}
\end{figure}

\begin{figure*}
\centering
\includegraphics[width=1.4\columnwidth]{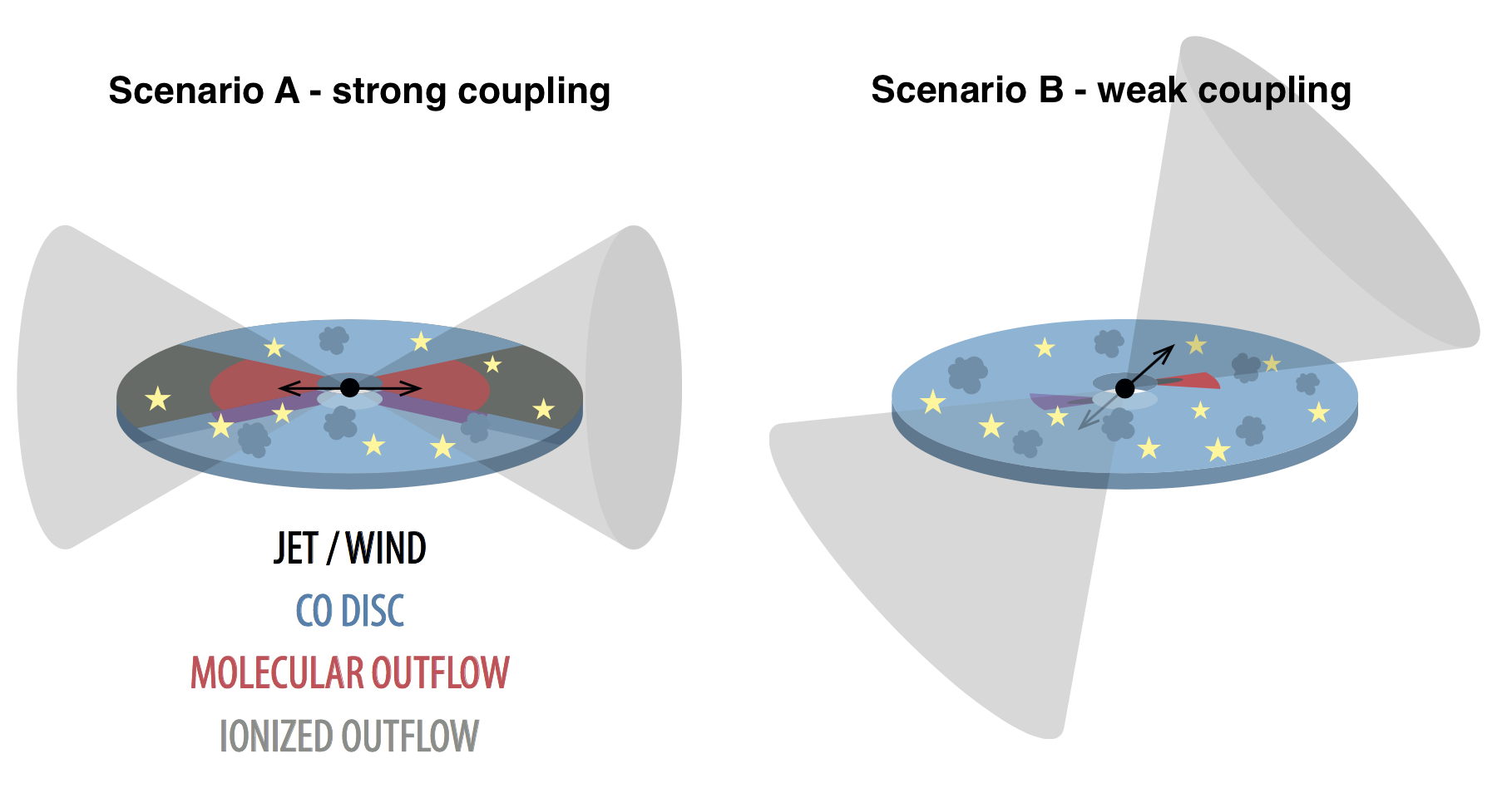}
\caption{AGN wind/jet/ionized outflow geometries considered in this work and the coplanar molecular outflows that they produce. The left panel corresponds to an example of strong coupling (scenario A), in which the wind/jet/ionized outflow are coplanar with the disk, launching a more massive molecular outflow. The right panel corresponds to an example of weak coupling (scenario B), in which the wind/jet/ionized outflow subtends a certain angle from the CO disk, having a small intersection region (shown in dark gray) and launching a more modest molecular outflow.} 
\label{scheme}
\end{figure*}

In order to interpret the molecular gas kinematics, which we describe in detail in Section \ref{individual}, we attempted to model them with a simple disk model using $^{\rm 3D}$BAROLO \citep{2015MNRAS.451.3021D}. The aim of this modeling is to help us identify noncircular motions that could be associated with inflows or outflows (see Fig. \ref{kinematics} and e.g., \citealt{2019MNRAS.489.1927S,2020arXiv200305663D}). 
The fits were done following the methodology described in \citet{2018ApJ...859..144A} and \citet{2020arXiv200305663D}. We first ran the model for each galaxy by fixing the kinematic center to the coordinates of the 1.3 mm continuum and the disk scale height to the default thin disk approximation describing a CO disk whose vertical structure is mostly unresolved at our spatial resolution.
We allowed the systemic and rotation velocities, velocity dispersion, disk inclination and PA to vary and used uniform weighting. We then ran the model again fixing the systemic velocity, inclination and PA to the average values derived from the first run, so only rotation velocity and velocity dispersion are allowed to vary. 
By subtracting the model velocity map from the observed mean velocity field, we obtained mean-velocity residual maps that we use to investigate deviations from circular motions. We did the same with the velocity dispersion maps. 


Finally, we produced position-velocity (PV) diagrams along the kinematic minor and major axis of the CO distribution using $^{\rm 3D}$BAROLO. These diagrams were extracted using a slit width of approximately the beam size. The PV diagrams along the minor axis are better suited for studying noncircular motions because along this axis the only gas motions with non-null projection in the plane of the galaxy are inward or outward radial motions (i.e., inflows or outflows). On the other hand, PV diagrams along the major axis serve to further characterize gas rotation. In the case of pure rotation, the PV diagram along the minor axis would just show emission around the systemic velocity with a width of tens of km~s$^{-1}$. This width is a combination of beam smearing and cloud-cloud velocity dispersion \citep{2014A&A...567A.125G}. 

In case of detecting radial motions along the minor axis, they can either be within the virial range (i.e., on the order of the observed circular motions) or exceed it. In the latter case we would be most likely witnessing outflowing molecular gas, as the velocities would be too large for an inflow. On the other hand, if the radial motions are on the order of the rotational velocities, we cannot assign them unambiguously to purely radial outflows without a further careful scrutiny of the gas kinematics. In that case, to interpret the CO(2$-$1) velocity residuals and PV diagrams we need to know the PA and orientation of the galaxies relative to the plane of the sky (i.e., to determine the near and far sides using optical and/or NIR data). This can be inferred from analysis of the morphology of the gas response (in terms of trailing spirals or leading edges of the bar), and/or observed colors/dust extinction. Fortunately, the majority of the QSO2s have been observed with the Hubble Space Telescope (HST) and/or have optical/NIR integral field observations that we used for gathering this information.

In the case of barred spirals such as J1100 and J1509, we also need to know the bar sizes and derive the best guesses for the position of their corotation resonances. The primary bar regulates the distribution and kinematics of the molecular gas on scales going from hundreds of pc (the gas follows the so-called x$_2$ orbits) to several kpc (x$_1$ orbits). We need to estimate the extent of the corotation region of the bar to disentangle whether the molecular gas motions being studied are inside/outside of this corotation region. The corotation radius can be estimated from the bar radius as R$_{\rm cr}\sim1.2\pm0.2\times R_{\rm bar}$ \citep{1992MNRAS.259..345A}. 

Once we determine which are the far and near sides of the disks, 
we can easily interpret our molecular gas observation in terms of radial motions. These radial motions will correspond to inflows or outflows depending on whether they are coplanar or vertical with the CO disks. In Figure \ref{kinematics} we summarize the methodology described above.

\begin{figure*}
\centering
\includegraphics[width=0.4\textwidth]{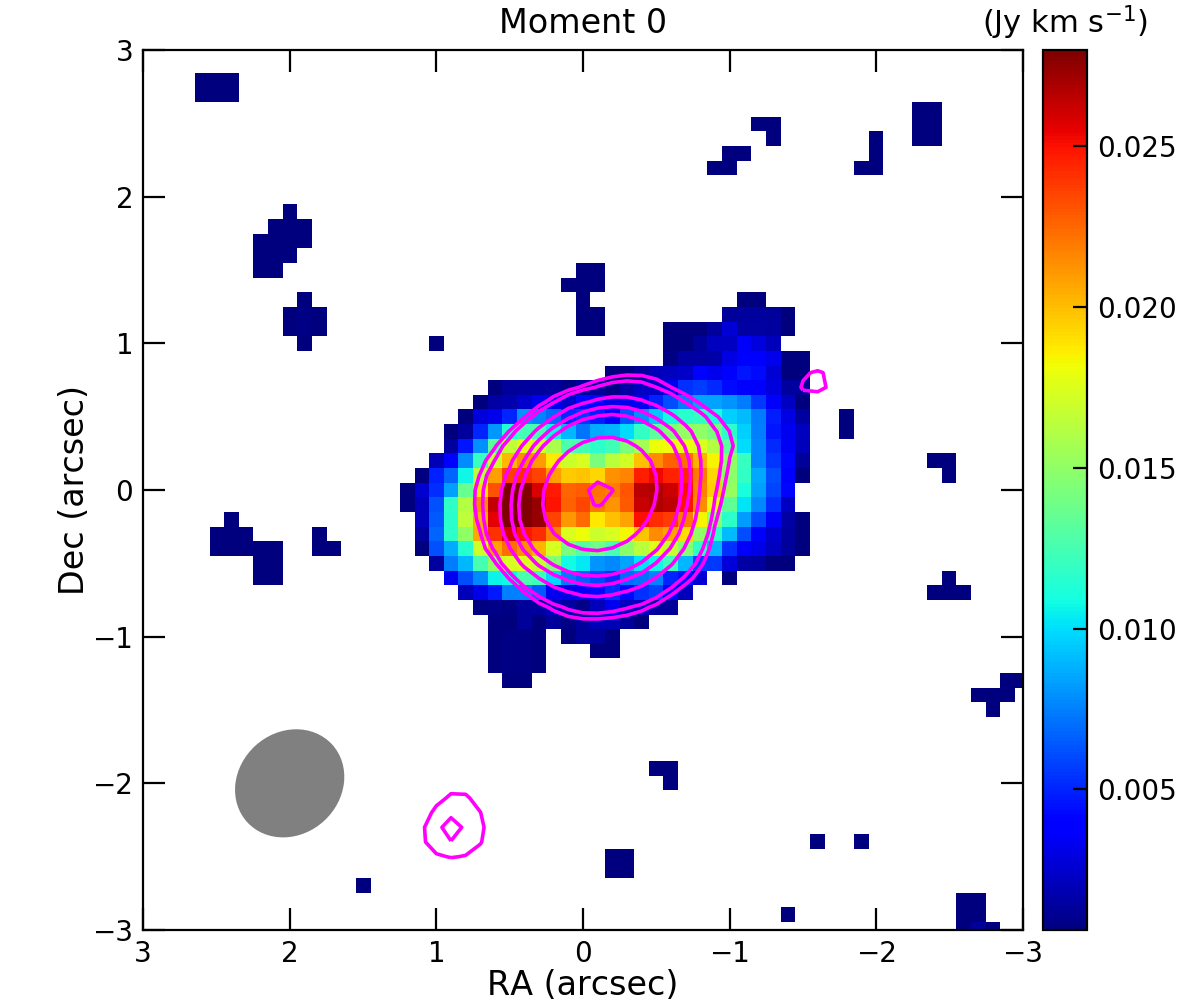}
\includegraphics[width=1.0\textwidth]{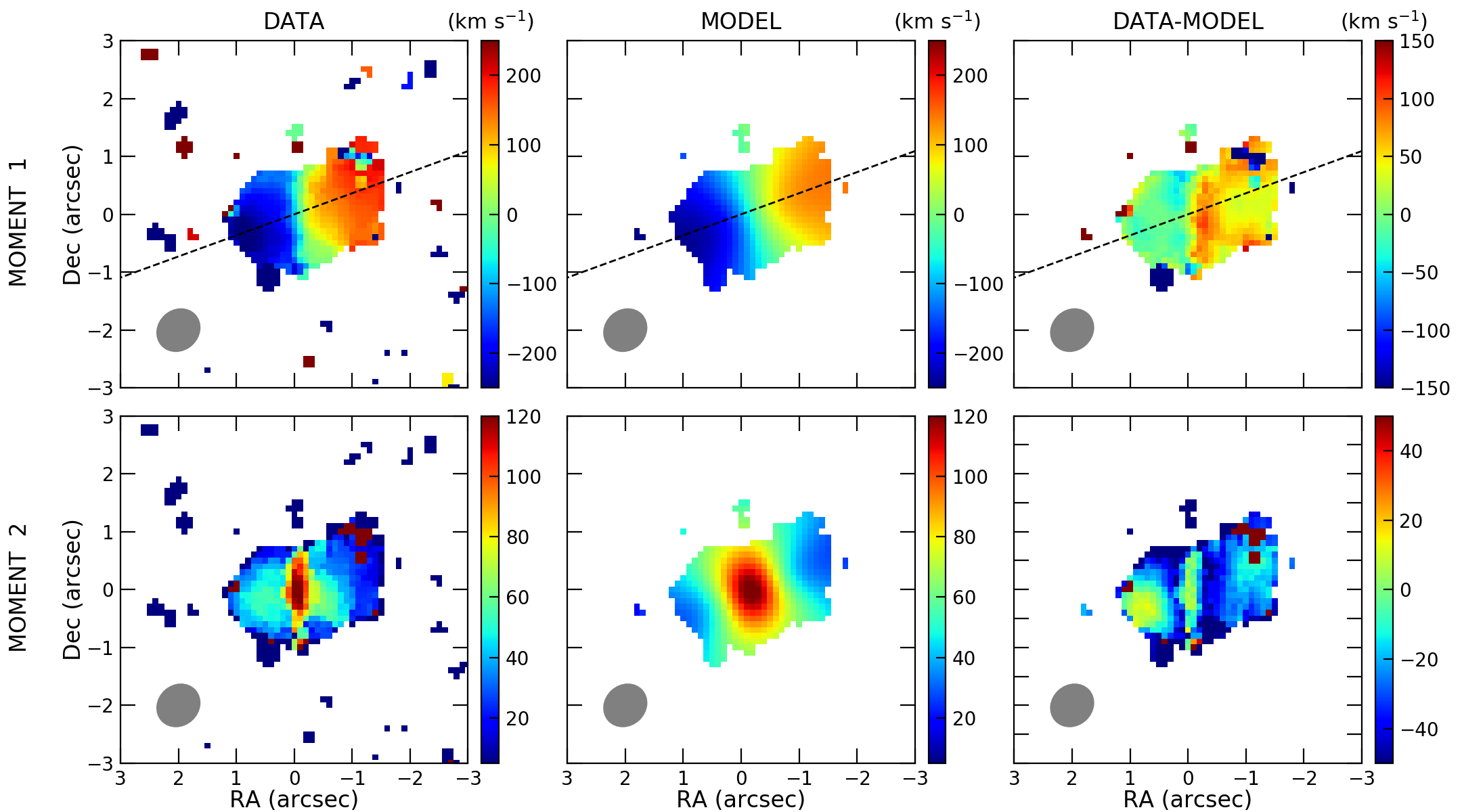}
\caption{CO(2$-$1) moment 0, 1, and 2 maps (i.e., flux, velocity, and velocity dispersion) of J1010. The $^{\rm 3D}$BAROLO models of the moment 1 and 2 maps and corresponding residuals are also shown. Continuum contours starting at 3$\sigma$ are shown in pink in the moment 0 map ($\sigma$=0.037 mJy/beam), and the beam size (0.80\arcsec$\times$0.69\arcsec) is shown in the bottom left corner of each panel. The major axis of the CO kinematics is shown as a black dashed line in the moment 1 maps. North is up and east to the left.}  
\label{fig2a}
\end{figure*}

In the case of quasars, in principle we expect any 
molecular outflow to be mostly coplanar with the CO disk (see Figure \ref{scheme}). This is supported by previous observational evidence in the case of spatially extended AGN-driven molecular outflows, for example NGC\,1068 \citep{2014A&A...567A.125G,2019A&A...632A..61G}, NGC\,3227 \citep{2019A&A...628A..65A}, IC\,5063 \citep{2015A&A...580A...1M,2018MNRAS.476...80M}, NGC\,5643 \citep{2018ApJ...859..144A,2021A&A...645A..21G}, NGC\,4388, NGC\,5506, and NGC\,7582 \citep{2021arXiv210410227G}. As shown in Figure \ref{scheme}, the AGN wind, jet and/or ionized outflow produces a radial expansion in the CO disk (i.e., a molecular outflow), which will be more or less massive depending on its orientation relative to the CO disk (strong and weak coupling, scenarios A and B in Figure \ref{scheme}). As we know from detailed studies of nearby AGN such as NGC\,1068, some of the molecular gas is forced to leave the plane of the galaxy and adopt a rather shell-like geometry (see \citealt{2019A&A...632A..61G}), but in order to interpret the CO kinematics here we consider that the bulk of the molecular outflows are either coplanar or vertical with the CO disk. 


To derive outflow mass rates (\.M$_{\rm out}$) we need to estimate outflow masses (M$_{\rm out}$) and assume a certain outflow geometry. The outflow masses were calculated by integrating the CO(2$-$1) emission along the minor axis and within the regions and velocities indicated in the PV diagrams shown in Section \ref{individual}. These outflow regions and velocities were selected from inspection of the PV diagrams along the minor axis and the velocity residual maps (see Section \ref{individual} for details). 
We assumed that the CO emission is thermalized and optically thick, so R$_{21}$=CO(2$-$1)/CO(1$-$0)=1 \citep{1992A&A...264..433B,2005ARA&A..43..677S}, and $\alpha_{\rm CO}$=0.8$\pm$0.5 M$_{\sun}(\rm K~km~s^{-1}~pc^2)^{-1}$ \citep{1998ApJ...507..615D}. The latter is more appropriate for outflowing molecular gas than the commonly assumed Galactic factor (see e.g., \citealt{2015A&A...580A...1M}). 

We then assume a time-averaged thin expelled shell geometry for deriving outflow mass rates:
\begin{equation}
    \dot{M}_{\rm out} = v_{\rm out} \times \frac{M_{\rm out}}{r_{\rm out}}\times {\rm tan (\alpha).}
    \label{eq1}
\end{equation}
This geometry is commonly adopted for molecular outflow calculations in the local universe (e.g., \citealt{2019A&A...632A..33A,2019MNRAS.483.4586F,2020A&A...633A.134L}), and it is more conservative than the multi-conical outflow geometry uniformly filled by outflowing clouds. This assumption for the outflow geometry implies a constant mass rate since the outflow started \citep{2020A&A...633A.134L}. In Equation \ref{eq1}, M$_{\rm out}$, v$_{\rm out}$ and r$_{\rm out}$ are the outflow mass, velocity and radius, and $\alpha$ is the angle between the molecular outflow and the line of sight. In the case of a coplanar molecular outflow, tan($\alpha$)=tan(90-i)=1/tan(i), where i is the inclination angle of the CO disk. Thus, Equation \ref{eq1} corresponds to the deprojected mass outflow rate, and the tan($\alpha$) factor results from deprojecting the outflow velocity, v$_{\rm out}$/cos($\alpha$), and the outflow radius, r$_{\rm out}$/sin($\alpha$).



\section{Results}
\label{results}

\subsection{Individual galaxies}
\label{individual}

We detect continuum emission (rest-frame $\lambda\sim$1.2--1.3 mm) at $>$3$\sigma$ in the seven QSO2s observed with ALMA, and CO(2$-$1) emission in five of them. The CO moment maps were generated by integrating over the spectral range where line emission is detected above 3$\sigma$ in the continuum-subtracted cubes. The continuum emission is shown as images for the QSO2s without CO detection (J0232 and J1152, for which the results are presented in Appendices \ref{appendixa} and \ref{appendixb}). For the other five QSO2s, it is shown as contours starting from 3$\sigma$ overlaid on the corresponding CO moment 0 maps.
To interpret the molecular gas kinematics, we followed the procedure described in Section \ref{methodology}. The regions with outflow motions have been identified in the corresponding PV diagrams.

\subsubsection{SDSS J101043.36+061201.4 (J1010)}
\label{J1010}

\begin{figure}
\includegraphics[width=1\columnwidth]{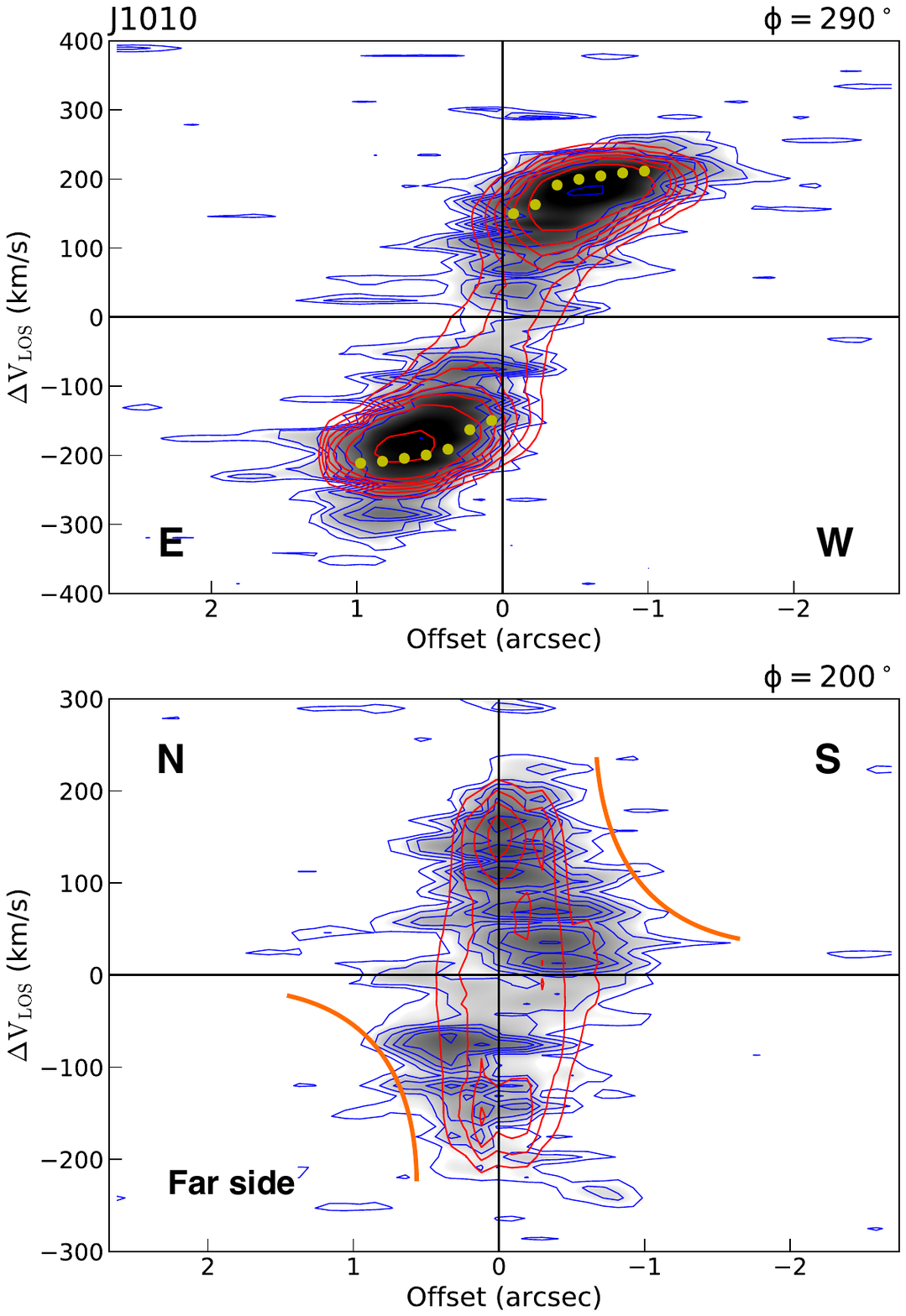}
\caption{PV diagrams along the CO kinematic major axis (PA=290\degr=-70\degr) and minor axis (PA=200\degr=20\degr) of J1010 extracted with $^{\rm 3D}$BAROLO. We used a slit width of $\sim$0.8\arcsec. Blue contours correspond to the observed CO(2$-$1) emission above 2$\sigma$ and red contours to the $^{\rm 3D}$BAROLO rotating disk model (inclination of i=36\degr)~shown in Figure \ref{fig2a}. Yellow dots are the average model velocities at different radii. Blue contours outside the boundaries of the model correspond to noncircular motions. Orange lines indicate the positive/negative pattern that, in the case of this QSO2, might be produced by either a coplanar inflow or a spatially extended vertical outflow.} 
\label{fig2b}
\end{figure}

The SDSS optical image of this QSO2 resembles an ETG morphology (S0a; \citealt{2013MNRAS.435.2835W}) clearly interacting with a small galaxy at $\sim$7\arcsec~(13 kpc) SW that it is also detected in CO (see Appendix \ref{appendixc}). The SFR estimated from the IR luminosity, of 30 M$_{\rm \sun}$~yr$^{-1}$, is comparable to those of the spiral galaxies J1100 and J1509, placing J1010 1.15 dex above the MS (see Section \ref{sample}). This is likely a result of the interaction with the small companion. Despite the high SFR, J1010 is well above the radio-IR correlation of star-forming galaxies (see Figure \ref{fig2}). \citet{2019MNRAS.485.2710J} estimated that only 2.6\% of the FIRST 1.4 GHz luminosity of J1010 can be accounted for by star formation. High angular resolution VLA data ($\sim$0.25\arcsec) at 6 GHz shows a compact and rounded morphology with a deconvolved major axis of $\sim$200 pc and PA$\sim$180\degr~\citep{2019MNRAS.485.2710J}. The ALMA 1.3 mm continuum image shows the same morphology (see pink contours in Figure \ref{fig2a}), with a deconvolved size of 0.25\arcsec$\times$0.24\arcsec~(450$\times$434 pc$^2$) and PA=41$\pm$69\degr. The position of the AGN is derived from the peak of this continuum emission.

The CO(2$-$1) moment maps of J1010 are shown in Figure \ref{fig2a}. The maximum radius of the CO emission is R$_{\rm CO}$=1.3\arcsec~(2.4 kpc). The moment 0 map reveals a double-peaked structure, with a separation of $\sim$0.7\arcsec~(1.25 kpc) between peaks in the E-W direction. The maximum of the 1.3 mm continuum lies in the middle of the two CO peaks. This morphology could be explained by a highly inclined disk with a ring-like in-plane distribution of molecular gas, but our $^{\rm 3D}$BAROLO modeling of the data is not compatible with such disk orientation (see below). Alternatively, this peculiar CO morphology could be the result of AGN feedback, either produced by molecular gas removal in the N-S direction, or by CO being excited to higher levels \citep{2019ApJ...875L...8R}. 
There is no optical counterpart to this double-peaked morphology. An optical image generated from a publicly available VLT/MUSE data cube with the same angular resolution as our ALMA data (FWHM$\sim$0.79\arcsec; PI: G. Venturi) shows a single nucleus coincident with the peak of the 1.3 mm continuum.

\paragraph{Kinematics.}

The kinematics of the ionized gas were studied by \citet{2014MNRAS.441.3306H} using Gemini/GMOS in integral field mode. They reported the presence of a compact rotating [OIII] disk with major axis PA$\sim$299\degr = -61\degr~in addition to broad [OIII] wings of FWHM$\sim$1000--1200 km~s$^{-1}$ blueshifted by -100 km~s$^{-1}$ (see also \citealt{2014MNRAS.440.3202V}). 
Using the MUSE datacube mentioned before, we detect broader and more intense [O III] emission lines to the NW, as well as maximum blueshifts. 
We also generated two continuum images centered at 6200 and 8500 \AA~to obtain a color map (see Appendix \ref{appendixc}). In both images we detect a dust lane SW from the nucleus, which produces redder colors. According to this, the N would be the far side of the galaxy. Thus, for the approaching side of the ionized outflow to be detected in the NW, it has to subtend a large angle relative to the galaxy/CO disk (scenario B in Figure \ref{scheme}). 




The CO velocity field shown in Figure \ref{fig2a} resembles rotation, blueshifted to the E and redshifted to the W. Indeed, this velocity map can be reproduced with a rotating disk of major axis PA$\sim$290\degr = -70\degr, in good agreement with the [OIII] kinematic major axis reported by \citet{2014MNRAS.441.3306H} and with the galaxy major axis (-67\degr; see Table \ref{tab2}). This is indicative of rotation-dominated gas kinematics. The inclination of the rotating disk model fitted to the CO data is i$\sim$36\degr, also consistent with the galaxy inclination (35$\degr$; see Table \ref{tab2}). Thus, despite the tidal interaction evidenced by the optical image, the optical and CO disks share similar kinematics. 
The velocity dispersion map peaks at the position of the nucleus, showing values of up to 120 km~s$^{-1}$. 
The $^{\rm 3D}$BAROLO model accounts to some extent for the large line width feature identified close to the minor axis region resulting from the beam smearing of the motions attributable to circular rotation and turbulence. This is better visualized in the modeled minor axis PV diagram shown in Figure \ref{fig2b}. 



In addition to ordered rotation along the major axis (v$_{\rm rot}\sim$200 km~s$^{-1}$; see top panel of Figure \ref{fig2b}), 
the PV diagram along the kinematic minor axis (PA$\sim$200\degr = 20\degr) reveals noncircular motions in the central 1.6\arcsec~(2.9 kpc) of J1010, mostly blueshifted to the N and redshifted to the S (see bottom panel of Figure \ref{fig2b}). 
This can be also seen in the residual velocity map shown in Figure \ref{fig2a}.  

Thus, in the case of this QSO2, for which the N is the far side, we are witnessing either a coplanar inflow or a vertical outflow (see Figure \ref{kinematics}). In the latter case, the outflow could be star-formation driven in the case of this interacting galaxy with a high SFR. An outflow could explain the double-peaked CO morphology shown in the moment 0 map. However, if this is the case, the vertical outflow must be spatially extended, as otherwise we would expect significant line splitting, something that we do not see in the PV diagrams.
Due to the limited angular resolution of the ALMA observations of this QSO2, we cannot fully resolve these radial motions. Higher angular resolution data are required to further investigate this interesting system.


\begin{figure*}
\centering
\includegraphics[width = 0.4\textwidth]{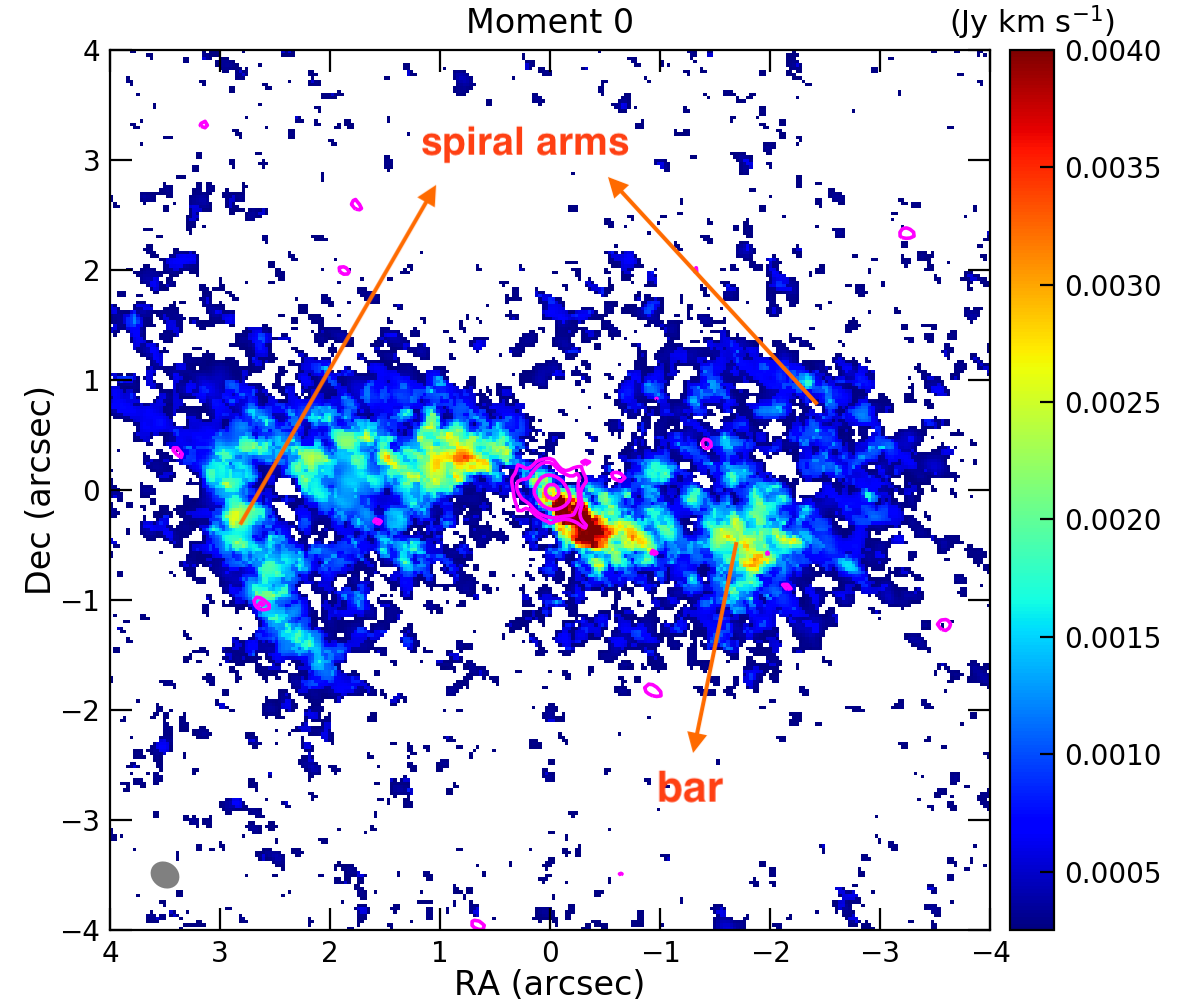}
\includegraphics[width = 1\textwidth]{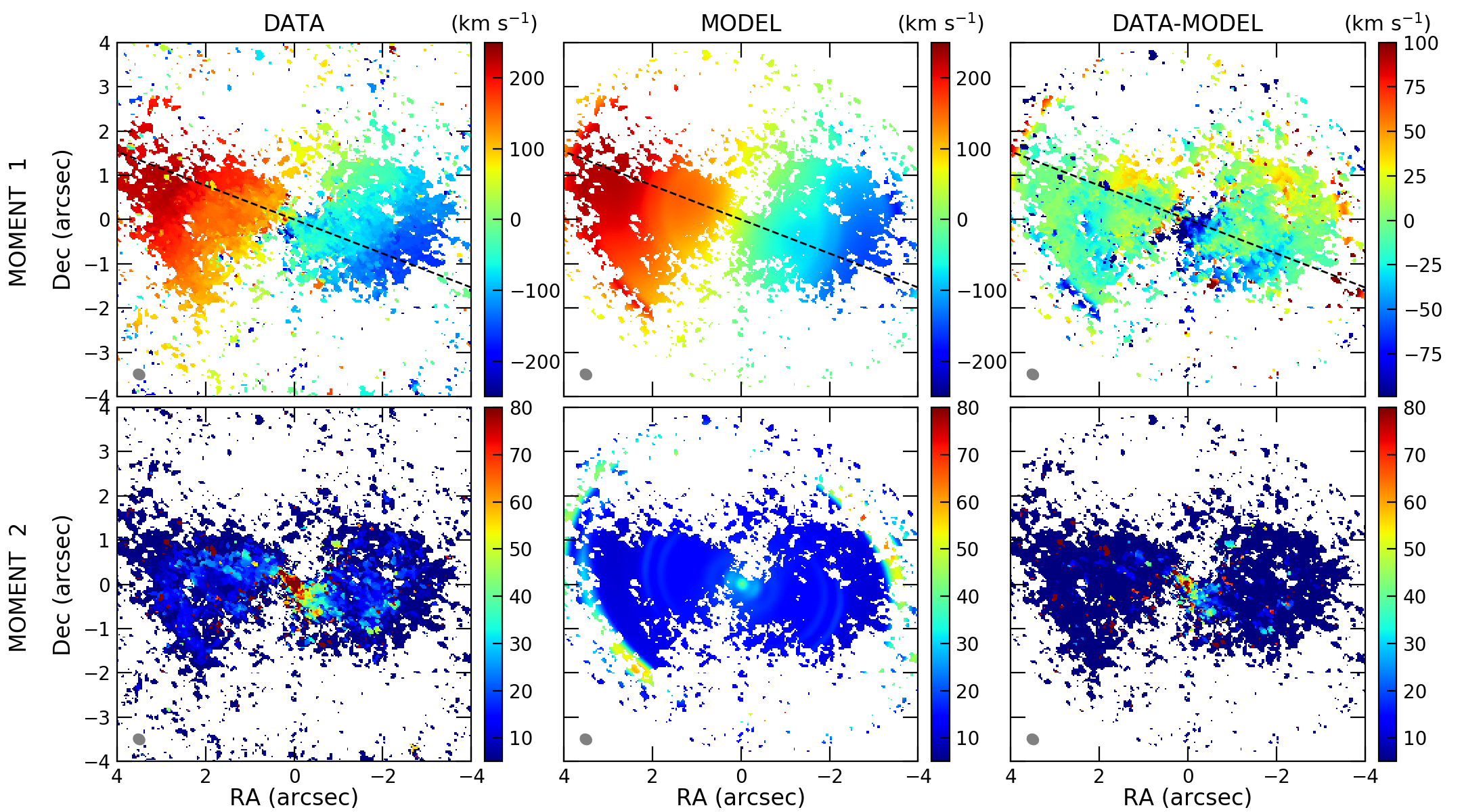}
\caption{Same as in Fig. \ref{fig2a} but for J1100. Continuum contours starting at 3$\sigma$ are shown in pink in the moment 0 map ($\sigma$=0.017 mJy/beam), and beam size is 0.24\arcsec$\times$0.20\arcsec.}
\label{fig3aa}
\end{figure*}

\begin{figure}
\includegraphics[width=1\columnwidth]{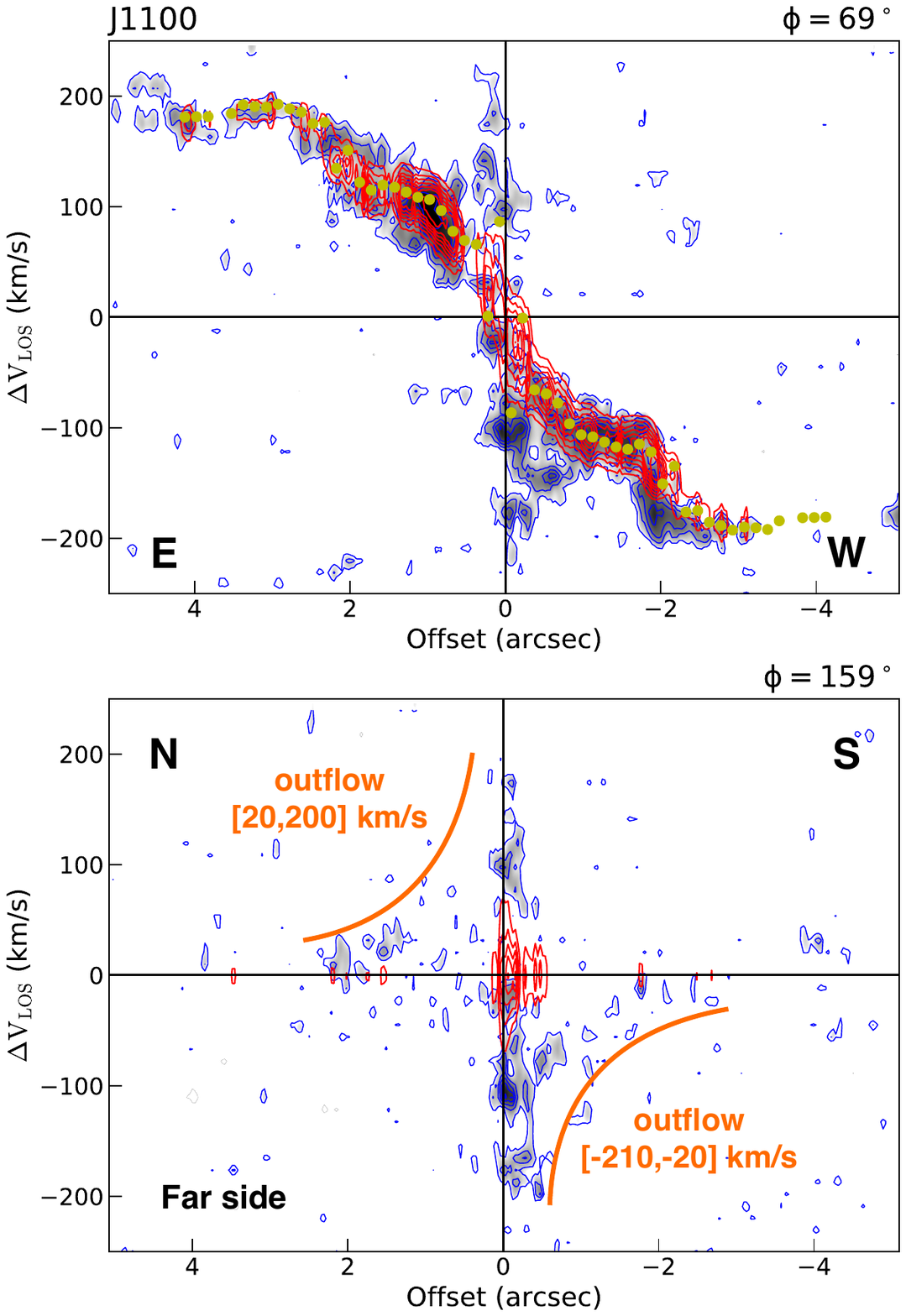}
\caption{Same as in Fig. \ref{fig2b} but for J1100. The CO kinematic major axis is PA=69\degr, the minor axis PA$\sim$159\degr=-21\degr~and the inclination i=38\degr. We used a slit width of $\sim$0.2\arcsec.}
\label{fig3cc}
\end{figure}

\subsubsection{SDSS J110012.39+084616.3 (J1100)}
\label{J1100}

J1100 has a SFR of 34 M$_{\sun}$~yr$^{-1}$ and it shows a clear radio-excess in Figure \ref{fig2}, with only $\sim$4\% of the 1.4 GHz luminosity being due to star formation as estimated by \citet{2019MNRAS.485.2710J}. VLA 6 GHz images of this QSO2 at the same angular resolution as our ALMA data ($\sim$0.2\arcsec) reveal a compact and rounded morphology, very similar to J1010, with a linear size of 800 pc and PA$\sim$170\degr~\citep{2019MNRAS.485.2710J}. Lower resolution VLA data do not show evidence for extended emission either. The ALMA 1.3 mm continuum emission of J1100 is shown as pink contours in the moment 0 map in Figure \ref{fig3aa}. It appears very compact at 3$\sigma$, with the emission peak indicating the AGN position. We measure a deconvolved size of 0.30\arcsec$\times$0.25\arcsec~(555$\times$463 pc$^2$) with PA=72$\pm$76\degr. 

The QSO2 host is a large and moderately inclined (i=38\degr) barred galaxy (SBb; \citealt{2013MNRAS.435.2835W}). The optical continuum SDSS and HST/ACS images available for this QSO2 show strong and wound spiral arms, and an undisturbed morphology \citep{2018ApJ...856..102F}. The [OIII] emission appears very compact in the HST/ACS images, nearly circular (1.1\arcsec~radius $\simeq$2 kpc), and showing a small tail to the SE. 
The ALMA CO(2$-$1) moment maps shown in Figure \ref{fig3aa} reveal a large disk of molecular gas 
with R$_{\rm CO}$=2.7\arcsec~($\sim$5 kpc). The CO morphology follows the stellar bar and the inner part of the spiral arms seen in the HST/ACS optical continuum image. From this image we estimated a bar radius of R$_{\rm bar}\sim$2.5\arcsec~(4.5 kpc) in the E-W direction (PA$\sim$85\degr). In the CO maps, from the west edge of the bar we see the beginning of one of the spiral arms, which shows negative velocities. From the eastern edge of the bar we see another spiral arm showing positive velocities. 


\paragraph{Kinematics.} The [OIII] kinematics are dominated by rotation, with a kinematic major axis PA$\sim$74\degr~\citep{2014MNRAS.441.3306H}.
Based on HST/STIS spectroscopy and using a slit orientation of -19\degr~(almost coincident with the galaxy minor axis; see Table \ref{tab2}), \citet{2018ApJ...856..102F} reported the presence of an [OIII] outflow of FWHM$\sim$1780 km~s$^{-1}$, velocities of up to -350 km~s$^{-1}$ and projected radius of $\sim$460 pc. The bulk of this outflowing gas is found toward the SE, in agreement with the Gemini/GMOS analysis of the [OIII] kinematics presented in \citet{2014MNRAS.441.3306H}.

The $^{\rm 3D}$BAROLO model of the moment 1 and 2 maps and corresponding residuals are shown in the middle and right panels of Figure \ref{fig3aa}. In the case of this QSO2 we fixed the inclination to match that of the galaxy (i$\sim$38\degr) to prevent the fitted model from having an unrealistic orientation of 70\degr. This happens because $^{\rm 3D}$BAROLO tries to fit the most conspicuous emission, which resembles an apparently more inclined disk. The kinematic major axis of the model (PA$\sim$69\degr) is very similar to the major axis of the galaxy (PA$\sim$67\degr; see Table \ref{tab2}) and of the [OIII] gas (PA$\sim$74\degr), consistent with rotation-dominated gas kinematics. 

As explained in Section \ref{methodology}, the bar regulates the CO distribution and kinematics within the corotation radius. In the case of J1100, R$_{\rm CR}\sim(1.2\pm0.2)\times R_{\rm bar}\sim$3\arcsec~(5.5 kpc). The morphology of the molecular gas within this radius is in agreement with the canonical gas response to a stellar bar with a radially extended inner Lindblad resonance (ILR) region. The gas appears to concentrate along the two leading edges of the stellar bar. This is the expected pattern when gas transits from the outer x$_1$ orbits to the inner x$_2$ orbits in the presence of an ILR region.
This can also be seen in the PV diagrams shown in Figure \ref{fig3cc}. Along the major axis (top panel), part of the gas follows the x$_1$ orbits of the bar, that is the smooth rotation pattern modeled with $^{\rm 3D}$BAROLO. The corotation region would correspond to offsets of $\pm$3\arcsec~(5.5 kpc), but we see gas rotation extending even further. In the PV diagram along the major axis we also see high-velocity gas ($\pm$200 km~s$^{-1}$) distributed in a nuclear disk (the inner 1\arcsec$\sim$1.8 kpc), which follows the x$_2$ orbits of the bar. 


The PV diagram along the minor axis (PA$\sim$159\degr = -21\degr; bottom panel of Figure \ref{fig3cc}) shows noncircular gas motions with maximum velocities of $\pm$200 km~s$^{-1}$. The bulk of these fast noncircular motions is found at $\sim$0.5\arcsec~(0.90 kpc) to the S (i.e., negative offsets), although forbidden velocities of less than 100 km~s$^{-1}$ are also seen to the N and S up to a radius of $\sim$2\arcsec~(3.7 kpc), following the classical positive/negative pattern. All these noncircular gas motions happen well inside the corotation radius of the bar and therefore could correspond to gas inflows. In order to assess whether or not this is the case, we first need to assume that the spiral arms trail galaxy rotation \citep{1982Ap&SS..86..215P}. In this case the galaxy rotation has to be clockwise. Considering this and the CO velocity field shown in Figure \ref{fig3aa}, the N has to be the far side. Thus, for the approaching side of the ionized outflow \citep{2014MNRAS.441.3306H,2018ApJ...856..102F} to be detected in the S, it must be almost coplanar with the galaxy disk.

Most of the molecular gas along the minor axis appears blueshifted in the S, although there is also redshifted gas in the N. This can be also seen from the residual velocity map shown in the middle right panel of Figure \ref{fig3aa}, and it is the opposite behavior of what it is expected from a coplanar inflow of gas induced by a stellar bar within the corotation radius. Instead, our analysis of the kinematics suggests the presence of a coplanar outflow with an average outflow velocity of $\pm$115 km~s$^{-1}$ and radius r$_{\rm out}\sim$0.7\arcsec~(1.3 kpc). These are the average values of the regions where the molecular gas shows velocities larger than those of the rotating disk model at a given radius. We discard the vertical inflow scenario (see Figure \ref{kinematics}) because this is highly unlikely in an undisturbed disk galaxy such as J1100. Besides, we do not see any evidence of line splitting.

By integrating the CO emission along the minor axis within these regions (see Figure \ref{fig3cc}), we estimate an outflow gas mass of 10.5$\times$10$^7$M$_{\rm \sun}$ and an outflow rate of 9.5 M$_{\rm \sun}$~yr$^{-1}$. Considering the deprojected outflow rate, 12.2 M$_{\rm \sun}$~yr$^{-1}$ and the SFR, of 34 M$_{\rm \sun}$~yr$^{-1}$, we estimate a mass loading factor of $\eta$=\.M$_{\rm out}$/SFR$\sim$0.3. Thus, the CO kinematics in the central kpc of J1100 are peculiar. Overall the molecular gas follows the canonical response to the bar (i.e., falling inward), but the quasar-driven wind and ionized outflow perturb the molecular gas in the disk and drive it outward.

\begin{figure*}
\centering
\includegraphics[width=0.4\textwidth]{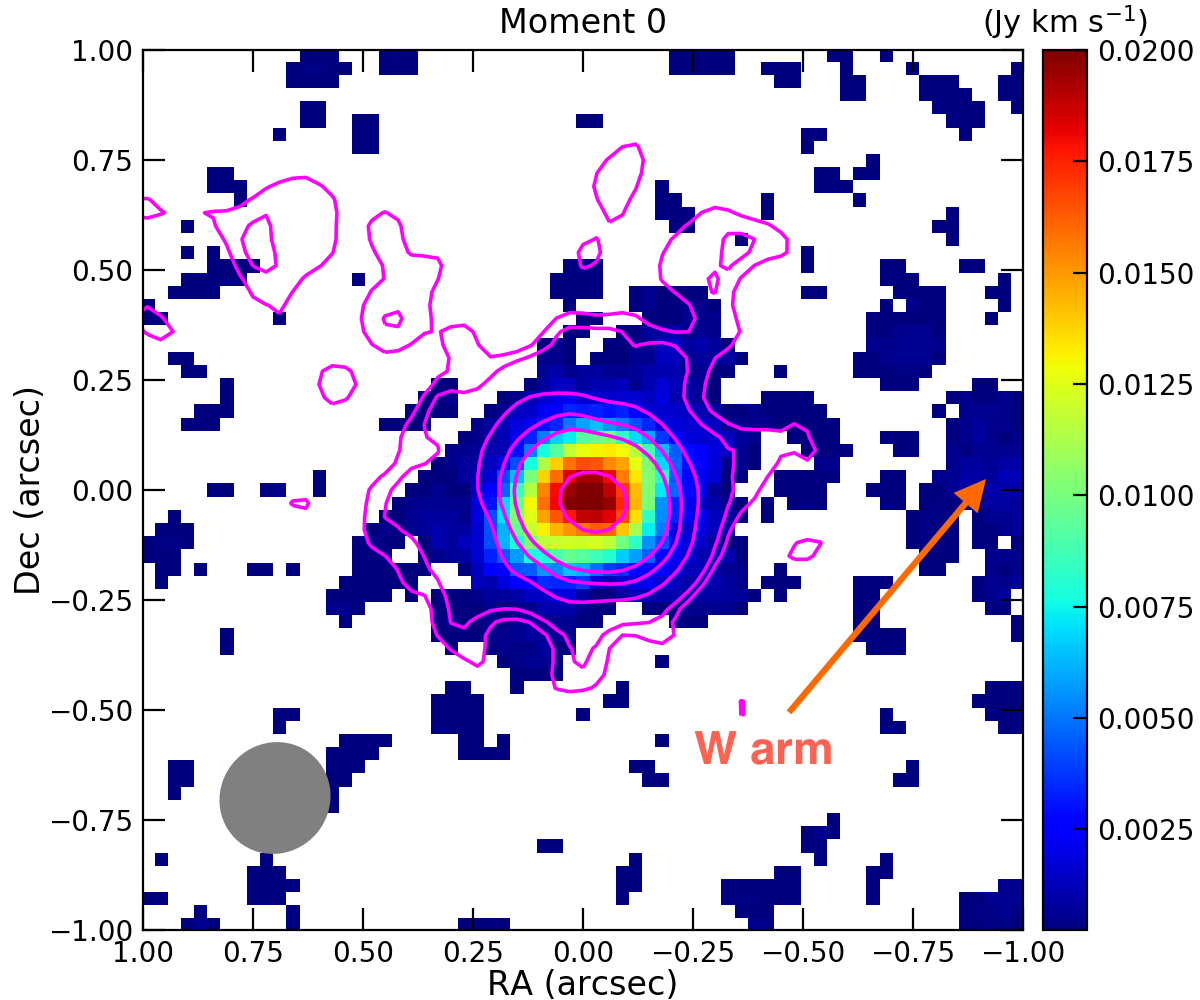}
\includegraphics[width=1.0\textwidth]{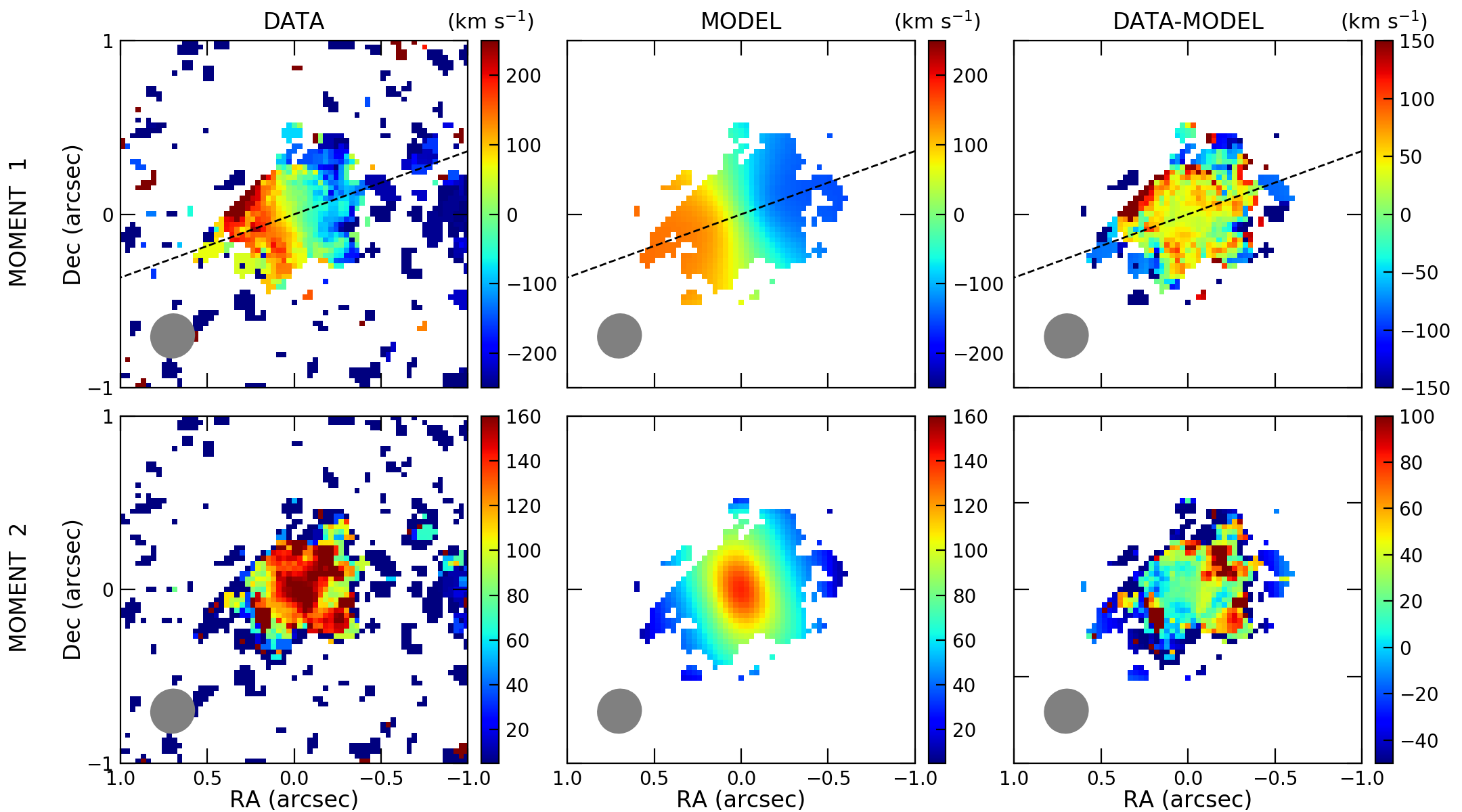}
\caption{Same as in Fig. \ref{fig2a} but for J1356 (N nucleus). Part of the W arm is also shown and labeled in orange in the moment 0 map. Continuum contours starting at 3$\sigma$ are shown in pink in the moment 0 map ($\sigma$=0.015 mJy/beam), and beam size is 0.25\arcsec$\times$0.24\arcsec.}
\label{fig5aa}
\end{figure*}

\begin{figure}
    \centering
    \includegraphics[width=1\columnwidth]{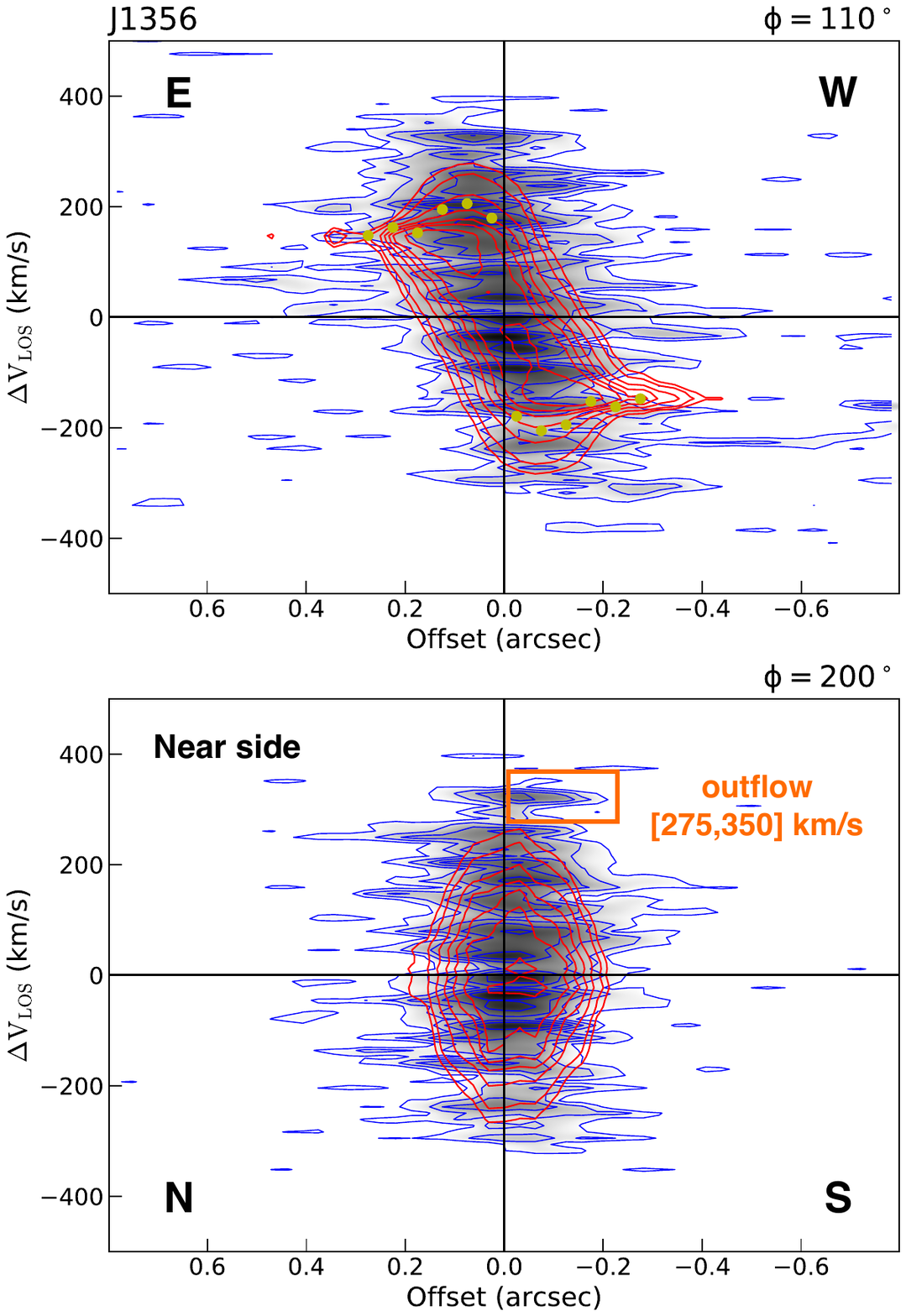}
    \caption{Same as in Fig. \ref{fig2b} but for J1356. The CO kinematic major axis is PA=110\degr, the minor axis PA=200\degr= 20\degr~and the inclination angle i=52\degr. We used a slit width of $\sim$0.25\arcsec. The orange box indicates the region and velocities used to estimate the mass of the outflowing gas, which in the case of this QSO2 does not show the characteristic positive/negative pattern.}
    \label{fig5cc}
\end{figure}

\subsubsection{SDSS J135646.10+102609.0 (J1356)}
\label{J1356}

This QSO2 is hosted in a spectacular merger system showing a completely distorted optical morphology. There are two nuclei (N and S), separated by only 1.1\arcsec~(2.4 kpc). The N nucleus is the QSO2, whose 
host galaxy is the dominant member of the merger in both the optical and molecular gas \citep{2014ApJ...790..160S}. It is a massive ETG with PA=156\degr~and i=55\degr~(see Table \ref{tab2}). 
Although the underlying stellar population is dominated by old stars \citep{2009ApJ...702..441G}, 
the SFR derived from its IR luminosity is the largest in our sample (69 M$_{\rm \sun}$~yr$^{-1}$; see Table \ref{tab2}), likely a consequence of the ongoing merger. Other components of the merging system, as identified from the ALMA and HST morphologies by \citet{2014ApJ...790..160S}, are the S nucleus and the western arm (W arm hereafter). The latter is a huge stellar feature containing a large percentage of the molecular gas in the galaxy. 

We detect the N nucleus in CO(2$-$1), which is shown in the top panel of Figure \ref{fig5aa}, as well as the W arm. Part of the latter is seen in the moment 0 map as blueshifted emission westward of the N nucleus. This feature extends up to a maximum distance of $\sim$2.3\arcsec~(5 kpc) from the quasar. By integrating the CO emission at velocities $<$0 km/s we detect the S nucleus and also the W arm (see Appendix \ref{appendixd} and \citealt{2014ApJ...790..160S}). The N galaxy represents $\sim$55\% of the total molecular gas mass in the system, the W arm $\sim$32\%, and the S nucleus $\sim$13\%.

The 1.3 mm continuum contours at 3$\sigma$ follow the CO emission, but they also show narrower structures extending up to $\sim$0.75\arcsec~(1.7 kpc) NE and $\sim$0.5\arcsec~(1.1 kpc) NW. 
These structures have different orientations than the major axis of the 6 GHz continuum emission (PA$\sim$20\degr, coincident with the kinematic minor axis of the CO disk; see below) measured from low-resolution VLA data ($\sim$1\arcsec; \citealt{2019MNRAS.485.2710J}). This 6 GHz emission has the appearance of a bent jet of 5.6 kpc to the SW (see Fig.~5 in the previously mentioned work). The VLA data at $\sim$0.25\arcsec~resolution shows a fairly compact morphology of $\sim$300 pc. From our ALMA continuum image we measure a deconvolved size of 0.27\arcsec$\times$0.20\arcsec~(600$\times$440 pc$^2$) with PA=61$\pm$17\degr. Using the VLA fluxes reported in \citet{2019MNRAS.485.2710J} and the ALMA 1.3 mm flux we measure a spectral index $\alpha$=-0.79.

\paragraph{Kinematics.}
Based on long-slit optical spectroscopy, \citet{2012ApJ...746...86G} reported the presence of a $\sim$20 kpc ionized outflow in the form of an expanding bubble. The base of this outflow would be located at $\sim$3.5\arcsec~(7.7 kpc) S of the QSO2 (see Figure 1 in \citealt{2014ApJ...790..160S}).  
Our analysis of the CO(2$-$1) data does not reveal a molecular counterpart to the ionized bubble, as it was also the case for the CO(1$-$0) and CO(3$-$2) transitions reported by \citet{2014ApJ...790..160S}.




The CO(2$-$1) moment 1 and 2 maps are shown in Figure \ref{fig5aa}. The N nucleus shows high average velocity dispersion values, of up to 160 km~s$^{-1}$. We fitted the CO kinematics with a rotating disk model of PA=110\degr~and i=52\degr. This indicates that the galaxy and CO disk would be coplanar, but the PAs do not coincide (see Table \ref{tab2}), which is not surprising in an ongoing major merger. \citet{2014MNRAS.441.3306H} analyzed the [OIII] kinematics of the central $\sim$3.5\arcsec$\times$5\arcsec~(7.7$\times$11 kpc$^2$) of the galaxy and reported a major axis with PA=225\degr=45\degr. These authors found an average blueshift of -215 km~s$^{-1}$ for the galaxy-integrated [OIII] profile, with a maximum blueshift of -600 km~s$^{-1}$ at $\sim$2\arcsec~(4.4 kpc) SW of the nucleus and a projected outflow radius of $\le$3.1 kpc. The outflow orientation coincides with the PA of the radio jet inferred from VLA data (20\degr), but it is possible that the blueshifted gas detected to the SW is associated with the W arm and the S nucleus, which are blueshifted relative to the systemic velocity (see Appendix \ref{appendixd}). We note that the analysis of the molecular gas kinematics that we are describing here focuses on the N nucleus. 

Using the HST/WFC3 images from \citet{2015ApJ...806..219C} we constructed a B-I color map of the galaxy. A dust lane crosses the N nucleus with a PA coincident with the CO major axis. On the small spatial scales considered in Figure \ref{fig5aa}, we detect redder optical colors NE of the nucleus (see Appendix \ref{appendixd}). Based on this, the N would be the near side. Since the ionized outflow is detected to the SW, it must subtend a large angle relative to the CO disk (scenario B in Figure \ref{scheme}, i.e., weak coupling).

The PV diagram along the minor axis (PA=20\degr; see bottom panel of Figure \ref{fig5cc}), shows tentative evidence of noncircular motions within the central $\sim$0.6\arcsec~(1.3 kpc) of the QSO2. 
We do not see the positive/negative pattern characteristic of outflowing/inflowing gas as in J1100, but this could be due to the complex gas kinematics of ongoing merger systems like this one. If the N is the near side, molecular gas outflowing in an almost coplanar geometry should be blueshifted to the N and redshifted to the S. The high-velocity gas with v$_{\rm max}\sim$350 km~s$^{-1}$ detected within a radius of 0.2\arcsec~(0.4 kpc) to the S would then correspond to outflowing gas (see also the redshifted residuals along the minor axis in Figure \ref{fig5aa}). This is supported by the order of magnitude of the radial velocities. A vertical inflow scenario is, in principle, also compatible with having redshifted gas detected in the far side (see Figure \ref{kinematics}). In a major merger, a vertical inflow driven by, for example, a tidal tail would be plausible, but in that case we would detect line splitting in the PV diagrams (see e.g., \citealt{2019A&A...632A..61G}), something that does not happen.

A high-velocity component of $\sim$400 km~s$^{-1}$, detected in CO(3$-$2) in the N nucleus, was reported by \citet{2014ApJ...790..160S}. They identified this with an outflow of M$_{\rm out}\sim$7$\times$10$^7$ M$_{\sun}$ by assuming L$^{\prime}_{\rm CO(3-2)}$=L$^{\prime}_{\rm CO(1-0)}$. This outflow mass is larger than our measurement from CO(2-1), of (1.4$\pm1.2)\times$10$^7$ M$_{\sun}$. We note, however, that this value has been obtained from integrating the high-velocity gas to the S and along the minor axis only (i.e., within the orange box in Figure \ref{fig5cc}). If we consider all the redshifted high-velocity gas, we measure a mass of 7.1$\times$10$^7$ M$_{\sun}$, as in \citet{2014ApJ...790..160S}. 
We also detect some high-velocity blueshifted gas in the PV diagram along the minor axis (see Figure \ref{fig5cc}), but not in the residual map shown in Figure \ref{fig5aa}. Thus, we prefer to be conservative and only use the high-velocity redshifted gas to the S along the minor axis to work out the outflow mass. Using this mass, the radius indicated above, and an average outflow velocity of 310 km~s$^{-1}$, we estimate an outflow rate of 10.0 M$_{\rm \sun}$~yr$^{-1}$. Considering the high SFR estimated for this QSO2, of 69 M$_{\rm \sun}$~yr$^{-1}$, and the deprojected \.M$_{\rm out}$=7.8 M$_{\rm \sun}$~yr$^{-1}$, the mass loading factor is $\eta\sim$0.1. 



\begin{figure*}
\centering
\includegraphics[width=0.4\textwidth]{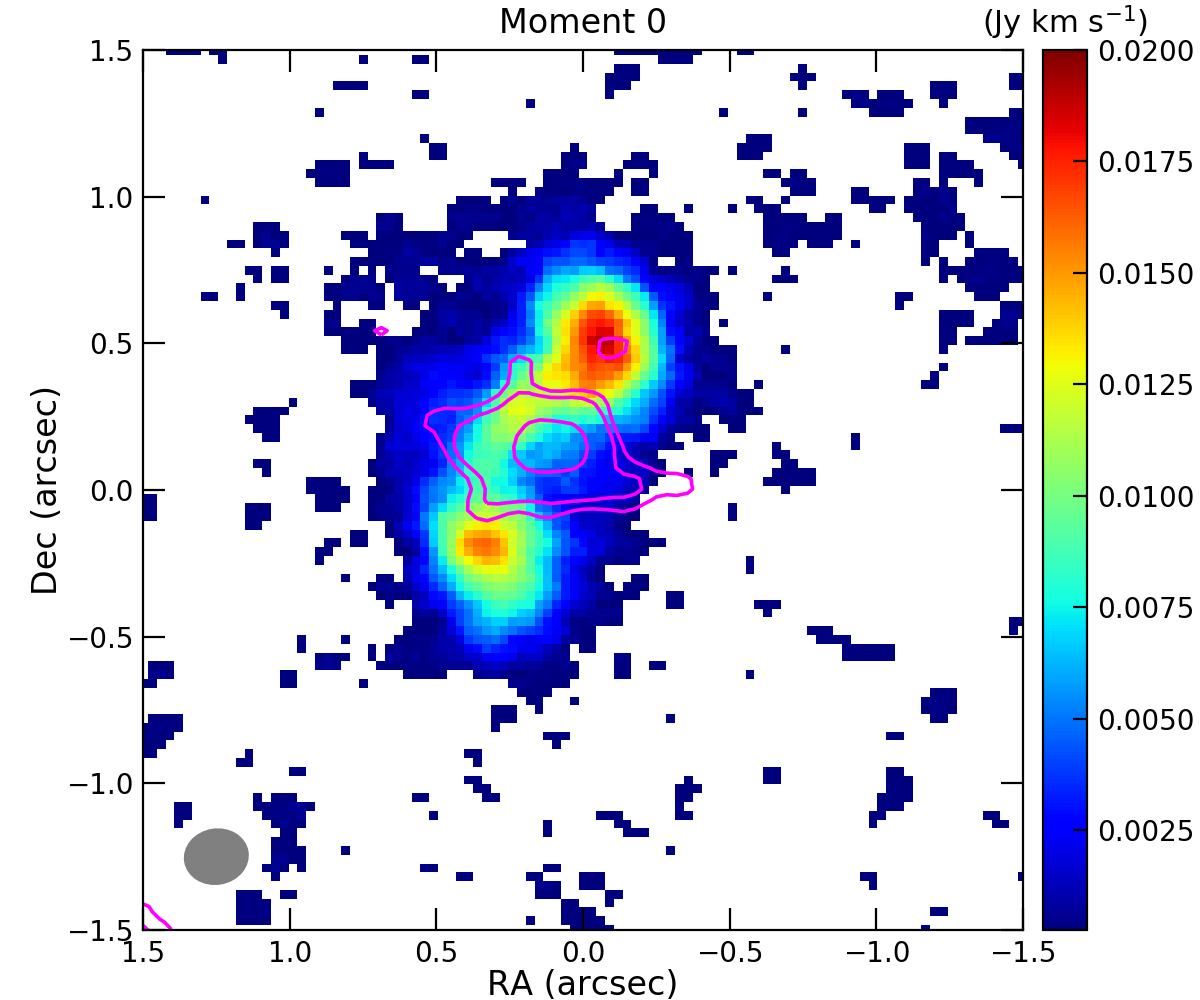}
\includegraphics[width=1.0\textwidth]{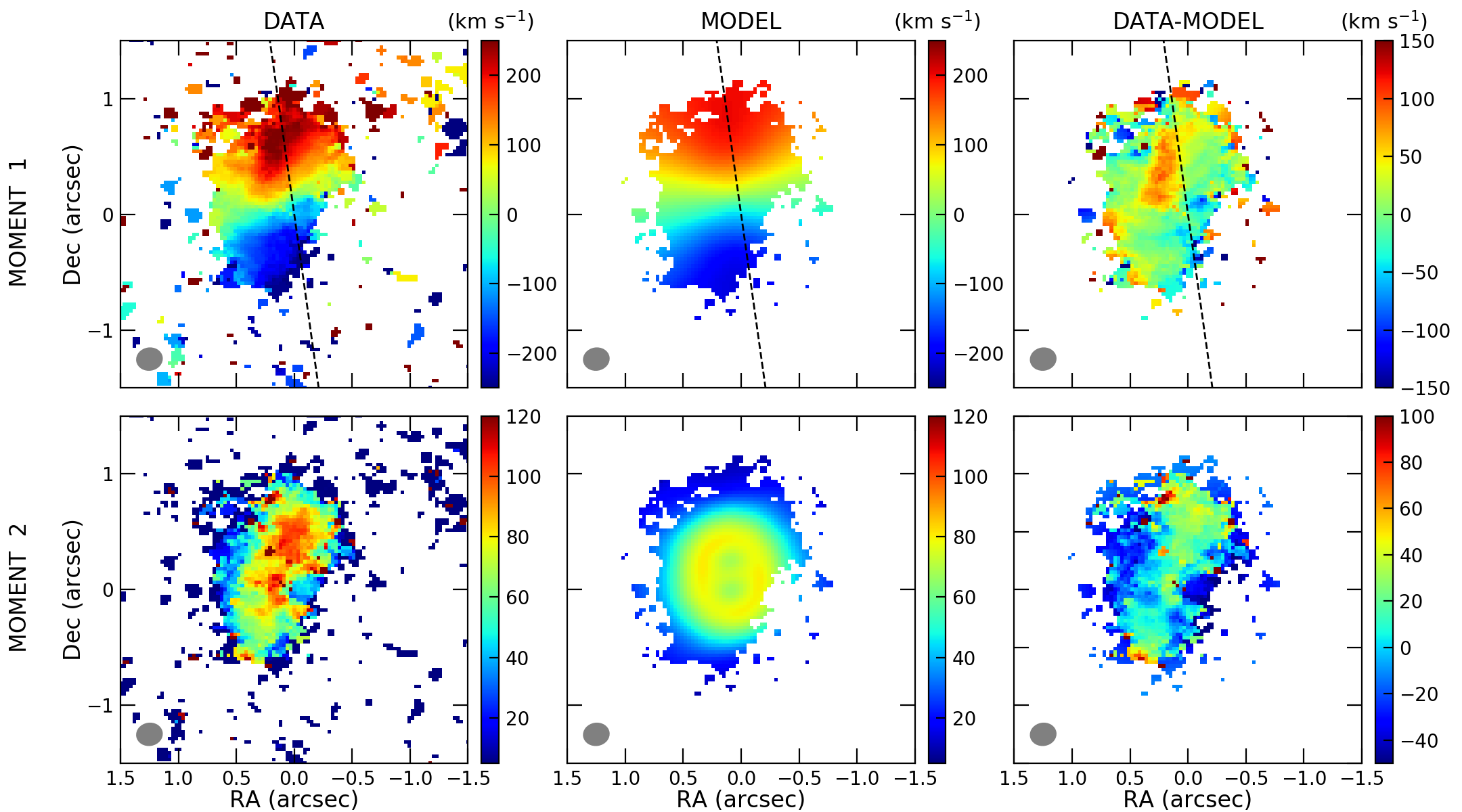}
\caption{Same as in Fig. \ref{fig2a} but for J1430. Continuum contours starting at 3$\sigma$ are shown in pink in the moment 0 map ($\sigma$=0.015 mJy/beam), and beam size is 0.21\arcsec$\times$0.18\arcsec.}
\label{fig6aa}
\end{figure*}

\begin{figure}
    \centering
    \includegraphics[width=1\columnwidth]{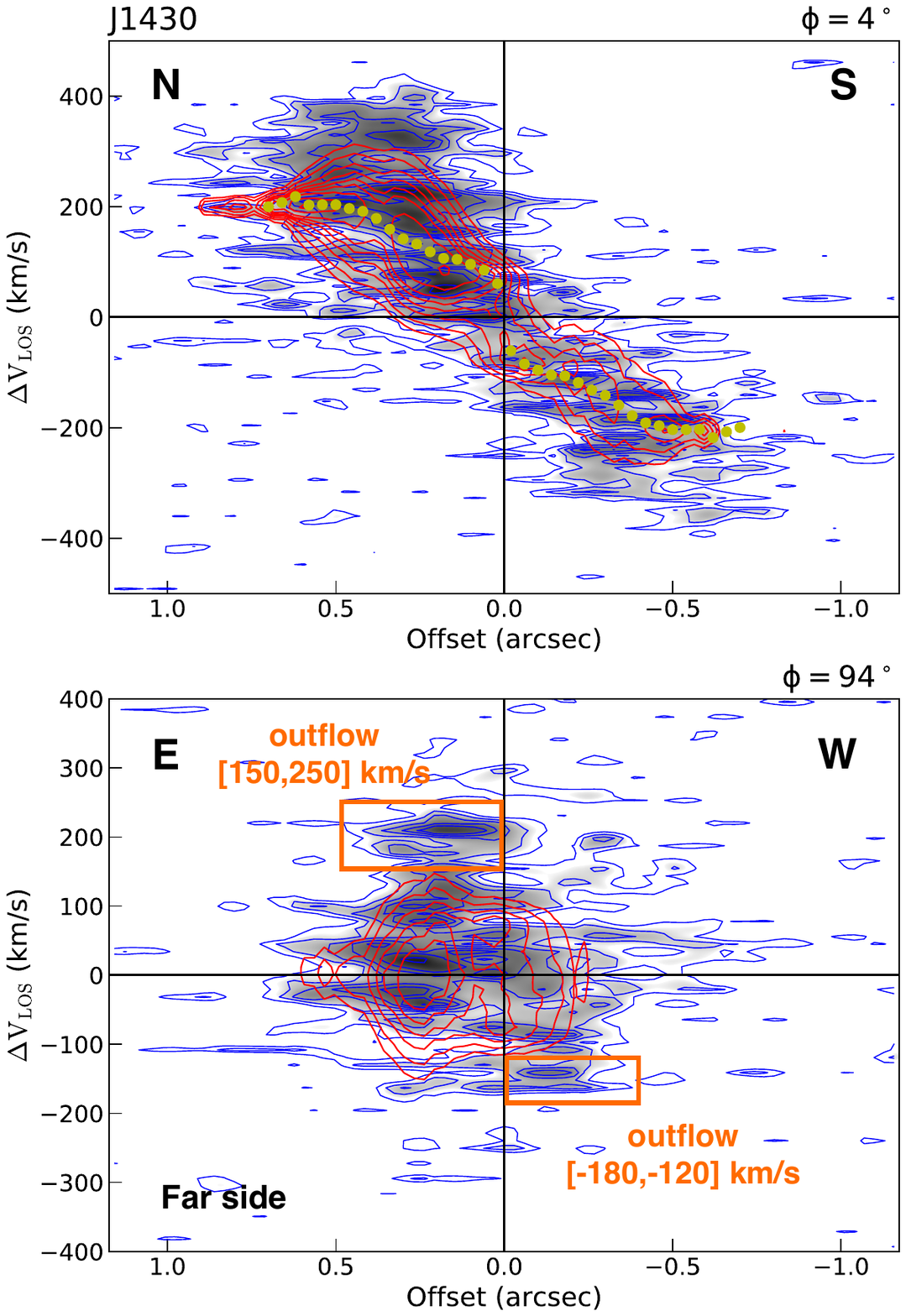}
    \caption{Same as in Fig. \ref{fig2b} but for J1430. The CO kinematic major axis is PA=4\degr, the minor axis PA=94\degr~and the inclination angle i=38\degr. We used a slit width of $\sim$0.2\arcsec. The orange boxes indicate the regions used to estimate the mass of the outflowing gas.}
    \label{fig6cc}
\end{figure}

\subsubsection{SDSS J143029.88+133912.0 (J1430; The Teacup)}
\label{Teacup}

Often referred to as the Teacup galaxy, J1430 has been widely studied in different wavelength ranges including the X-rays \citep{2018ApJ...856L...1L}, optical \citep{2012MNRAS.420..878K,2015ApJ...800...45H,2018MNRAS.474.2302V}, NIR \citep{2017MNRAS.470..964R} and radio \citep{2019MNRAS.485.2710J}. The host galaxy is bulge-dominated, showing clear signatures of a past interaction. These include a spectacular system of concentric shells and nuclear dust lanes (see Figure \ref{fig1} and \citealt{2015AJ....149..155K}). 

Using VLA data at 0.3\arcsec~resolution, \citet{2015ApJ...800...45H} reported the presence of two compact steep radio sources: one corresponding to the QSO2 nucleus (HR-A) and another (HR-B) at $\sim$0.5\arcsec~(0.8 kpc) NE (PA$\sim$60\degr). These authors proposed that this could be a compact radio jet that might be accelerating the ionized gas in the central kpc and possibly driving the large-scale radio bubbles. Our ALMA data reveal continuum emission at 1.3 mm peaking in the middle of two CO(2$-$1) blobs, as can be seen from the top panel of Figure \ref{fig6aa}. This continuum emission looks more jet-like than in the case of the other QSO2s in our sample and it has an almost E-W orientation, perpendicular to the CO emission. From the analysis of the continuum image we derive a deconvolved size of 0.37\arcsec$\times$0.23\arcsec~(590$\times$370 pc$^2$) and PA=80$\pm$12\degr, almost coincident with the major axis of the 1.4 GHz continuum emission measured from FIRST (77\degr; \citealt{2014MNRAS.441.3306H}).

The morphology of the cold molecular gas at 0.2\arcsec~resolution looks very different to that of the warm molecular gas observed with VLT/SINFONI at 0.5\arcsec~resolution \citep{2017MNRAS.470..964R}. Instead of the single-peaked disk-like structure observed in H$_2$, the CO moment 0 map shown in Figure \ref{fig6aa} shows a double-peaked morphology, with the two peaks separated by $\sim$0.8\arcsec~(1.3 kpc) with PA$\sim$-10\degr. This galaxy has been observed in the optical with HST (angular resolution of 0.1\arcsec) and it shows a single nucleus in both continuum and [OIII] \citep{2012MNRAS.420..878K,2015ApJ...800...45H}. 

\paragraph{Kinematics.} The moment 1 map shows a distorted rotation pattern, redshifted and blueshifted to the N and S, respectively (see Figure \ref{fig6aa}). This is also the case for the H$_2$ \citep{2017MNRAS.470..964R} and [OIII] velocity fields \citep{2014MNRAS.441.3306H}. The moment 2 map reveals higher values of the velocity dispersion (100--120 km~s$^{-1}$) across $\sim$1\arcsec~(1.6 kpc) in the direction perpendicular to the jet. This enhancement of the velocity dispersion has been observed in ionized gas in four nearby AGN from the MAGNUM survey \citep{2021A&A...648A..17V}, and interpreted as due to the action of the jets perturbing the gas in the galaxy disk. 

The $^{\rm 3D}$BAROLO model of the moment 1 and 2 maps and corresponding residuals are shown in the middle and bottom rows of Figure \ref{fig6aa}. Since it is difficult to define a clear major axis of the CO distribution in the case of this QSO2, we first let the PA vary freely and fixed the inclination to 38\degr~(i.e., coincident with the galaxy inclination measured from optical images; see Table \ref{tab2}). This is a post-merger system and the inclination of the CO and stellar disk might be different, but if we try to fix the PA to the average value of the previous step (PA=4\degr) and let the inclination vary, we get values between i=35\degr and 41\degr. These values produce lower residuals than higher/lower inclinations, so we can safely assume i=38\degr~for the CO disk. The kinematic major axis of the CO disk (PA=4\degr) is different from the [OIII] major axis derived from the analysis of Gemini/GMOS data (PA=27\degr; \citealt{2014MNRAS.441.3306H}) and from the galaxy's PA of -19\degr. This is likely due to the past merger and/or to the action of the jet on the ionized and molecular gas.

Ionized outflows detected in the form of blueshifted broad components of optical and NIR emission lines have been reported for the Teacup \citep{2015ApJ...800...45H,2017MNRAS.470..964R}. A spatially resolved outflow with radial extent of $\sim$1 kpc was found by \citet{2017MNRAS.470..964R} using the Paschen $\alpha$ and [Si VI] lines in the NIR. They reported maximum outflow velocities of -1100 and -900 km~s$^{-1}$ and PAs of 75\degr and 72\degr~respectively. The approaching side of the ionized outflow is on the NE side, within $\sim$1\arcsec~(1.6 kpc), as also found for [OIII] in the optical \citep{2014MNRAS.441.3306H,2015ApJ...800...45H}. Considering this, if the ionized outflow subtends a large angle relative to the CO disk (scenario B in Figure \ref{scheme}), the E would be the far side. This is confirmed by the presence of strong dust lanes westward of the nucleus \citep{2015AJ....149..155K}. These dust lanes produce redder colors on that side of the galaxy, as shown by the r-i color map shown in Appendix \ref{appendixe}.


The PV diagram along the minor axis (see bottom panel of Figure \ref{fig6cc}) shows strong noncircular motions within r$_{\rm out}\sim$0.3\arcsec~(0.5 kpc) of the galaxy. These motions have maximum velocities of 250 and -180 km~s$^{-1}$, mainly redshifted to the E and blueshifted to the W. This can be seen also from the residual map shown in Figure \ref{fig6aa}. Since the E is the far side, this corresponds to either outflowing gas in the CO disk plane or vertical inflowing gas. As in the case of J1356, we favor the first scenario based on the high radial velocities, of up to 250 km~s$^{-1}$, and the lack of line splitting in the PV diagrams. 

By integrating the high-velocity CO(2$-$1) emission along the minor axis and within the regions labeled in Figure \ref{fig6cc}, we measure an outflow mass of M$_{\rm out}$=3.12$\times$10$^7$ M$_{\rm \sun}$ and an outflow rate of 12.3 M$_{\rm \sun}$~yr$^{-1}$. Considering the deprojected outflow mass rate, of \.M$_{\rm out}$=15.8 M$_{\rm \sun}$~yr$^{-1}$ and the SFR, of 12 M$_{\rm \sun}$~yr$^{-1}$, the mass loading factor is $\eta\sim$1.3, the largest in our sample.

We did not detect a warm molecular counterpart of this outflow using NIR data from VLT/SINFONI \citep{2017MNRAS.470..964R}. Unlike the nuclear Paschen $\alpha$ and [Si VI] lines, for which both narrow and broad Gaussian components were necessary to reproduce the line profiles, the H$_2$ could be fitted with a single Gaussian component, although slightly blueshifted (-50 km~s$^{-1}$) relative to the narrow component of Pa$\alpha$. 
We note that the cold molecular outflow is barely resolved even at the high angular resolution of the ALMA data, making it challenging to detect its warm molecular gas counterpart with the 0.5\arcsec~resolution of the SINFONI observations. Furthermore, cold molecular gas is much more abundant than the warm molecular gas traced with the NIR H$_2$ lines, which might be another reason for detecting the molecular outflow in CO but not in H$_2$. A detailed comparison of the warm and cold molecular gas in this galaxy will be the subject of forthcoming work (Audibert et al. in preparation). 
 

The combined action of the cold molecular outflow and the jet could explain the double-peaked morphology shown in the flux map of Figure \ref{fig6aa}. 
This morphology is similar to that of the Seyfert 2 galaxy NGC\,2110, which also has a radio jet oriented perpendicularly to an area depleted of CO(2--1). The molecular gas might have being pushed outward, resulting in a depletion of the central region. Alternatively, the molecular gas in the region more affected by the jet might have been excited to higher-J transitions \citep{2019ApJ...875L...8R}. 
Thus, the Teacup could be another example of AGN feedback shaping the molecular gas reservoir in the central kpc of an AGN (see also \citealt{2019ApJ...875L...8R,2021A&A...645A..21G,2021arXiv210410227G}).

\subsubsection{SDSS J150904.22+043441.8 (J1509)}
\label{J1509}

\begin{figure*}
\centering
\includegraphics[width=0.4\textwidth]{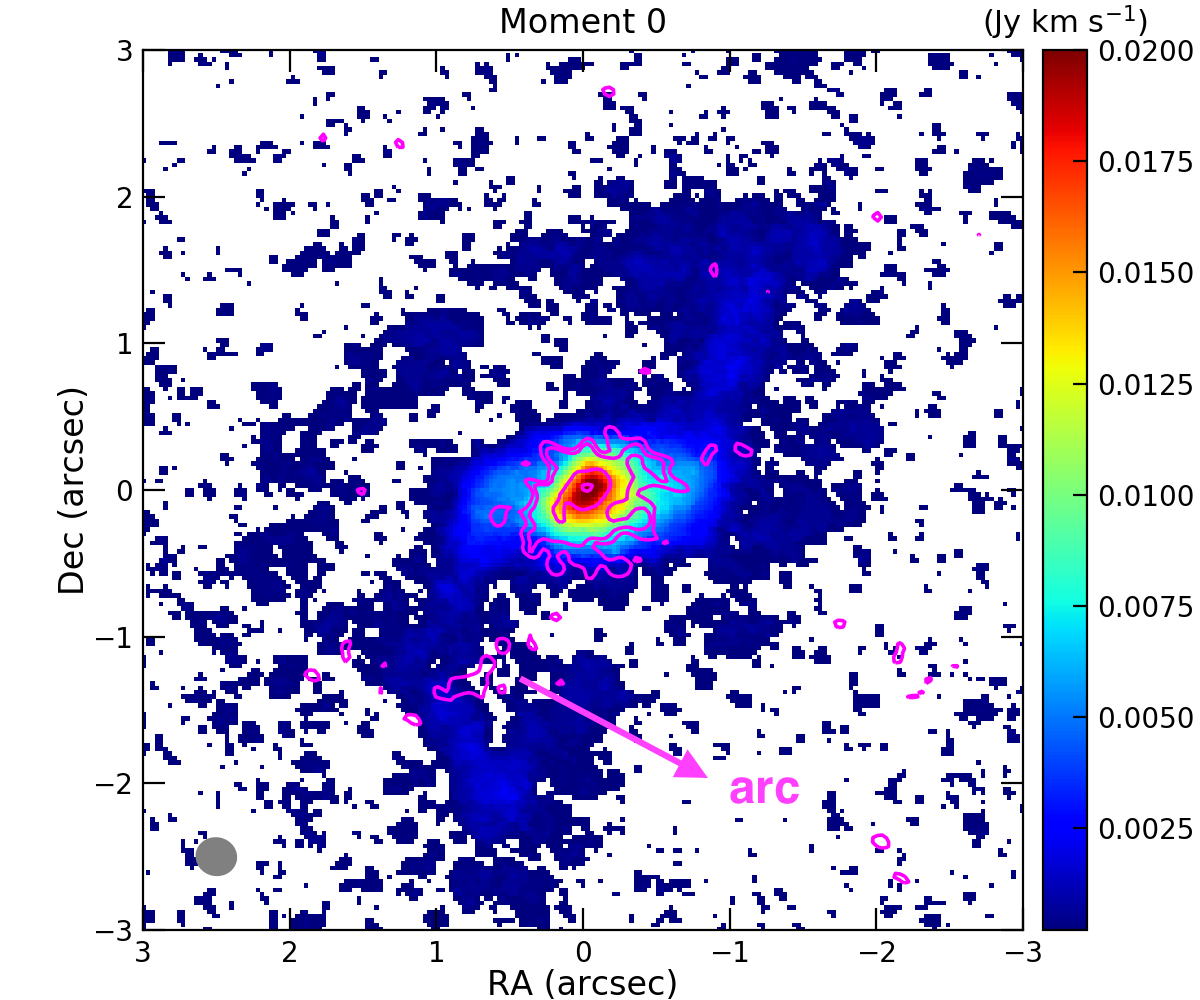}
\includegraphics[width=1\textwidth]{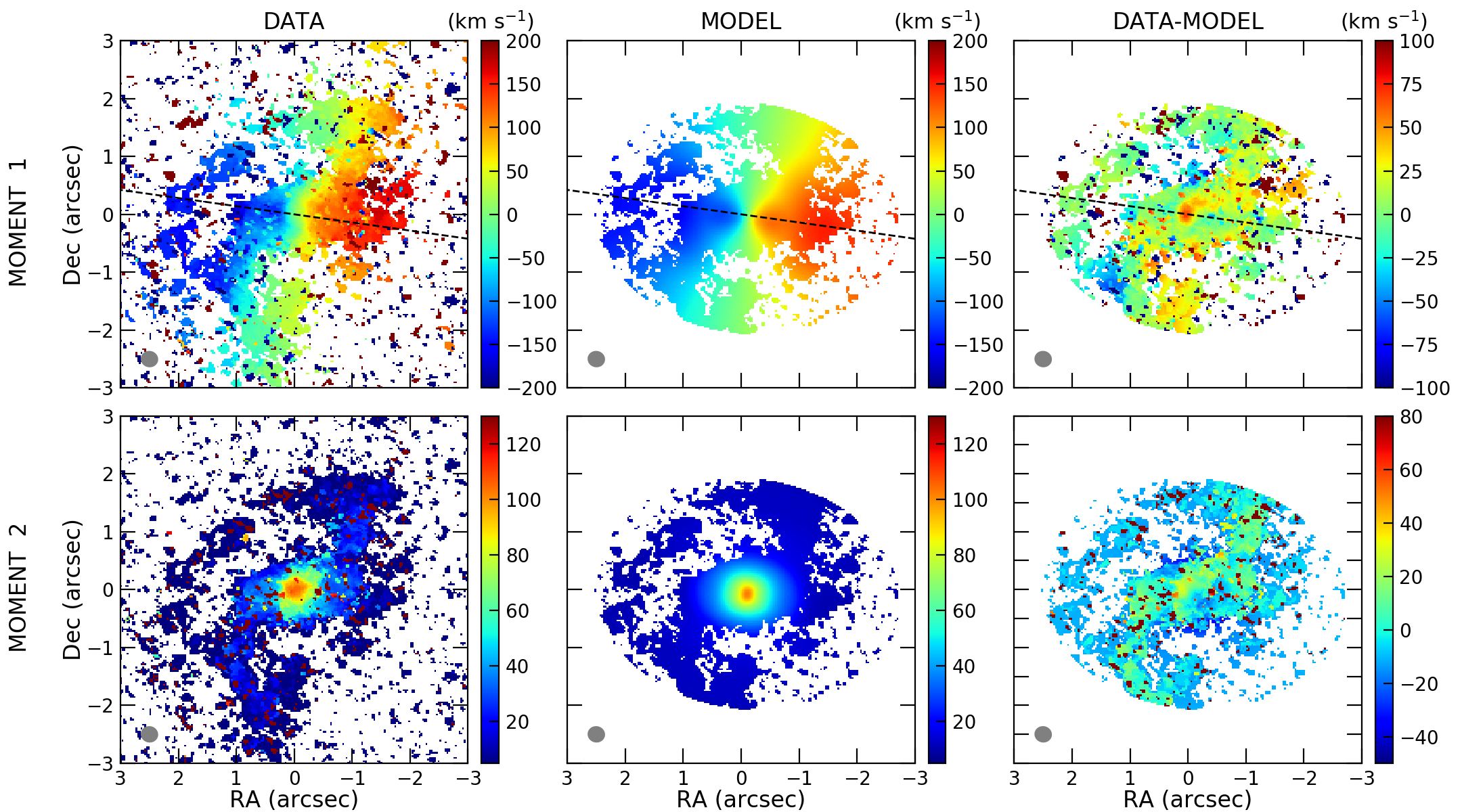}
\caption{
Same as in Fig. \ref{fig2a} but for J1509. Continuum contours starting at 3$\sigma$ are shown in pink in the moment 0 map ($\sigma$=0.013 mJy/beam), and beam size is 0.26\arcsec$\times$0.24\arcsec.} \label{fig7a}
\end{figure*}

\begin{figure}
\includegraphics[width=1\columnwidth]{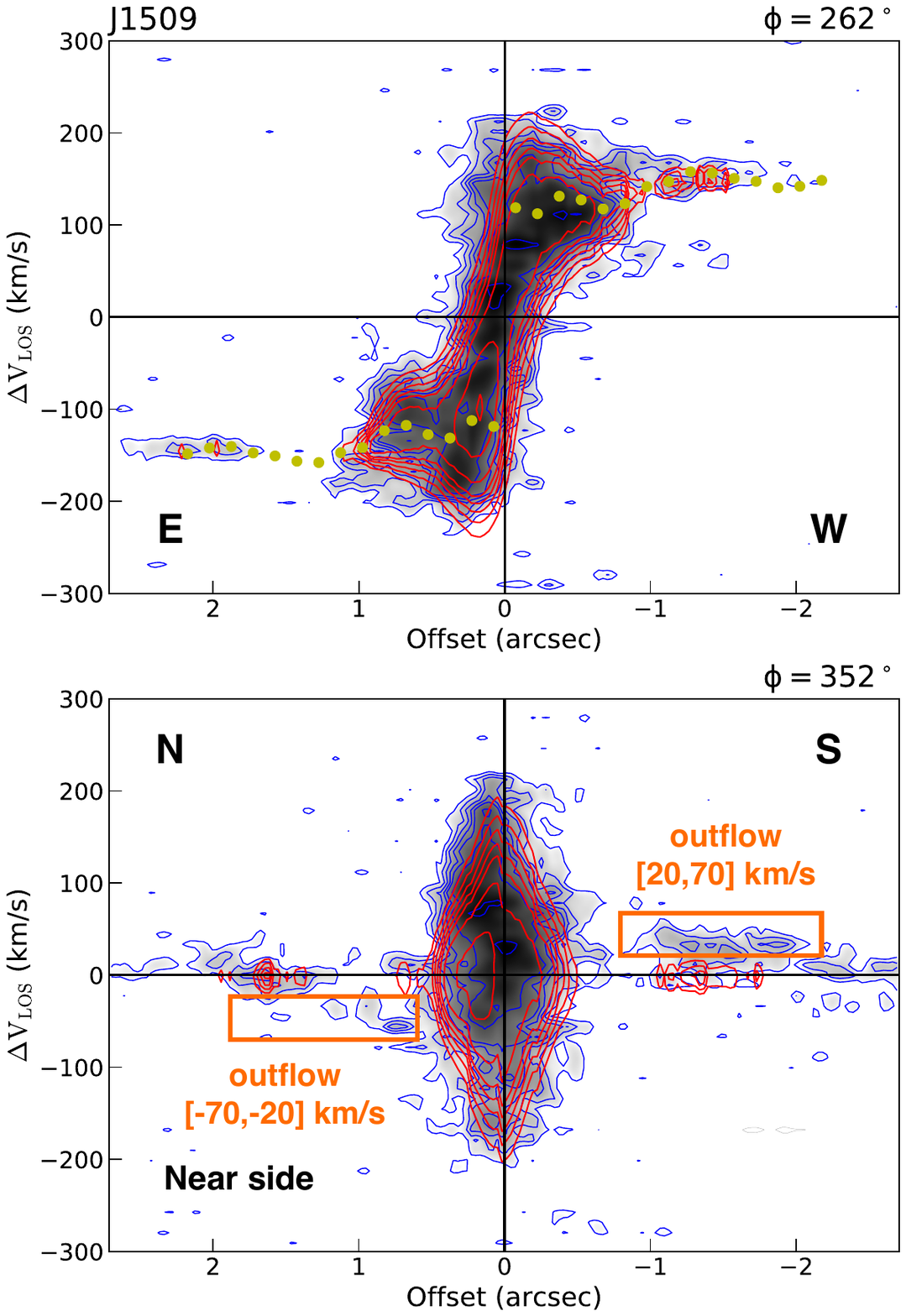}
\caption{Same as in Fig. \ref{fig2b} but for J1509. The CO kinematic major axis is PA=262\degr= 82\degr, the minor axis PA=352\degr= -8\degr~and the inclination i=43\degr. We used a slit width of $\sim$0.25\arcsec. The orange boxes indicate the regions and velocities used to estimate a lower limit to the mass of the outflowing gas. The gas shows the characteristic positive/negative pattern but we do not resolve the high-velocity gas within the outflow.}  
\label{fig7b}
\end{figure}

This galaxy is classified as a barred spiral (SBa) in Galaxy Zoo \citep{2013MNRAS.435.2835W}, with PA=94\degr~and inclination i=44\degr~(see Table \ref{tab2}). From the r-band SDSS image we measure a bar orientation of $\sim$-30\degr~and R$_{\rm bar}\sim$2.25\arcsec~(5.6 kpc). 

Unlike the majority of QSO2s in our sample, there are no high angular resolution optical and radio data of J1509 to be compared with the ALMA continuum and molecular gas distribution. The 1.3 mm continuum emission of J1509 is quite peculiar (see pink contours in Figure \ref{fig7a}). It follows the inner oblong structure of molecular gas, but from this central region an arc-like structure directed to the SE that does not follow the CO distribution is exhibited. By fitting a 2D Gaussian to the continuum emission at $\ge3\sigma$ we measure a deconvolved size of 0.74\arcsec$\times$0.55\arcsec~ (1.5$\times$1.1 kpc$^2$) with PA=135$\pm$22\degr. This structure 
cannot be associated either with dust or free-free emission, since it would otherwise follow the CO morphology. It should then correspond to synchrotron emission, and the most likely explanation for the peculiar morphology is a radio jet that in its interaction with the surrounding environment has been forced to bend. Alternatively, if the jet precesses about a defined axis it can result in the jet being curved as observed in the plane of the sky \citep{1984RvMP...56..255B}. The presence of a jet is further supported by the spectral index measured from the 1.4 GHz and 200 GHz fluxes, which corresponds to a steep spectrum ($\alpha$=-0.64; see Section \ref{continuum}). 


The moment 0 map in Figure \ref{fig7a} shows an oblong structure in CO, almost in the E-W direction and with a major axis of $\sim$1.7\arcsec~(3.4 kpc). Outside this compact elongated gas disk there are two spiral arms that develop inside the stellar bar out to r$\sim$2\arcsec~(4 kpc) to the NW and SE (PA$\sim$-30\degr). This feature mimics the leading edges signature typical of the gas response to the bar in the presence of an extended ILR region. The leading edges are joint at the northern end of the stellar bar by a gas arc at r$\sim$2\arcsec. This arc lacks a southern counterpart, which gives it the appearance of an asymmetric ring. 




\paragraph{Kinematics.}

In \citet{2019MNRAS.487L..18R} we analyzed a NIR long-slit spectrum (slit PA=-16\degr, almost coincident with the minor axis of the CO disk) of this QSO2 obtained with the instrument EMIR on the 10.4 m Gran Telescopio CANARIAS (GTC). We detected blueshifted ionized and warm molecular gas within radial sizes of 1.3$\pm$0.2 and 1.5$\pm$0.2 kpc. The maximum velocity that we measured for the warm molecular outflow is -750 km~s$^{-1}$, with a FWHM$\sim$1500 km~s$^{-1}$. For the ionized gas, using the Pa$\alpha$ and Br$\delta$ lines, we measured maximum velocities of -1200 km~s$^{-1}$ and FHWMs$\sim$1800 km~s$^{-1}$.
We also detected a blueshifted broad component in the coronal line of [Si VI]$\lambda$1.963 $\mu$m, with v$_{\rm max}$=-850 km~s$^{-1}$ and FWHM=1500 km~s$^{-1}$. This emission line can only be produced by AGN photoionization or shocks \citep{2020ApJ...895L...9R}. Thus, the NIR data clearly shows that we are witnessing a multiphase AGN-driven outflow in J1509.

The CO velocity field (moment 1 map in Figure \ref{fig7a}) shows a rotating distribution, blueshifted to the E and redshifted to the W. Considering this, and assuming that the spiral arms trail galaxy rotation, rotation has to be clockwise and thus, the N is the near side. Our $^{\rm 3D}$BAROLO models of the moment 1 and 2 maps and corresponding residuals are shown in the middle and right panels of Figure \ref{fig7a}. The orientation of the kinematic major axis (PA=82\degr) and inclination of the CO disk (i=43\degr) are very similar to those measured from the optical image (see above), indicating rotation-dominated CO kinematics. 
The stellar bar regulates the CO distribution and kinematics within the corotation radius, which we estimate as R$_{\rm CR}\sim(1.2\pm0.2)\times R_{\rm bar}\sim$3.3\arcsec. 
The PV diagram along the major axis displayed in Figure \ref{fig7b} shows that the bulk of the gas is rotating within the inner 2\arcsec~(4 kpc). Beyond this region, we also see rotation up to distances of $\sim$2.7\arcsec~to the E and $\sim$2\arcsec~to the W. This corresponds to gas transiting from the x$_1$ to the x$_2$ orbits of the bar. 

The gas beyond the central oblong structure is concentrated along the two leading edges of the stellar bar and it follows the canonical response. If we look at the residual velocities in Figure \ref{fig7a}, once we subtract our rotating disk model we see mostly blueshifted velocities to the N and redshifted to the S along the minor axis. These residuals correspond to outflowing gas in a mostly coplanar geometry, considering that the N is the near side. We also see gas falling onto the galaxy center (i.e., the redshifted blob in the center of the velocity residual map). The PV diagram along the minor axis (PA=-8\degr; bottom panel of Figure \ref{fig7b}) shows evidence for outflowing gas at low velocities (up to $\pm$70 km~s$^{-1}$), 
showing the characteristic positive/negative velocity pattern to the S and N.
This low-velocity gas is clearly resolved up to r$\sim$2\arcsec~(4 kpc), but we do not resolve the high-velocity gas in spite of the good angular resolution (see the $^{\rm 3D}$BAROLO contours in the bottom panel of Figure \ref{fig7b}). 

Thus, the CO kinematics of this QSO2 suggest a competition between the gas that follows the canonical response to the bar and consequently falls inward, and the gas that is being pushed away by the jet, wind and/or ionized outflow in a coplanar geometry. We discard the vertical inflow scenario to explain the blueshifted/redshifted gas to the N/S (see Figure \ref{kinematics}) because 1) it is highly unlikely in an undisturbed disk galaxy, and 2) we do not observe any signature of line splitting. 
The case of J1509 is different from the Teacup because we see evidence for both a warm and cold molecular outflows, as found in the case of the Seyfert 2 galaxy IC\,5063 \citep{2014Natur.511..440T,2015A&A...580A...1M}, where the jet and the multiphase outflow are coplanar with the galaxy and CO disks \citep{2018MNRAS.476...80M}. This could also be the case for J1509, and in this case we should see the blueshifted side of the ionized outflow in the N side, being almost coplanar with the H$_2$/CO disk (i.e., scenario A in Figure \ref{scheme}).


In \citet{2019MNRAS.487L..18R} we estimated a warm molecular outflow mass of (1.0$\pm$0.2)$\times$10$^4M_{\sun}$, assuming local thermal equilibrium. Using the conversion factor of 6$\times$10$^{-5}$ reported by \citet{2014A&A...572A..40E} for two nearby LIRGs observed in the NIR and millimeter ranges we estimated a total molecular gas mass in the outflow of (1.7$\pm$0.4)$\times$10$^8M_{\sun}$. This is consistent with the lower limit of M$_{\rm out}\geq$6.8$\times$10$^7 M_{\rm \sun}$ that we measure by integrating the CO(2$-$1) emission along the minor axis and within the regions indicated in Figure \ref{fig7b}. These regions are the ones where the molecular gas shows higher velocities than the rotating disk model at these radii. The cold molecular outflow is therefore slower and less turbulent than its warm molecular and ionized counterparts, but dominant in terms of mass. From this lower limit of the outflow mass, we estimate an outflow rate $\geq$1.03 M$_{\rm \sun}$~yr$^{-1}$. Considering the deprojected value, of $\geq$1.11 M$_{\rm \sun}$~yr$^{-1}$, and the SFR derived from the IR luminosity, of 34 M$_{\rm \sun}$~yr$^{-1}$, we measure a mass loading factor of $\eta\geq$0.03.




\subsection{Continuum emission of the QSO2s}
\label{continuum}

\begin{table*}
\caption{Properties measured from the CO(2$-$1) emission line of the QSO2s.}
\centering
\begin{tabular}{lccccccccrr}
\hline
\hline
ID  	    & $\nu_{\rm CO}$  &     rms$_{\rm10\ km~s^{-1}}$  & S$\Delta$v$_{\rm CO}$ & FWHM & $L^{\prime}_{\rm CO(2-1)}\times$10$^{-9}$ & \multicolumn{2}{c}{R$_{\rm CO}$} & M$_{\rm H_2}\times$10$^{-9}$ & t$_{\rm dep}$ & f$_{\rm H_2}$\\
            &  (GHz) &   (mJy beam$^{-1}$)            &  (Jy km s$^{-1}$) & (km s$^{-1}$) & (K km s$^{-1}$ pc$^2$)  &(\arcsec) & (kpc) & (M$_{\sun}$) & (Myr) & \\
\hline
J0232  & \dots &   0.39 &  $<1.3^*$      & 430$^*$    & $<0.15$       & 1.52 & 2.9$^*$   & $<$0.67        & $<$220     & $<$0.01     \\
J1010  & 209.912 &   0.81 & $7.92\pm0.96$  & $490\pm26$ & $0.89\pm0.11$ & 1.33 & 2.4       & 3.87$\pm$1.59     & 130     & 0.04        \\
J1100  & 209.541 &   0.43 & $30.6\pm3.4$   & $360\pm15$ & $3.63\pm0.40$ & 2.70 & 5.0       &  15.8$\pm$6.3     & 460     & 0.15        \\
J1152  & \dots &   0.85 & $<1.5^*$       & 430$^*$    & $<0.08$       & 2.10 & 2.9$^*$   & $<$0.37        & $<$100     & $<$0.01     \\
J1356  & 205.210 &   0.41 & $14.7\pm2.3$   & $320\pm70^{\dagger}$& $2.65\pm0.42$ & 2.28&5.0& 11.5$\pm$5.2      & 170     & 0.06        \\
J1356N & 205.210 &   0.41 & $8.14\pm0.91$  & $582\pm24$ & $1.47\pm0.16$ & 0.77 & 1.7       & 6.39$\pm$2.57     & \dots  & \dots       \\
J1430  & 212.459 &   0.39 & $16.9\pm1.8$   & $485\pm18$ & $1.43\pm0.15$ & 0.81 & 1.3       & 6.24$\pm$2.48     & 520    & 0.04        \\
J1509  & 207.327 &   0.42 & $27.5\pm3.6$   & $339\pm26$ & $4.04\pm0.53$ & 1.97 & 4.0       & 17.6$\pm$7.4      & 520     & 0.20        \\ 
\hline	   					 			    								      
\end{tabular}
\tablefoot{For each target we report the observed CO(2$-$1) frequency, the rms representative of the spectral region next to CO (for a channel width of 10 km~s$^{-1}$), integrated CO flux, FWHM of the line profile, CO luminosity and maximum spatial extent of the CO emission at 3$\sigma$ measured from the AGN position. * Upper limits at 3$\sigma$ assuming FWHM=430 km~s$^{-1}$ and a circular aperture of R$_{\rm CO}$=2.9 kpc (mean values of the three ETGs with CO detection). $\dagger$ The asymmetric CO emission line profile of J1356 can be better fitted with two components of FWHM=184$\pm$25 and 550$\pm$65 km s$^{-1}$. The last three columns list the gas masses estimated by assuming R$_{\rm 21}$=1 and $\alpha_{\rm CO}$=4.35$\pm$1.30 M$_{\sun}(\rm K~km~s^{-1}~pc^2)^{-1}$, corresponding depletion timescales (t$_{\rm dep}$=M$_{\rm H_2}$/SFR) and H$_2$ gas fractions (f$_{\rm H_2}$=M$_{\rm H_2}$/M$_{\rm *}$).}
\label{tab4}
\end{table*}


As can be seen from Section \ref{individual} and Appendices \ref{appendixa} and \ref{appendixb}, most of the continuum maps are either compact (J0232, J1010, and J1100; having deconvolved sizes of 0.2, 0.4, and 0.5 kpc, respectively) or slightly asymmetric (J1152 and J1356; 0.3$\times$0.2 and 0.6$\times$0.4 kpc$^2$), with the peak indicating the AGN position. This is not the case for the Teacup (J1430) and J1509, which show jet-like morphologies with sizes and PAs of 0.6$\times$0.4 kpc$^2$ and $\sim$80\degr, and 1.5$\times$1.1 kpc$^2$ and $\sim$135\degr, respectively (see Table \ref{tab3}). These millimeter morphologies are similar to the centimeter morphologies depicted from 6 GHz VLA data at 0.25\arcsec~resolution for the four QSO2s in common with \citet{2019MNRAS.485.2710J}. J1010 and J1100 appear compact in the VLA images, whilst J1356 and J1430 are extended. According to these authors, the extended radio structures most likely correspond to jets (see Table \ref{tab3}). In the case of J1356 the structures that we detect at 3$\sigma$ in our ALMA continuum image have different PA than the radio axis derived from VLA data (see Section \ref{J1356} and Table \ref{tab3}). For J1430, the PA from the ALMA continuum map is very similar to the PAs measured from the VLA data at 6 GHz (60\degr) and also from FIRST (77\degr; \citealt{2014MNRAS.441.3306H}). All the continuum sizes, PAs, flux densities and corresponding rms values measured from our ALMA data are reported in Table \ref{tab3}, together with the PAs and morphologies measured from the VLA observations of the QSO2s \citep{2019MNRAS.485.2710J}. 

For J1010, J1100, J1356, and J1430 we can calculate spectral indices using the ALMA 200-220 GHz observed fluxes and the 1.5, 5.2, and 7.2 GHz VLA fluxes reported in \citet{2019MNRAS.485.2710J}. For the other three sources we used the 1.4 GHz FIRST integrated fluxes instead. We note that the difference in angular resolution between the FIRST and ALMA data could have an impact on the determination of the spectral index of the QSO2s with extended radio morphologies (i.e., J1509 and, to a lesser extent, J1152). We obtain spectral indices ($\alpha$, with S$_{\nu}\propto\nu^{\alpha}$) between -0.55 and -0.90, which correspond to steep spectrum radio sources (see Table \ref{tab3}). Thus, even the QSO2s with compact radio morphologies in the sample (J0232, J1010 and J1100) have steep spectra, similar to compact HERGs of intermediate radio power at z$<$0.1 \citep{2020MNRAS.494.2053P}. This suggests that the observed millimeter continuum emission of the QSO2s most likely corresponds to synchrotron radiation from particles accelerated by shocks and/or small-scale jets. 



\begin{figure*}
\centering
\includegraphics[width=2\columnwidth]{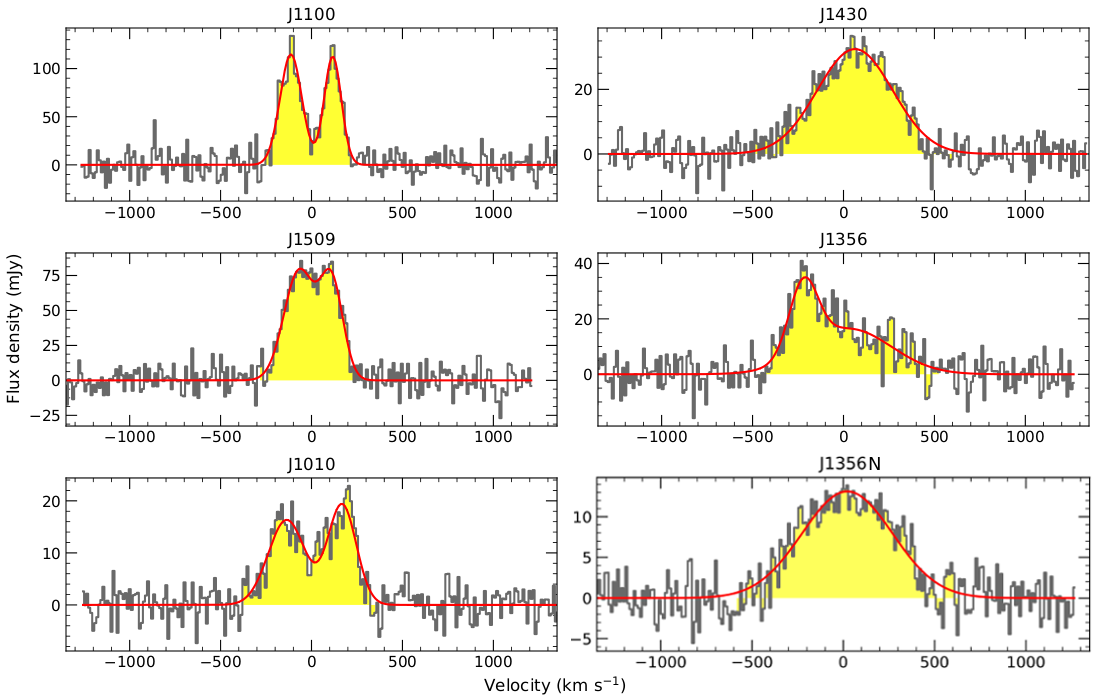}
\caption{Integrated spectra of the five QSO2s with CO(2$-$1) detection. J1010, J1100 and J1509 show double-peaked CO profiles characteristic of rotating disks. J1356 and J1430 show very broad CO profiles, indicating the presence of different emission-line components. For J1356 we also include the spectrum of the N nucleus only (J1356N). In all panels, v=0 km~s$^{-1}$ corresponds to the CO frequencies listed in Table \ref{tab4}.} 
\label{fig3bis}
\end{figure*}

\subsection{Molecular gas content of the QSO2s}
\label{molecular}

\subsubsection{CO morphologies}

We detect CO(2$-$1) emission in the five QSO2s included in Section \ref{individual}, and the CO morphologies are very diverse. The two spiral galaxies, J1100 and J1509, are the most extended, showing CO emission at $\geq$3$\sigma$ up to distances of 5 and 4 kpc from the AGN position, respectively. Their CO emission follows the spiral arms and bars, as is normally the case for this type of galaxy (see e.g., Figure 1 in \citealt{2017ApJ...846..159B}). 
In the case of J1356, if we only consider the N nucleus of J1356 we measure R$_{\rm CO}\sim$1.7 kpc, but we detect CO at $\geq$3$\sigma$ in the W arm up to a distance of 5 kpc from the QSO2 nucleus (see Table \ref{tab4}). 
Finally, J1010 and J1430 show double-peaked morphologies that do not have optical counterpart (see top panels of Figures \ref{fig2a} and \ref{fig6aa}), with peak separations of 1.25 and 1.3 kpc, respectively, and almost aligned with the CO kinematic major axis. In the case of J1430, the jet is perpendicular to the two peaks, whereas in the case of J1010 the continuum morphology appears rather compact and round (0.4$\times$0.4 kpc$^2$). This CO morphologies could have been produced by the action of AGN feedback in the central kpc of these QSO2s (see Sections \ref{J1010} and \ref{Teacup}).

The integrated CO spectra are shown in Figure \ref{fig3bis}. We extracted them from the line emitting regions at $\geq$3$\sigma$ in the combined data. 
The two spiral galaxies and J1010 show double-peaked CO profiles, typical of rotating disks. J1430 shows a broad single-peaked CO line profile with a full width at half maximum (FWHM) of 485 km~s$^{-1}$. Finally, J1356 shows an asymmetric CO profile, with a prominent red wing and a blueshifted peak. This profile corresponds to emission from the N nucleus (broad CO component) and the W arm (blueshifted peak). Since J1356 is an ongoing merger system, in Table \ref{tab4} we also report the CO(2$-$1) flux measured for the N nucleus only, which is the one shown in Figures \ref{fig5aa} and \ref{fig5cc}.
For the red ETGs, J0232 and J1152, we estimated 3$\sigma$ upper limits for the fluxes by assuming FWHM=430 km~s$^{-1}$ and considering a circular aperture of R$_{\rm CO}$=2.9 kpc, which are the average values of the three ETGs with CO detections (J1010, J1356, and J1430).
All the CO(2$-$1) fluxes are reported in Table \ref{tab4}. Corresponding errors include the uncertainty associated with the measurement of the spectra and the 10\% of flux calibration error. 

Finally, we measured the flux of the central kpc of the galaxies (r=0.5 kpc; except for J1010, for which we used r=0.72 kpc instead because of its lower angular resolution) to estimate the percentage of the total emission that it represents. We find that the central kpc of the spiral galaxies contains $\sim$5--12\% of the total molecular gas, whereas in the interacting, merging and post-merger systems it represents between 18 and 25\% (in the case of J1356 it is 32\% if we consider the N nucleus only; i.e., J1356N in Table \ref{tab4}).
Merging systems are thus more centrally concentrated than the spiral galaxies in our sample. Large nuclear molecular gas concentrations of between 38 and 75\% were also reported for a sample of four PG-quasars at z$\sim$0.06 and L$_{\rm bol}\sim$10$^{45}$ erg~s$^{-1}$ observed in CO(2$-$1) with ALMA \citep{2020ApJ...898...61I}. These concentrations correspond to the gas in the central 700 pc as compared with the inner 2 kpc of the PG-quasars.
Larger AGN and control samples need to be observed with high angular resolution millimeter data to investigate any relation between the distribution of molecular gas and galaxy morphology, AGN luminosity, Eddington ratio and/or outflow properties \citep{2021arXiv210410227G}.

\begin{figure*}
\centering
\includegraphics[width=1.4\columnwidth]{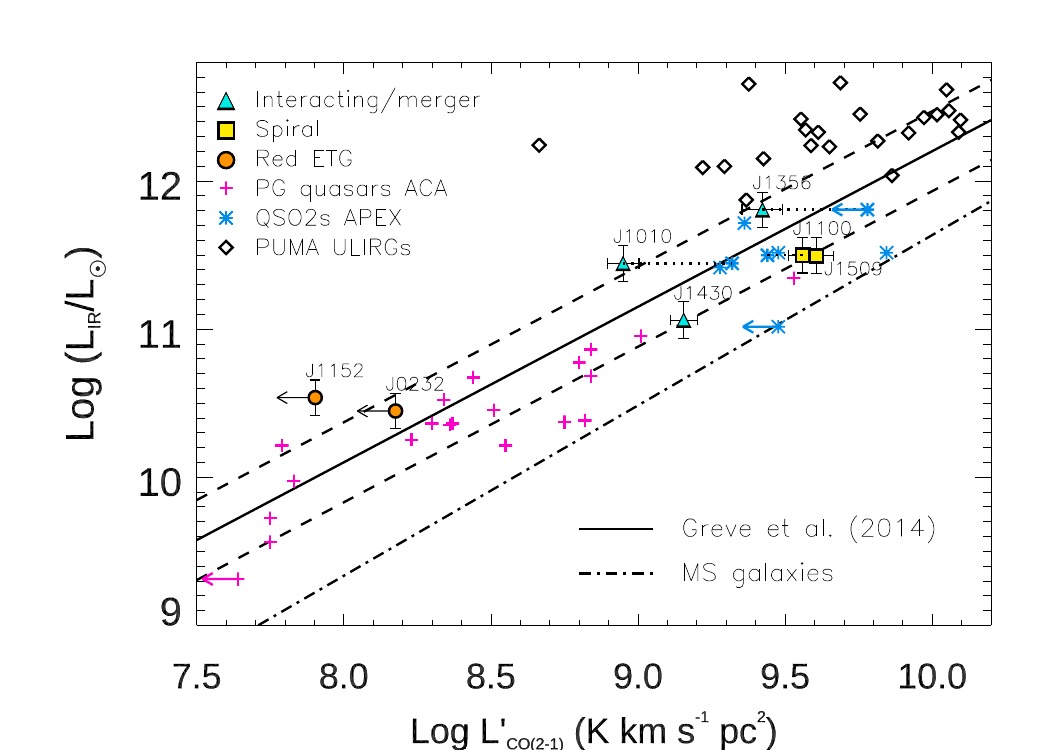}
\caption{IR (8--1000 $\mu$m) versus CO(2$-$1) luminosities. The QSO2s in our sample are shown with same colors and symbols as in Figure \ref{fig2}. The solid line is the relation (log L$_{\rm IR}$ = 1.05 log L$^{\prime}_{\rm CO}+1.7$) found for LIRGs and ULIRGs at z$\leq$0.1 and dusty star forming galaxies at z$>$1 from \citet{2014ApJ...794..142G}. The dashed lines indicate the correlation scatter, of 0.27 dex. The dot-dashed line is the fit of star-forming galaxies with z=0--2 from \citet{2010MNRAS.407.2091G}. This relation (log L$_{\rm FIR}$ = 1.15 log L$^{\prime}_{\rm CO}+0.02$) has a scatter of 0.31 dex and it corresponds to CO(1$-$0) instead of CO(2$-$1), so we assume r$_{\rm 21}$=1 and use L$_{\rm IR}$=1.3$\times$L$_{\rm FIR}$ to estimate total IR luminosity as indicated in \citet{2010MNRAS.407.2091G}. Pink plus symbols are the PG quasars at z$<$0.1 from \citet{2020ApJS..247...15S}, open diamonds the PUMA ULIRGs at z$<$0.16 from \citet{2021A&A...651A..42P}, and blue asterisks the QSO2s at z$\sim$0.1 from \citet{2020MNRAS.498.1560J}. The APEX L$^{\prime}_{\rm CO(2-1)}$ of J1010 is higher than the corresponding ALMA value because it includes two companion galaxies, it is lower in the case of J1100, equal for J1430, and for J1356 is an upper limit.} 
\label{fig4}
\end{figure*}





\subsubsection{Integrated CO luminosities}
\label{integrated}

Using the fluxes reported in Table \ref{tab4} and following equation 3 in \citet{2005ARA&A..43..677S} we calculated the CO luminosities 
reported in Table \ref{tab4}. 
In Figure \ref{fig4} we show L$_{\rm IR}$ (8--1000 $\mu$m) versus L'$_{\rm CO(2-1)}$ for our QSO2s, with different symbols and colors indicating whether the galaxies are spirals, red ETGs or interacting, merging and post-merger systems. Quasars of roughly the same bolometric luminosity (log L$_{\rm bol}$=45.7--46.3 erg~s$^{-1}$) and hosted in galaxies of similar stellar masses (log M$_*$=10.9--11.3 M$_{\rm \sun}$) have CO luminosities spanning over more than one order of magnitude. This is in contradiction with the idea that all low-redshift, optically-selected quasars reside in gas-rich host galaxies and not in ellipticals (e.g., \citealt{2003ApJ...585L.105S,2020ApJS..247...15S}).

For comparison, in Figure \ref{fig4} we show the IR-CO(2$-$1) relation of \citet{2014ApJ...794..142G}, derived from a sample of nearby LIRGs and ULIRGs at z$\leq$0.1 and submillimeter galaxies at z$>$1. We also include ALMA Compact Array (ACA) measurements for PG quasars at z$<$0.1 with $<$L$_{\rm bol}$ $>$=44.27$\pm$0.36 erg~s$^{-1}$ from \citet{2020ApJS..247...15S}. These bolometric luminosities are closer to the values measured for Seyfert galaxies than for quasars, and indeed, they have lower IR and CO luminosities than the five QSO2s with CO(2$-$1) detections. The beamsize of the ACA CO(2$-$1) observations is 7.4\arcsec$\times$4.8\arcsec~(13.7$\times$8.9 kpc$^2$ at z=0.1), and the IR luminosities were estimated from SED fitting. 
We also include in Figure \ref{fig4} the nine QSO2s at z$\sim$0.1 from \citet{2020MNRAS.498.1560J}. These targets have [OIII] luminosities above 10$^{8.4}L_{\rm \sun}$, 1.4 GHz luminosities $\geq$10$^{23.5}$~W~Hz$^{\rm -1}$ and ionized outflows with FWHM\ga700 km~s$^{-1}$. These criteria exclude QSO2s in red ETGs like J0232 and J1152, which show slower ionized outflows and lower radio luminosities. Four of our five radio luminous QSO2s are in \citet{2020MNRAS.498.1560J} and thus they have APEX CO(2$-$1) fluxes measured in an aperture of $\sim$28\arcsec~(52 kpc at z=0.1). With ALMA we measure exactly the same flux for J1430, and a higher flux for J1100 (see Figure \ref{fig4}). For J1356 an APEX upper limit is reported in \citet{2020MNRAS.498.1560J}. Finally, for J1010 the APEX CO(2$-$1) flux is overestimated by a factor of 2 because the large aperture includes two companion galaxies that also emit in CO (see Appendix \ref{appendixc} and Table \ref{tab4}). 
We inspected the optical morphologies of the other five QSO2s studied in \citet{2020MNRAS.498.1560J} using color-combined SDSS images and all of them are also disks and/or interacting
galaxies. This is why they all show large CO and IR luminosities in Figure \ref{fig4}, occupying the same region as our spirals, interacting, and merging QSO2s. 
As we discuss in Section \ref{reservoirs}, selecting luminous QSO2s that have strong ionized outflows and high radio powers biases the samples to have large molecular gas reservoirs and high SFRs. 

These gas-rich QSO2s lie close to the \citet{2014ApJ...794..142G} relation, which is reasonable considering that all of them are LIRGs. Indeed, most of the galaxies at z$\leq$0.1 in \citet{2014ApJ...794..142G} are LIRGs. For further comparison, in Figure \ref{fig4} we also include the PUMA sample of ULIRGs at z$<$0.16 from \citet{2021A&A...651A..42P}. These 23 ULIRGs, observed with ALMA in CO(2$-$1) at an angular resolution of $\sim$400 pc, are above the Greve relation. Thus, according to their IR and CO luminosities, QSO2s show intermediate values between those of MS galaxies (dot-dashed line in Figure \ref{fig4}) and nearby ULIRGs. 



\subsubsection{Molecular gas masses}

Assuming that the CO emission is thermalized and optically thick, the CO luminosity is independent of J and of rest frequency, and thus the brightness temperature ratio, R$_{21}$=CO(2$-$1)/CO(1$-$0)=1 \citep{1992A&A...264..433B,2005ARA&A..43..677S}. Since we do not have CO(1$-$0) fluxes measured for our QSO2s, we assume R$_{21}$=1 instead of using average values from the literature, which usually range between 0.5 and 1.2 (see e.g., \citealt{2020ApJS..247...15S} and references therein) and are missing for the particular case of QSO2s. The only exception is J1356, for which a total CO(1$-$0) luminosity of (1.03$\pm$0.32)$\times10^9$ K km s$^{-1}$ pc$^2$ is reported in \citet{2014ApJ...790..160S}, measured from ALMA Cycle 0 data. This value is considerably lower than the CO(2$-$1) luminosity reported in Table \ref{tab4}, probably due to the best sensitivity of our combined ALMA Cycle 6 data. 

In order to estimate molecular gas masses, we also need to assume a CO-to-H$_2$ conversion factor ($\alpha_{\rm CO}$). For easier comparison with the literature, we chose the Milky Way value $\alpha_{\rm CO}$=4.35$\pm$1.30 M$_{\sun}(\rm K~km~s^{-1}~pc^2)^{-1}$ from  \citet{2013ARA&A..51..207B}.  
The same or similar values have been used to estimate the gas masses of the COLD GASS \citep{2012ApJ...758...73S,2016MNRAS.462.1749S} and ATLAS$^{\rm 3D}$ surveys \citep{2011MNRAS.414..940Y}, the type-1 quasars in \citet{2017MNRAS.470.1570H} and the QSO2s in \citet{2020MNRAS.498.1560J}. We discuss results from these surveys in Section \ref{reservoirs}. 
Moreover, the Galactic factor is very close to the peak value of the distribution of conversion factors estimated for the xCOLD GASS high-mass sample (log M$_*$/M$_{\rm \sun}>$10.0; see Figure 7 in \citealt{2017MNRAS.470.4750A}). Furthermore, this $\alpha_{\rm CO}$ distribution is narrow, indicating that molecular gas scaling relations should not change substantially for massive galaxies at low-redshift. Corresponding gas masses (M$_{\rm H_2}$), depletion timescales due to star formation (t$_{\rm dep}$=M$_{\rm H_2}$/SFR) and molecular gas fractions (f$_{\rm H_2}$=M$_{\rm H_2}$/M$_{\rm *}$) for the QSO2s in our sample are given in Table \ref{tab4}.

\begin{figure*}
\centering
\includegraphics[width=1.45\columnwidth]{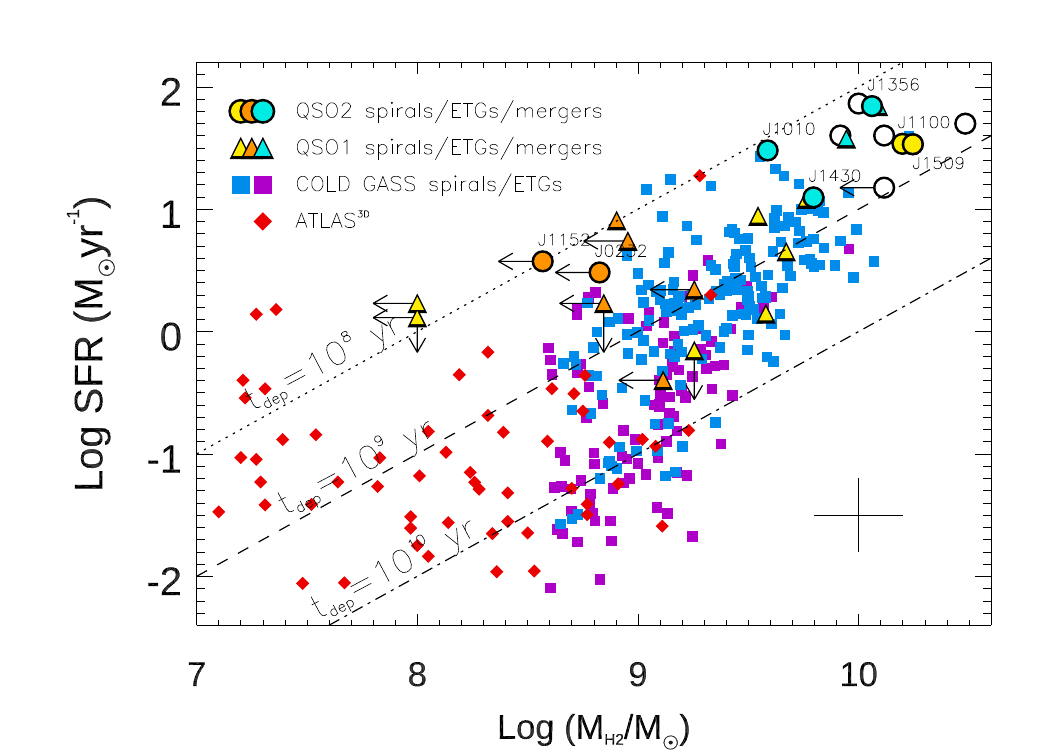}
\caption{SFR versus molecular gas masses for the QSO2s, shown as circles of the same colors as in Fig. \ref{fig2}. Average errors are indicated in the bottom right corner. Dotted, dashed and dot-dashed lines correspond to constant depletion times of 10$^8$, 10$^9$, and 10$^{10}$ yr. Data from different samples are shown as follows: ATLAS$^{\rm 3D}$ are red diamonds, COLD GASS are blue/purple squares (disk/bulge-dominated galaxies), type-1 QSOs from \citet{2017MNRAS.470.1570H} are triangles of the same colors as our QSO2s (orange=bulge-dominated, yellow=disk-dominated, light turquoise=major mergers), and QSO2s from \citet{2020MNRAS.498.1560J} are open circles.}  \label{fig_discussion1}
\end{figure*}

The two spiral galaxies have molecular gas masses of $\sim$2$\times$10$^{10}$~M$_{\rm \sun}$ and the interacting, merging and post-merger galaxies $\sim$4-11$\times$10$^{9}$~M$_{\rm \sun}$. These total gas masses are in good agreement with the values reported in the literature for small samples of QSO2s at z$<$0.3, of $\sim$7-25$\times$10$^{9}$~M$_{\rm \sun}$ \citep{2012ApJ...753..135K,2013MNRAS.434..978V,2020MNRAS.498.1560J}, and larger than the gas masses measured for COLD GASS (purple and blue squares in Figure \ref{fig_discussion1}). On the other hand, the red ETGs have gas masses $<$4-7$\times$10$^{8}$~M$_{\rm \sun}$, consistent with the values reported for bulge-dominated type-1 quasars in \citet{2017MNRAS.470.1570H} and also with those of the ETGs in the ATLAS$^{\rm 3D}$ survey (red diamonds in Figure \ref{fig_discussion1}).

\section{Discussion} 
\label{Discussion}

\subsection{The molecular gas reservoirs of nearby QSO2s}
\label{reservoirs}



\citet{2017MNRAS.470.1570H} studied the CO(1$-$0) and CO(2$-$1) emission of 14 nearby type-1 AGN at z$<$0.2 with log L$_{\rm bol}$=44--46 erg~s$^{-1}$ using NOEMA data\footnote{The host galaxies have M$_*\sim$10$^{11}M_{\rm \sun}$ and they lie in the SFR MS. Their H$\alpha$-based SFRs range from 0.4 to 12 M$_{\rm \sun}$yr$^{-1}$, except for two major merger systems, which have SFRs of 38 and 69 M$_{\rm \sun}$yr$^{-1}$.}. Although some of the bolometric luminosities are more Seyfert- than QSO-like, in the following we refer to them as type-1 QSOs. \citet{2017MNRAS.470.1570H} claimed that galaxy morphology has an influence on the CO content of quasars. Whilst disk-dominated and merging quasars show gas masses typical of star-forming galaxies of the same stellar mass (M$_{\rm H_2}\sim$10$^{9-10}$ M$_{\rm \sun}$), bulge-dominated quasars have M$_{\rm H_2}\le$10$^9$ M$_{\rm \sun}$ (see Figure \ref{fig_discussion1}). This dependence of gas mass on galaxy morphology is also apparent when comparing the disk-dominated massive galaxies in the COLD GASS survey and the bulge-dominated galaxies in the COLD GASS and ATLAS$^{\rm 3D}$ surveys, also shown in Figure \ref{fig_discussion1}.

As can be seen from the same figure, we also find that our interacting, merging and post-merger QSO2s and the spirals have the largest gas masses (4--18$\times$10$^9$ M$_{\rm \sun}$), occupying the upper right corner of Figure \ref{fig_discussion1}. This is also the case of the 5 QSO2s in \citet{2020MNRAS.498.1560J} not included in our sample, shown as open circles. On the other hand, the red ETGs have smaller CO masses ($<$6.7$\times$10$^8$ M$_{\rm \sun}$). We note, however, that J1152 is a post-merger ETG, showing bright tidal features similar to these observed in J1430. The difference between both J0232 and J1152 and the other ETGs in our sample are their redder optical colors, lower SFRs (3 and 3.7 M$_{\rm \sun}$yr$^{-1}$), lower radio luminosities (log L$_{\rm 1.4 GHz}\le$23 W~Hz$^{-1}$) and slower ionized outflows (see Table \ref{tab3}). 

The red ETGs, interacting and merging QSO2s show the shortest depletion times ($<$220 Myr), as also found by \citet{2017MNRAS.470.1570H} for their sample of type-1 quasars using H$\alpha$-based SFRs (see Figure \ref{fig_discussion1}). In the bulge-dominated quasars, the SFR is enhanced relative to non-active ETGs of the same stellar mass (e.g., COLD GASS bulge-dominated galaxies, shown as purple squares in Figure \ref{fig_discussion1}), and the molecular gas masses are small (gas fractions f$_{\rm H_2}<$0.01), leading to short t$_{\rm dep}$. However, as mentioned in Section \ref{sample}, the IR-based SFRs that we measure for the red ETGs are likely overestimated, which would translate into longer depletion timescales. In these objects we might be witnessing either later-stage or less intense AGN feedback, in which outflows are weaker, the molecular gas reservoirs smaller, and the last bursts of star formation are taking place. 


For the five QSO2s with CO detections, we measure M$_{\rm H_2}$=4--18$\times$10$^{9}$ M$_{\rm \sun}$, t$_{\rm dep}$=130--520 Myr and f$_{\rm H_2}$=0.04--0.20. 
These values are similar to those reported for the QSO2s observed with APEX by \citet{2020MNRAS.498.1560J} 
if we use R$_{21}$=1 to convert their CO(2$-$1) luminosities into CO(1$-$0) values: M$_{\rm H_2}$=6--25$\times$10$^{9}$ M$_{\rm \sun}$, t$_{\rm dep}$=100--800 Myr and f$_{\rm H_2}$=0.08--1.0. These gas fractions are higher than ours, {bf possibly} due to the different methods employed to estimate the stellar masses (SED fitting versus NIR magnitudes). Indeed, for the four QSO2s that we have in common, the stellar masses reported by \citet{2020MNRAS.498.1560J} are lower. Potentially, our stellar masses could be contaminated with AGN emission, but they are consistent with those reported by \citet{2019ApJ...873...90S}, also derived from SED fitting (see Section \ref{sample}). 
These values of M$_{\rm H_2}$, t$_{\rm dep}$ and f$_{\rm H_2}$ are representative of disk-dominated, interacting and merging QSO2s, but not of QSO2s in red ETGs such as J0232 and J1152. QSO2s with fast ionized outflows and high radio luminosities (log L$_{\rm 1.4 GHz}>$23.5 W~Hz$^{-1}$) have large reservoirs of molecular gas and high SFRs, as also discussed in \citet{2020MNRAS.498.1560J}. 
Thus, even though AGN feedback is taking place in the form of ionized and molecular gas outflows (see Section \ref{outflows}), there is still plenty of molecular gas and star formation in the spirals and merging QSO2s. 

The gas fractions measured for the five QSO2s with CO(2$-$1) detections (f$_{\rm H_2}$=0.04--0.20) are in between those reported in \citet{2016MNRAS.462.1749S} for massive galaxies (log M$_*$/M$_{\rm \sun}$=10.6--11.2) with specific-SFRs (sSFRs)\footnote{Our QSO2s have log sSFRs=[-10.4, -9.4] yr$^{-1}$.} of log sSFR=-9.95$\pm$0.01 yr$^{-1}$, f$_{\rm H_2}$=0.13$\pm$0.01, and those of galaxies in the MS, f$_{\rm H_2}$=0.03--0.04. J1100 and J1509 (i.e., the spirals) show gas fractions of 0.15 and 0.20, whilst the merging QSO2s have f$_{\rm H_2}$=0.04--0.06. 
The spirals and J1430 have t$_{\rm dep}\sim$500 Myr, and J1010 and J1356 have t$_{\rm dep}\sim$130--170 Myr. These depletion times are shorter than those of massive galaxies (log M$_*$/M$_{\sun}$>10.8) in the MS (1.2-1.5 Gyr; \citealt{2016MNRAS.462.1749S}) and also of massive galaxies with -10.4<log sSFR<-9.6 (i.e., similar to the QSO2s), of $\sim$1 Gyr. 

The interacting and merging QSO2s show the shortest depletion times. This is also the case for the two quasars in major mergers in \citet{2017MNRAS.470.1570H}, shown as turquoise triangles in Figure \ref{fig_discussion1}, and for the interacting QSO2s in \citet{2020MNRAS.498.1560J}, as e.g. J0945+1737, J1000+1242, and J1316+1753, all of them having t$_{\rm dep}<$300 Myr. Even shorter depletion timescales due to SFR only, of tens of Myr, are estimated for ULIRGs (e.g., \citealt{2014A&A...562A..21C}). 
As mentioned in Section \ref{integrated}, our QSO2s show intermediate properties between ULIRGs and MS galaxies in terms of molecular gas content and SFR.


\citet{2017ApJS..233...22S} compared the molecular gas content of BPT-selected active and non-active galaxies in the xCOLD GASS sample with log M$_*$/M$_{\rm \sun}>$10 and matched in sSFR. They found slightly lower molecular gas fractions in AGN (f$_{\rm H_2}\sim$0.014) than in the matched non-AGN sample (f$_{\rm H_2}\sim$0.021). Similar molecular gas contents measured from single-dish radio telescopes were reported by \citet{2018MNRAS.473.5658R} for the LLAMA sample of low-to-intermediate luminosity AGN and matched control sample galaxies, and also by \citet{2020ApJ...899..112S} for a representative sample of 40 z$<$0.3 PG quasars. Nevertheless, for the few luminous AGN in xCOLD GASS, \citet{2017ApJS..233...22S} reported higher gas fractions than those of weaker AGN and similar to sSFR-matched inactive galaxies. Recently, \citet{2021ApJS..252...29K} found that X-ray selected AGN in nearby massive galaxies have higher gas fractions than inactive galaxies matched in stellar mass. Moreover, they find that the more luminous the AGN and the higher its Eddington ratio, the larger the molecular gas mass. 
This is consistent with luminous AGN being more frequently found in more massive star-forming and merging/interacting galaxies (e.g., \citealt{2015MNRAS.452.1841S}), although here we showed that this is not the case for all luminous quasars.

These and other works searched for differences between the gas content of AGN and non-active galaxies in the hope of finding evidence for quenching induced by AGN feedback. However, most of these comparisons show that the AGN gas fractions are equal or larger than those of non-active galaxies matched in stellar mass, indicating that when the AGN is on, generally there is gas and star formation (although this depends on galaxy morphology; see for example the QSO2s in red ETGs). This does not mean that AGN feedback is not having an effect on the molecular gas reservoirs of galaxies, but considering the short AGN lifetimes, and the possibility that every massive galaxy goes through one or several active phases, spotting the action of feedback on the total gas fractions becomes challenging. 
The region we should focus on if we aim to evaluate the effect of the current episode of nuclear activity is the central kpc. This is the region having a dynamical timescale comparable to the periods of nuclear activity and the outflows ($\sim$1--100 Myr; see e.g., \citealt{2004cbhg.symp..169M}). Furthermore, reduced gas fractions relative to star-forming regions in the same host galaxies have been recently measured in the central kpc of AGN using CO observations \citep{2021MNRAS.tmpL..48E}. In our QSO2s, we find smaller concentrations of molecular gas in the central kpc of the spiral galaxies, which are the ones showing more massive molecular outflows and higher Eddington ratios (see Section \ref{outflows}). In the merging QSO2s, we see tentative evidence of AGN feedback modifying the distribution of CO in the central 1-2 kpc of the galaxies (i.e., double-peaked CO morphologies in J1010 and J1430). This could be consistent with an inside-out quenching scenario (e.g., \citealt{2018RMxAA..54..217S}).



In summary, our ALMA data of nearby QSO2s, along with the comparison with the QSO2s in \citet{2020MNRAS.498.1560J} and the QSO1s in \citet{2017MNRAS.470.1570H}, reveal the existence of at least two different scenarios regarding the molecular gas content.
First, the disk-dominated and merging quasars have large molecular gas masses and extended CO morphologies. The gas fractions (f$_{\rm H_2}$=0.04-0.20) are consistent with those of massive non-active galaxies with similar sSFRs, but the depletion timescales of the QSO2s (t$_{\rm dep}$=100--500 Myr) are shorter. The merging QSO2s show more concentrated CO morphologies than the spirals. 
Second, the quasars in red ETGs show smaller molecular gas reservoirs, slower ionized outflows and lower radio luminosities. These red ETGs might be undergoing the last episodes of star formation, leading to short depletion timescales of $<$200 Myr. High resolution millimeter observations of a larger sample of quasars are required to confirm/discard these trends.

\subsection{Cold molecular outflows}
\label{outflows}

\begin{table*}
\caption{Disc model and cold molecular outflow properties.}
\centering
\begin{tabular}{lccccccccccc}
\hline
\hline
ID&  \multicolumn{2}{c}{Disk model}  & \multicolumn{7}{c}{Cold molecular outflow}  \\
 & PA & i & \multicolumn{2}{c}{r$_{\rm out}$} & v$_{\rm out}$ &  S$\Delta v_{\rm CO}$ &  M$_{\rm out}$  & \.M$_{\rm out}$ & t$_{\rm dyn}^{\rm out}$ & t$_{\rm dep}^{\rm out}$ & $\eta$  \\
 & (deg) &  (deg)  & (\arcsec) &  (kpc)   & (km~s$^{-1}$)    & (Jy km s$^{-1}$)   &  (10$^7$ M$_{\rm \sun}$) & (M$_{\rm \sun}$~yr$^{-1}$) & (Myr) & (Gyr) & \\
\hline
J1100 &  69 & 38 & 0.7$\pm$0.3 & 1.3$\pm$0.5   &115$\pm$95  &  1.11$\pm$0.13 & 10.5$\pm$7.8 &  12.2$\pm$9.0 & 11.0 & 1.3 & 0.3\\
J1356 & -70 & 52 & 0.20$\pm$0.05  & 0.4$\pm$0.2       & 310$\pm$40     &  0.10$\pm$0.02 & 1.4$\pm$1.2 &  7.8$\pm$6.9 & 1.4 & 0.8 & 0.1 \\
J1430 &   4 & 38 & 0.3$\pm$0.1   & 0.5$\pm$0.2        & 185$\pm$65       &  0.46$\pm$0.07 & 3.1$\pm$2.4 & 15.8$\pm$12.2 & 2.5 & 0.4 & 1.3\\
J1509 &  82 & 43 & $\leq$1.50 & $\leq$3.00 &$\geq$45   &  $\geq$0.58    & $\geq$6.8 &  $\geq$1.1 & $\leq$66 & $\leq$16 & $\geq$0.03\\
\hline	   					 			    								      
\end{tabular}
\tablefoot{Columns 2 and 3 correspond to the PA and inclination of the fitted disk model. Columns 4, 5, and 6 are the projected outflow radii and velocity estimated from the residual maps and PV diagrams along minor axis. Deprojected outflow velocities and radii can be estimated as v$_{\rm out}/\rm sin(i)$ and r$_{\rm out}/\rm cos(i)$, i being the CO disk inclination. Columns 7, 8, and 9 are the outflow fluxes, masses, and deprojected mass rates calculated as described in Section \ref{methodology}. Mass errors include the uncertainty in $\alpha_{\rm CO}$=0.8$\pm$0.5 M$_{\sun}(\rm K~km~s^{-1}~pc^2)^{-1}$, estimated from \citet{1998ApJ...507..615D}. 
Columns 10, 11, and 12 are the dynamical time of the outflows (t$_{\rm dyn}^{\rm out}$=r$_{\rm out}$/v$_{\rm out}$), depletion timescales due to the outflow (t$_{\rm dep}^{\rm out}$=M$_{\rm H_2}$/\.M$_{\rm out}$), and mass loading factors ($\eta$=\.M$_{\rm out}$/SFR). 
}
\label{tab6}
\end{table*}

\begin{table*}
\caption{Molecular outflow properties for different types of AGN.}
\centering
\begin{tabular}{lcccccc}
\hline
\hline
Object & log L$_{\rm bol}$ & r$_{\rm out}$ &  v$_{\rm max}$ &  \.M$_{\rm out}$         &  M$_{\rm out}$/M$_{\rm tot}$ & Refs \\
     &  (erg~s$^{-1}$)       &  (kpc)       &   (km~s$^{-1}$) & (M$_{\rm \sun}$~yr$^{-1}$) & & \\
\hline
Seyferts         & 43.2-44.2 & 0.03--0.5   & 90--200   & 0.3--5    &   0.001--0.01 &  1,2,3  \\
Jetted Seyferts  & 44.6-44.9 & 0.1--0.5    & 300--650  & 15--30    &   0.03-0.05   &  4,5    \\
QSO2s            & 45.7-46.3 & 0.4--1.3    & 200--350  & 8--16 &   0.002--0.007  &  6      \\
AGN ULIRGs       & 44.7-45.7 & 0.5--1.2    & 600--1200 & 60--400   &  0.02--0.27   &  7,8    \\
PDS\,456         & 47.0      & 1.2--5      & 1000      & 290       &  0.12         &  9      \\
\hline	   					 			    								      
\end{tabular}
\tablefoot{L$_{\rm bol}$ corresponds to AGN bolometric luminosity. Outflow gas masses and corresponding rates have been recalculated when necessary assuming $\alpha_{\rm CO}$=0.8 M$_{\sun}(\rm K~km~s^{-1}~pc^2)^{-1}$ and the outflow geometry corresponding to Equation \ref{eq1}.}
\tablebib{(1) \citet{2019A&A...628A..65A}; (2) \citet{2020arXiv200305663D}; (3) \citet{2021A&A...645A..21G}; (4) \citet{2014A&A...567A.125G}; (5) \citet{2015A&A...580A...1M}; (6) This work; (7) \citet{2010A&A...518L.155F}; (8) \citet{2014A&A...562A..21C}; (9) \citet{2019A&A...628A.118B}.}
\label{tab7}
\end{table*}

From our analysis of the CO kinematics, we find that the bulk of the molecular gas is rotating in the five QSO2s with CO(2$-$1) detection. In addition to this, we detect noncircular gas motions that are on the order of the circular motions. Therefore, after close inspection of the PV diagrams and residual maps shown in Section \ref{individual}, we find that
four of the QSO2s show evidence of coplanar outflow motions (J1100, J1356, J1430, and J1509). The alternative interpretation are vertical inflows, a scenario that is not justified in the central regions of molecular gas disks, where the inflow motions are known to take place inside the plane (driven by angular momentum transport by bars and/or spirals). 
Only in the case of J1356, which is a completely distorted merging system, might it be possible to have a vertical inflow driven by, for example, a tidal tail, but in that case we would observe line splitting in the PV diagrams, something that does not happen. In the case of J1010 the PV diagram along the minor axis indicates that we are witnessing either a coplanar inflow or vertical outflow. Vertical molecular outflows are usually starburst-driven, as in the case of M82 \citep{2002ApJ...580L..21W,2015ApJ...814...83L}. J1010 has strong star formation, and it is an interacting galaxy, making it possible to have a vertical outflow. However, such a vertical outflow would produce line splitting in the corresponding PV diagrams unless spatially extended beyond the projected radius of the disk \citep{2019A&A...632A..61G}. In the case of J1010 the inflow/outflow is barely resolved, so we cannot rule out the vertical outflow scenario.


For the coplanar molecular outflows we measure projected radii r$_{\rm out}\sim$0.4--1.3 kpc, velocities v$_{\rm out}\sim$115--310 km~s$^{-1}$, and masses M$_{\rm out}\sim$1.4--10.5$\times$10$^7$ M$_{\sun}$ (see Table \ref{tab6}). They are compact, we barely resolve some of them despite the high angular resolution of the data. For example, in the case of J1509 we do not resolve the high-velocity gas within the outflow, but we can estimate a lower limit of its mass from the low-velocity gas (see Figure \ref{fig7b}). 
The outflow mass rates range between 8 and 16 M$_{\rm \sun}$~yr$^{-1}$, and $\geq$1.1 M$_{\rm \sun}$~yr$^{-1}$ for J1509 
(see Table \ref{tab6}). Considering the SFRs reported in Table \ref{tab2}, we measure mass loading factors $\eta$=\.M$_{\rm out}$/SFR$\sim$0.1--1.3. Except in the case of J1430, the SFRs are larger than the outflow rates, indicating that these molecular outflows might not be efficient in removing molecular gas ($\eta\leq$1) despite the high AGN luminosity of the QSO2s \citep{2014A&A...562A..21C}. Using the galaxies total molecular gas masses we estimate depletion timescales associated with the outflows (t$_{\rm dep}^{\rm out}$) that range between 400 Myr and 1.3 Gyr (see Table \ref{tab6})\footnote{In the case of J1356 we used the gas mass measured in the N galaxy (J1356N in Table \ref{tab4}), but if we consider the total molecular gas mass in the merger we get t$_{\rm dep}^{\rm out}\sim$1.5 Gyr.}. These are longer depletion timescales than those associated with SFR (100--500 Myr). 

Taken at face value, the low mass loading factors and long depletion timescales that we find for the QSO2s could be indicating that the molecular outflows are not AGN-driven but star formation-driven. However, the fact that precisely in these four QSO2s the outflows are coplanar with the CO disks is indicative of AGN-driven outflows \citep{2021arXiv210410227G}. 
Moreover, we note that our molecular outflow masses are rather conservative, as we only consider noncircular gas motions consistent with outflowing gas along the minor axis (see Sections \ref{methodology} and \ref{individual}). For example, in the case J1356, \citet{2014ApJ...790..160S} reported an outflow mass rate of 115 M$_{\rm \sun}$~yr$^{-1}$ (using Eq. \ref{eq1}), whereas here we are measuring 7.8 M$_{\rm \sun}$~yr$^{-1}$. As indicated in Section \ref{J1356}, if we consider all the high-velocity gas (v$_{\rm out}\ge$300 km~s$^{-1}$) that we detect in CO(2$-$1), we measure the same outflow mass as \citet{2014ApJ...790..160S}, of $\sim$7$\times$10$^{7}$ M$_{\sun}$. However, according to our analysis of the kinematics, this mass would include rotation and/or tangential noncircular motions. Apart from the larger outflow mass, \citet{2014ApJ...790..160S} used the highest outflow velocity (v$_{\rm out}$=500 km~s$^{-1}$) and r$_{\rm out}$=0.3 kpc to compute the outflow mass rate.


\subsubsection{QSO2 molecular outflows in a broader context}

The outflow radii, velocities and mass rates of the QSO2s are intermediate between those of the cold molecular outflows reported for Seyfert galaxies and those of AGN in ULIRGs (see Table \ref{tab7}). 
In Seyfert galaxies, the outflow masses represent 0.1--1\% of the total gas mass in the galaxies, very similar to the QSO2s (0.2--0.7\%; see Table \ref{tab7}). 
It is noteworthy that the jetted Seyfert galaxies NGC\,1068 and IC\,5063 \citep{2014A&A...567A.125G,2015A&A...580A...1M} show faster and more massive molecular outflows. 
The spatial coincidence between the ionized/molecular gas outflow and the radio jet in these two Seyfert galaxies indicates that the molecular gas is being pushed by the combined action of the AGN-driven wind and the jet. These are examples of strong coupling driving massive molecular outflows, which represent between 3--5\% of the total gas mass in the galaxies. 


The other extreme in this comparison of outflow properties are AGN in ULIRGs (e.g., \citealt{2010A&A...518L.155F,2014A&A...562A..21C,2019MNRAS.483.4586F}) and extremely powerful quasars such as PDS\,456 \citep{2019A&A...628A.118B,2020A&A...635A..47H}. These 
objects show the most powerful AGN-driven molecular outflows in the local universe, having radii of 0.5--1.2 kpc, maximum velocities of 600--1200 km~s$^{-1}$ and accounting for 2--27\% of the total molecular gas mass (see Table \ref{tab7}). The AGN bolometric luminosities of these ULIRGs are slightly lower than those of our QSO2s (10$^{44.7-45.7}$ erg~s$^{-1}$), and their morphologies are consistent with major mergers, having SFRs=6.5--140 M$_{\rm \sun}~$yr$^{-1}$. Thus, as discussed in Section \ref{reservoirs}, galaxy morphology and offset from the star-formation MS are linked to the molecular gas content of galaxies, and apparently, also to the strength of the molecular outflows. 
The molecular outflow radii of the ULIRGs can extend to a few kpc when low surface brightness components are considered \citep{2013A&A...549A..51F,2019ApJ...871...37H,2020A&A...633A.163C}. Indeed, it could be possible that having observed our QSO2s in CO(2$-$1) prevents us from detecting a more diffuse neutral or molecular gas phase of the outflows. The neutral component might be detected in HI (n$_H\sim$100 cm$^{-3}$), and the diffuse molecular component in CI(1$-$0) and CI(2$-$1), which have critical densities of 500 and 1000 cm$^{-3}$ \citep{2004MNRAS.351..147P}, or CO(1$-$0), with n$_{H}$=440 cm$^{-3}$ for a T$_{\rm kin}$=20 K \citep{2010ApJ...718.1062Y}.

\begin{figure}
\centering
\includegraphics[width=1.03\columnwidth]{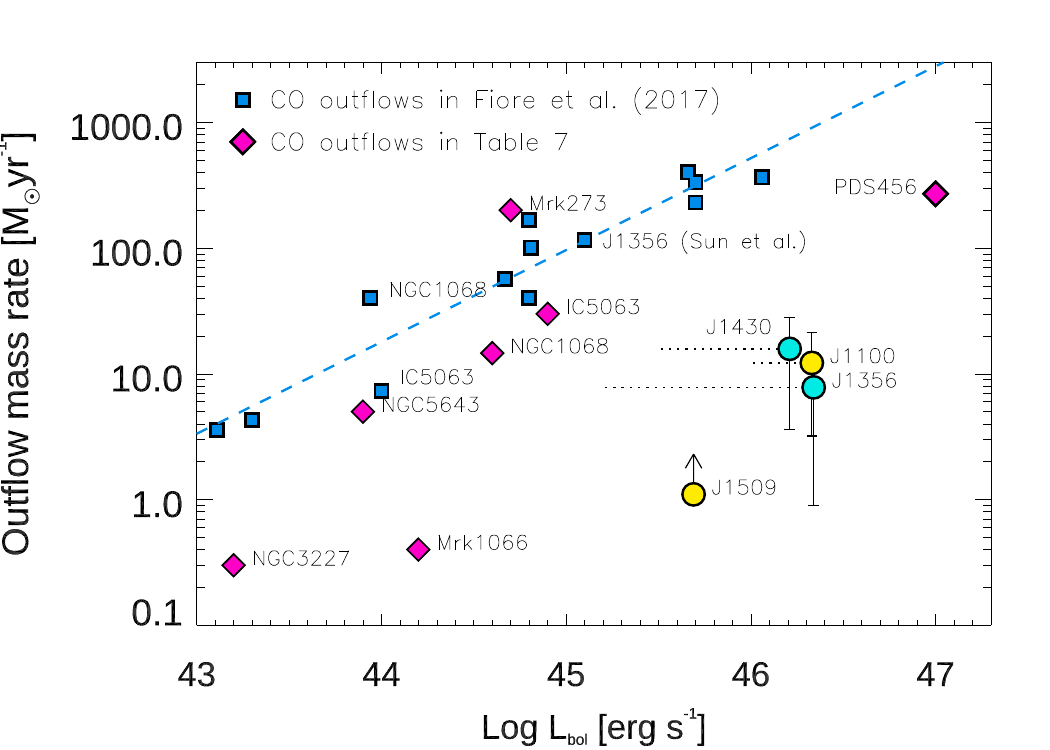}
\caption{Molecular mass outflow rate as a function of AGN luminosity. Circles correspond to the QSO2s. Horizontal dotted lines indicate the position that they would occupy if we use the more conservative L$_{\rm bol}$ values derived from SED fitting from \citet{2019MNRAS.485.2710J}. Blue squares are the CO outflows of Seyferts and ULIRGs compiled by \citet{2017A&A...601A.143F}, but dividing the outflow rates by a factor of 3 to match our outflow geometry (see Eq. \ref{eq1}). Blue dashed line is the linear fit of the blue squares. Pink diamonds are the galaxies included in Table \ref{tab7} that are not in \citet{2017A&A...601A.143F} except NGC\,1068 and IC\,5063, for which they report different AGN luminosities and outflow rates.}  \label{fiore}
\end{figure}

If we plot the mass outflow rates measured for the QSO2s in a \.M$_{\rm out}$ versus L$_{\rm bol}$ diagram (see Figure \ref{fiore}), the QSO2s are significantly below the linear fit of the CO outflow mass rates compiled by \citet{2017A&A...601A.143F}. According to this observational scaling relation, at the average luminosity of the QSO2s (log L$_{\rm bol}\sim$46 erg~s$^{-1}$), we would expect outflow rates of $\sim$500 M$_{\rm \sun}$~yr$^{-1}$. Even if we use the more conservative values of L$_{\rm bol}$ of J1100, J1356, and J1430, derived from SED fitting by \citet{2019MNRAS.485.2710J}, we would expect \.M$_{\rm out}$ values ranging from $\sim$100 to 500 M$_{\rm \sun}$~yr$^{-1}$.
This suggests that is not only AGN luminosity what defines how powerful molecular outflows are. Indeed, the powerful quasar PDS\,456 (log L$_{\rm bol}$=47 erg~s$^{-1}$ and \.M$_{\rm out}\sim$290 M$_{\rm \sun}$~yr$^{-1}$; \citealt{2019A&A...628A.118B}) also has a much lower mass outflow rate than what it would be expected from extrapolating this scaling relation ($\sim$2800 M$_{\rm \sun}$~yr$^{-1}$). The Seyfert galaxies NGC\,3227 and Mrk\,1066, included in Table \ref{tab7}, also fall well below the blue dashed line. In fact, this scaling relation might be the upper boundary of the \.M$_{\rm out}$ versus L$_{\rm bol}$ relation, according to Figure \ref{fiore}, mostly driven by the powerful outflows of the ULIRGs and the jetted-Seyferts. Other factors different from AGN luminosity, including jet power, coupling between wind/jet/ionized outflows and CO disks, and amount/geometry of dense gas in the nuclear regions might be also important for efficiently driving massive molecular outflows. 


Focusing on the individual outflow and galaxy properties of the QSO2s, we find that 
the spiral galaxies, J1100 and J1509, have more massive outflows, of M$_{\rm out}$=10.5 and $\geq$6.8$\times$10$^7$ M$_{\sun}$, respectively (0.7\% and $\geq$0.4\% of the total gas mass), than J1356 and J1430. For them we measure M$_{\rm out}$=1.4 and 3.1$\times$10$^7$ M$_{\sun}$ (0.2\% and 0.5\% of the total gas mass). From the analysis of the kinematics, described in Section \ref{individual}, we found that the ionized gas outflows reported for J1100 and J1509 must be almost coplanar with the galaxy disks (i.e., more favorable orientation for dragging molecular gas outward; scenario A in Figure \ref{scheme}), whereas those of J1010, J1430, and J1356 would subtend a relatively large angle (scenario B in Figure \ref{scheme}). This could explain the more massive molecular outflows measured for the spiral galaxies, whose ionized outflows, in combination with the jets and winds, might be contributing to drag a larger amount of molecular gas outward in an almost coplanar geometry. We have to bear in mind, however, that the merging QSO2s show more chaotic kinematics than the spirals, and thus, their molecular outflows could have a more 3D geometry than those of the spirals. In that case, the outflow masses of J1356 and J1430 would be underestimated from the methodology used here. A more detailed study of the molecular gas kinematics of these QSO2s will be the subject of a forthcoming work (Audibert et al., in prep.). 

The more massive outflows measured for the spirals, with r$_{\rm out}\sim$1.3 and $\leq$3 kpc, could be responsible for the smaller molecular gas concentrations that we measure in the central kpc (r=0.5 kpc) of the galaxies (5--12\%), in comparison with the merging systems (18--25\%). The radii and dynamical timescales of the QSO2 outflows (r$_{\rm out}$/v$_{\rm out}$; see Table \ref{tab6}), between 1.4 and 11 Myr (and $\leq$66 Myr for J1509), are consistent with them being driven by the current AGN episode and having an effect on the central kpc of the galaxies. 
On smaller spatial scales (the inner $\sim$200 pc), a radial redistribution of the molecular gas induced by AGN feedback has been reported by \citet{2021arXiv210410227G} from a comparison between low and intermediate-luminosity Seyferts (NUGA and GATOS samples). The most luminous Seyferts, which also have the highest Eddington ratios and show evidence for molecular outflows, have smaller molecular gas concentration in the inner 50 pc of the galaxies relative to the inner 200 pc. On this respect, it is noteworthy that the Eddington ratios of the two QSO2s in spiral galaxies are higher (f$_{\rm Edd}$=1.1 and 0.6; see Table \ref{tab1}) than for the merging systems (f$_{\rm Edd}$=0.15, 0.09, and 0.4). Larger samples of AGN are required to continue exploring these tantalizing trends.

\section{Conclusions}
\label{conclusions}

We have explored for the first time the cold molecular gas content traced by the 2--1 line of CO, and adjacent continuum emission, of a sample of nearby (z$\sim$0.1) and luminous L$_{\rm [OIII]}>$10$^{8.5}L_{\rm \sun}$ QSO2s at an angular resolution of $\sim$0.2\arcsec~(370 pc). The ALMA observations permit us to study the molecular gas content, morphology and kinematics of these QSO2s. The main results are summarized as follows. 

   \begin{enumerate}

  \item Quasars of roughly the same bolometric luminosity (log L$_{\rm bol}$=45.7--46.3 erg~s$^{-1}$) and hosted in galaxies of similar stellar masses (log M$_*$=10.9--11.3 M$_{\rm \sun}$) have molecular gas masses ranging from 4--18$\times$10$^{9}$~M$_{\rm \sun}$ to $<$4-7$\times$10$^{8}$~M$_{\rm \sun}$. This is in contradiction with the idea that all low-redshift, optically-selected quasars reside in gas-rich host galaxies and not in ellipticals.  
  
  \item Galaxy morphology and color, radio luminosity, and outflow properties are related with the total molecular gas content of quasars. QSO2s in disks and/or merging systems with high radio luminosities and fast ionized outflows have larger gas fractions than red ETGs with lower radio luminosities and slower ionized outflows. 
  
  
  \item QSO2s show intermediate properties between those of local ULIRGs and MS galaxies regarding molecular gas content and SFRs. They have depletion times associated with star formation ranging from $\sim$100 to 500 Myr.
   
  \item The CO(2-1) morphologies of QSO2s in the merging systems are more centrally concentrated than those in the spiral galaxies. The central kpc (r=0.5 kpc) of the galaxies contains $\sim$5--12\% of the total molecular gas in the case of the spirals, and between 18 and 25\% in the interacting, merging and post-merger systems.  
  
  \item We find more massive and extended molecular outflows in the spiral galaxies than in the merging systems. The spirals have ionized outflows that are mostly coplanar with the CO disks/galaxies, whereas in the case of the merging systems the ionized outflows would subtend a larger angle relative to the CO disks/galaxies. Furthermore, the QSO2s in spirals have larger Eddington ratios than the merging QSO2s. This could explain the smaller molecular gas concentrations in the central kpc of the spirals. 
  
  \item The outflow mass rates that we measure (8--16 M$_{\rm \sun}$~yr$^{-1}$) are much lower than expected from observational scaling relations with AGN luminosity (e.g., \citealt{2017A&A...601A.143F}). This implies that is not only the AGN luminosity that defines how powerful molecular outflows are. Other factors including jet power, coupling between wind, jet, and/or ionized outflows and the CO disks, and amount/geometry of dense gas in the nuclear regions might be also relevant.  
  
  \item We use the total molecular gas masses and deprojected outflow mass rates to estimate depletion timescales associated with the outflows. We measure t$_{\rm dep}^{\rm out}$ ranging between 400 Myr and 1.3 Gyr. These are larger depletion timescales than those associated with SFR, leading to low mass loading factors of $\eta\sim$0.1--1.3. Except in the case of the Teacup, star formation is more effectively consuming molecular gas than the outflows.
  
  \item Taken at face value, the low mass loading factors and long depletion timescales that we find for the QSO2s could be indicating that the molecular outflows are not AGN-driven but star formation-driven. However, the fact that in four of the five QSO2s these outflows are coplanar with the CO disks is indicative of AGN-driven outflows.
     \end{enumerate}

We do not find evidence for a significant impact of the QSO2s' molecular outflows, or more generally, of quasar feedback, on the total molecular gas reservoirs and star formation rates. However, they appear to be modifying the distribution of cold molecular gas in the central kiloparsec of the galaxies. Similar high-resolution mm observations as those presented here of a larger sample of QSO2s are required to confirm this.


\begin{acknowledgements}
This paper makes use of the following ALMA data: ADS/JAO.ALMA\#2018.1.00870.S. ALMA is a partnership of ESO (representing its member states), NSF (USA) and NINS (Japan), together with NRC (Canada), MOST and ASIAA (Taiwan), and KASI (Republic of Korea), in cooperation with the Republic of Chile. The Joint ALMA Observatory is operated by ESO, AUI/NRAO and NAOJ.
CRA and MB thank Anna L\'opez (IRAM) and Luke Maud (ESO) for their support with the data processing. CRA thanks R. Morganti, J. Falc\'on-Barroso, and J. A. Acosta-Pulido for useful discussions. CRA also thanks B. Husemann for kindly providing the data to produce Figure 17 and G. P\'erez D\' iaz (SMM IAC) for designing Figure 4. 
CRA acknowledges financial support from the European Union's Horizon
2020 research and innovation programme under Marie Sk\l odowska-Curie
grant agreement No 860744 (BiD4BESt). CRA and PSB acknowledge support from the project 
``Feeding and feedback in active galaxies'', with reference
PID2019-106027GB-C42, funded by MICINN-AEI/10.13039/501100011033. 
CRA and AA acknowledge the projects ``Quantifying the impact of quasar feedback on galaxy evolution'', with reference EUR2020-112266, funded by MICINN-AEI/10.13039/501100011033 and the European Union NextGenerationEU/PRTR, and from the Consejer\' ia de Econom\' ia, Conocimiento y Empleo del Gobierno de 
Canarias and the European Regional Development Fund (ERDF) under grant ``Quasar feedback and molecular gas reservoirs'', with reference ProID2020010105, ACCISI/FEDER, UE.
MB and CF acknowledge support from PRIN MIUR project ``Black Hole winds and the Baryon Life Cycle of Galaxies: the stone-guest at the galaxy evolution supper'', contract 2017PH3WAT. AAH and SGB acknowledge financial support from grant PGC2018-094671-B-I00 (MCIU/AEI/FEDER,UE). SGB acknowledges support from the research project PID2019-106027GA-C44 of the Spanish MICINN-AEI. 
AAH work was done under project No. MDM-2017-0737 Unidad de Excelencia ``María de Maeztu''- Centro de Astrobiolog\' ia (INTA-CSIC). MPS acknowledges support from the Comunidad de Madrid through the Atracci\'on de Talento Investigador Grant 2018-T1/TIC-11035 and PID2019-105423GA-I00 (MCIU/AEI/FEDER,UE). JP acknowledges support from STFC [ST/V000624/1]. We finally thank the anonymous referee for her/his constructive report.
\end{acknowledgements}

\bibliographystyle{aa} 
\bibliography{biblio} 

\begin{appendix}

\section{SDSS J023224.24-081140.2 (J0232)}
\label{appendixa}

This QSO2 is hosted in the most compact galaxy of our sample, with a major axis of 6\arcsec~(11 kpc) 
measured from the r-band SDSS image (see Table \ref{tab2}). Its optical morphology resembles an undisturbed elliptical or lenticular galaxy (ETG; see Figure \ref{fig1}). From the SDSS magnitudes we measure M$_u$-M$_g$=1.45, which is closer to the typical colors of red-sequence galaxies. Together with J1152 it is one of the least powerful radio and FIR emitters in the sample (see Figure \ref{fig2}). The SFR that we estimate from the extrapolated FIR luminosity (see Table \ref{tab2} for details), of 1.7 M$_{\sun}$~yr$^{-1}$, places it in the MS of local SDSS DR7 star-forming galaxies (see Section \ref{sample}). 
Regarding the ionized gas kinematics, the [OIII] lines in the SDSS spectrum show a blue wing that can be reproduced with a Gaussian of FWHM$\sim$770 km~s$^{-1}$, blueshifted by -370 km~s$^{-1}$ relative to the narrow component \citep{2014MNRAS.440.3202V}.

\begin{figure}[!h]
\FloatBarrier
\centering
\includegraphics[width=0.85\columnwidth,angle=-90]{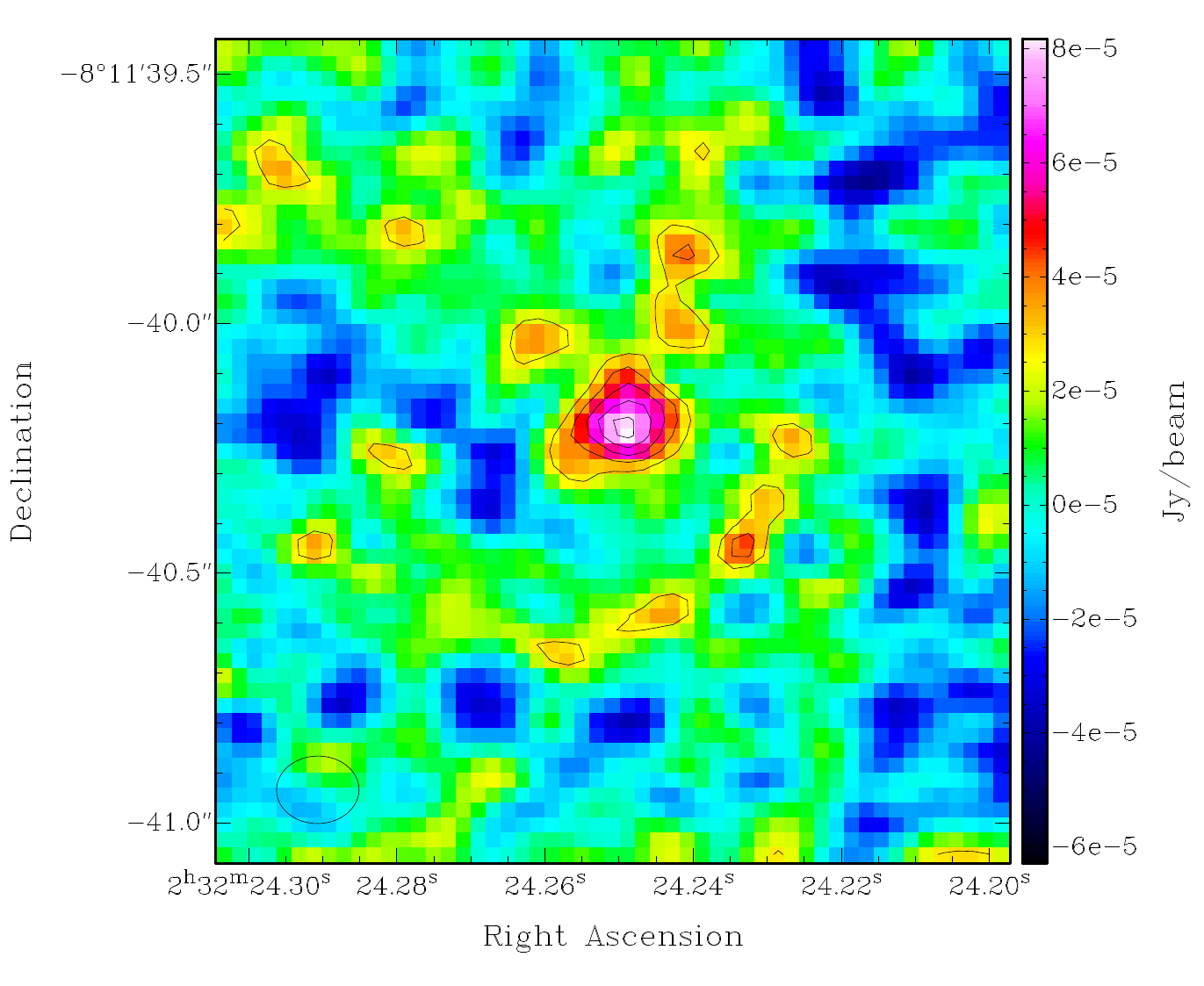}
\caption{1.48 mm (203 GHz) continuum map of J0232 with contours at 2, 3, 4, 5, and 6$\sigma$ overlaid in black ($\sigma$=0.013 mJy/beam). The beam size (0.18\arcsec$\times$0.15\arcsec) is shown in the bottom left corner. North is up and east to the left.} 
\label{fig1aa}
\end{figure}

Our ALMA observations reveal compact 1.3 mm continuum emission at $\ge$3$\sigma$, with a deconvolved size of 0.13\arcsec$\times$0.09\arcsec~(240$\times$166 pc$^2$) and PA=145$\pm$50\degr~(see Figure \ref{fig1aa} and Table \ref{tab3}). Using our continuum flux and the FIRST 1.4 GHz flux we measure a spectral index $\alpha$=-0.67, which corresponds to a steep spectrum. 
We do not detect CO(2$-$1) emission above 3$\sigma$, but from this non-detection we can estimate an upper limit on the gas mass by assuming R$_{\rm 21}$=1 and $\alpha_{\rm CO}$=4.35 M$_{\sun}(\rm K~km~s^{-1}~pc^2)^{-1}$, which yields M$_{\rm H_2}<$ 6.7$\times 10^8 M_{\rm \sun}$ (see Section \ref{molecular} for details). 

\section{SDSS J115245.66+101623.8 (J1152)}
\label{appendixb}

The host galaxy of this QSO2 is classified as an 
elliptical galaxy in the Galaxy Zoo project \citep{2013MNRAS.435.2835W}, based on the SDSS images (see Figure \ref{fig1}). \citet{2018ApJ...856..102F} reported no signs of galaxy interactions from the analysis of the HST/ACS continuum image. However, deeper optical imaging observations obtained in January 2020 with the Wide Field Camera on the 2.5 m Isaac Newton Telescope (La Palma, Spain), reveal spectacular shells and a tidal tail of several kpc \citep{2021MNRAS.tmp.2947P}. These features are the result of a past merger or interaction with another galaxy, and they coincide with the faint structures toward the NE (shell) and SW (fan and tail) shown in the SDSS image (see Figure \ref{fig1}). Finally, from inspection of the HST/ACS image of J1152 we detect a dust lane crossing the galaxy W of the nucleus. The HST/ACS [OIII] image reveals a spectacular biconical morphology (PA=10\degr) with bubbles and ripples showing a remarkable resemblance to the Teacup galaxy \citep{2012MNRAS.420..878K}. 
HST/STIS spectroscopy with a slit PA=10\degr~reveals mainly rotation-dominated kinematics (positive velocities to the N and negative to the S) with a high-amplitude rotation curve (peak velocities $\pm$200 km~s$^{-1}$). noncircular motions are restricted to a radius of 130 pc NE of the nucleus, with FWHM=360 km~s$^{-1}$ and velocities of up to -300 km~s$^{-1}$ \citep{2018ApJ...856..102F}. This is the slowest ionized outflow in our sample  (see Table \ref{tab4a}).

\begin{figure}[!ht]
\FloatBarrier
\centering
\includegraphics[width=0.8\columnwidth,angle=-90]{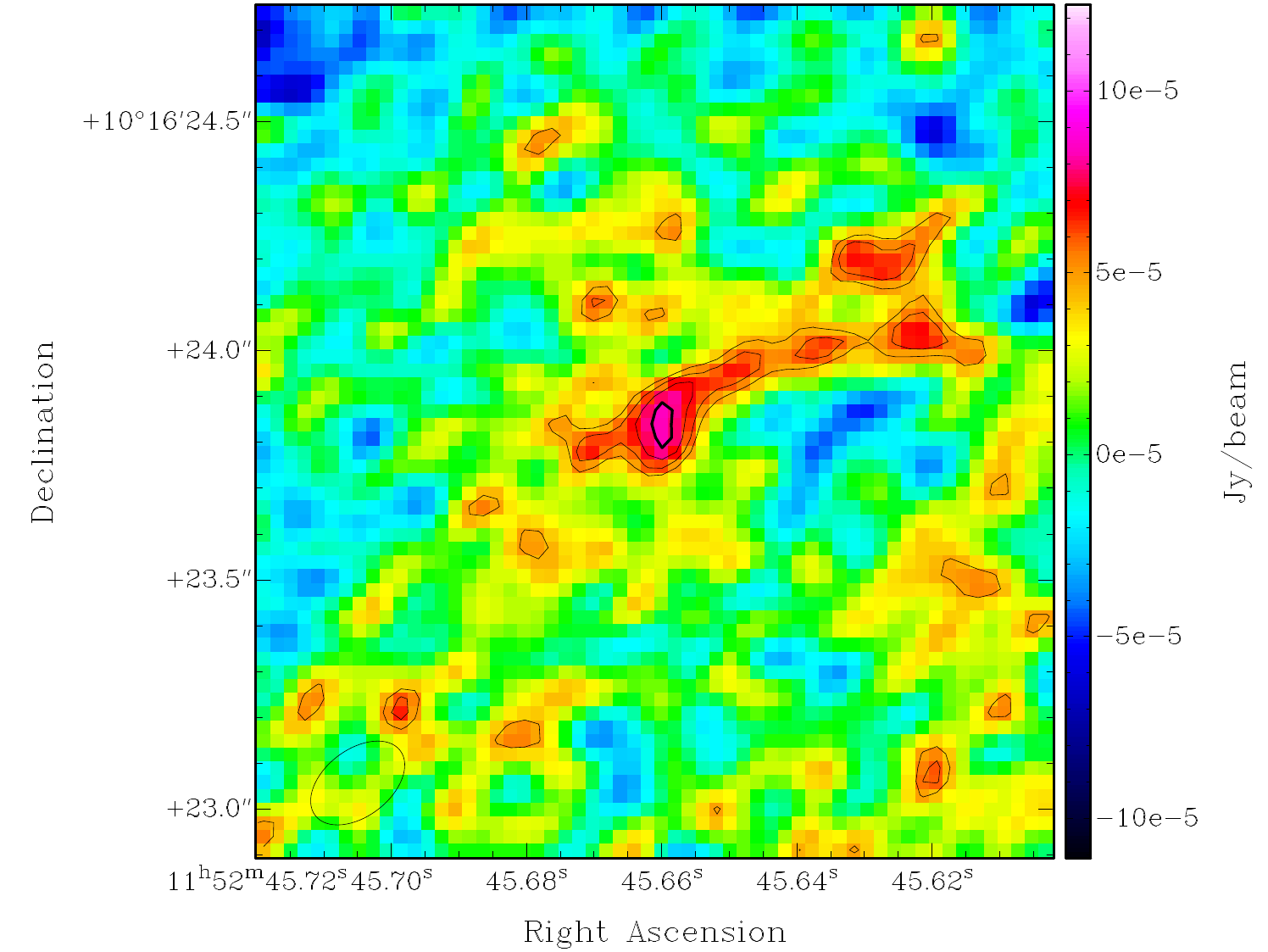}
\caption{1.35 mm (222 GHz) continuum map of J1152 with contours at 2, 2.5, 3, and 3.5$\sigma$ in black ($\sigma$=0.023 mJy/beam). The beam size (0.23\arcsec$\times$0.15\arcsec) is shown at the bottom left corner. North is up and east to the left.} 
\label{fig4aa}
\end{figure}

This QSO2 was observed in band 6 with ALMA because of its redshift (z=0.07). We detect continuum emission at $>$3$\sigma$, although weaker than in the other QSO2s (except J0232). The continuum in the case of J1152 is centered at 1.2 mm (246.7 GHz) rest-frame because of the spectral setup chosen. At $\ge$3$\sigma$ the continuum appears compact, although there is a fainter elongated structure toward the NW detected at 2.5$\sigma$ (see Figure \ref{fig4aa}), but similar structures at the same level are detected across the whole field-of-view. We measure a deconvolved size of 0.20\arcsec$\times$0.12\arcsec~(267$\times$160 pc$^2$) with PA=142$\pm$80\degr. 
In the cm regime, J1152 has the lowest 1.4 GHz luminosity in our sample, but it still shows a radio excess in the radio-FIR diagram shown in Figure \ref{fig2}. 
Using our ALMA continuum flux and the FIRST 1.4 GHz flux we measure a spectral index $\alpha$=-0.65, very similar to J0232.

The host galaxy appears depleted of CO(2$-$1), which is somehow surprising considering the bright morphological features detected in the optical image, which are the product of a past gas-rich interaction with another galaxy (see \citealt{2011MNRAS.410.1550R} and references therein). 
As we did for J0232, from this non-detection we can estimate an upper limit to the gas mass of M$_{\rm H_2}<$ 3.7$\times 10^8 M_{\rm \sun}$, by assuming R$_{\rm 21}$=1 and $\alpha_{\rm CO}$=4.35 (see Section \ref{molecular} for details).

\section{SDSS J101043.36+061201.4 (J1010)}
\label{appendixc}

\begin{figure}[!h]
\centering
\includegraphics[width=0.80\columnwidth,angle=-90]{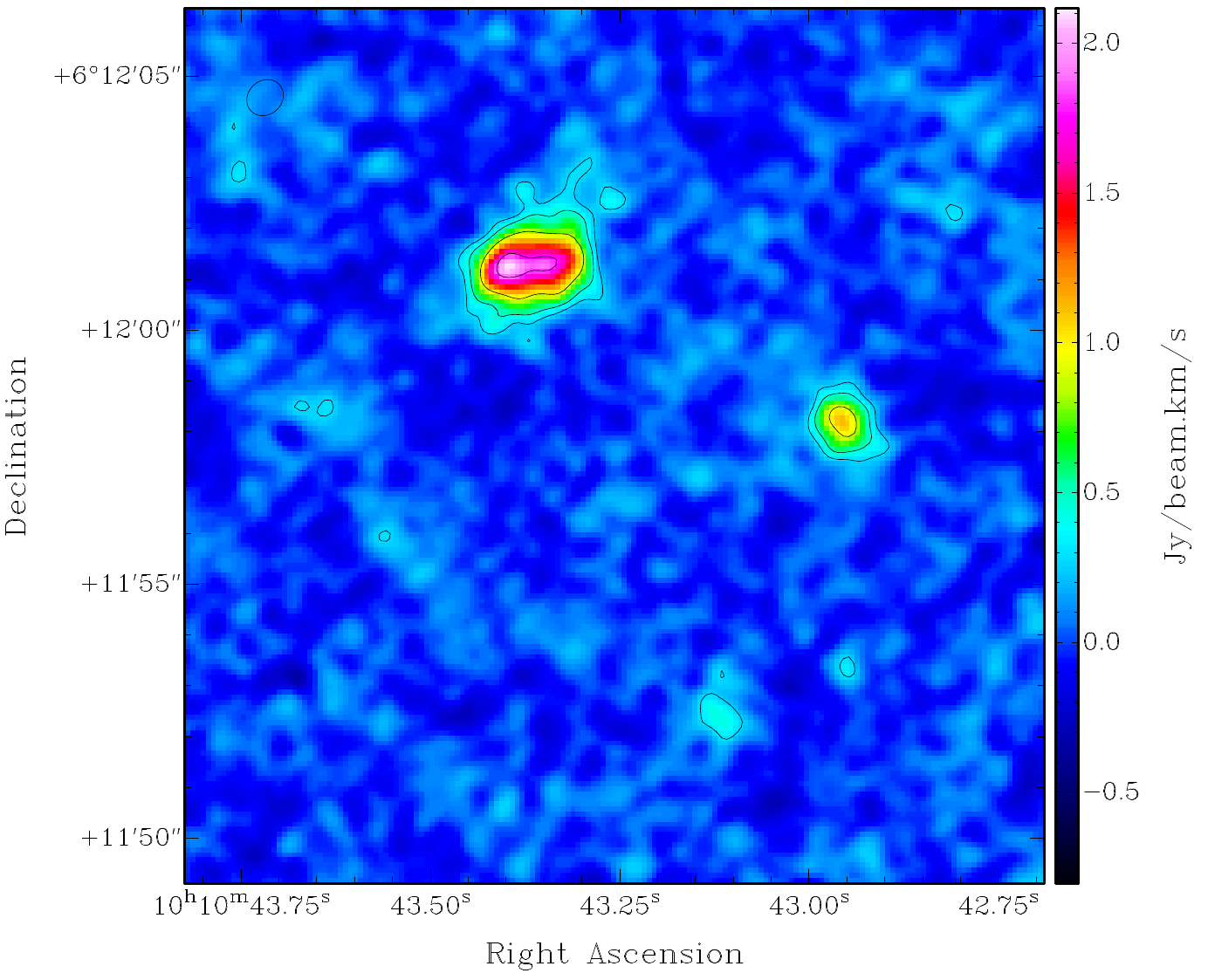}
\caption{Large-scale CO(2$-$1) map of J1010 and two fainter companion sources at the same redshift. Contours correspond to 3, 5, 10, and 20$\sigma$, with $\sigma$=0.09 Jy/beam. The beam size (0.80\arcsec$\times$0.69\arcsec) is shown at the top left corner of the figure. North is up and east to the left.} 
\label{figA1}
\end{figure}

\begin{figure}[!ht]
\FloatBarrier
\centering
\includegraphics[width=0.85\columnwidth]{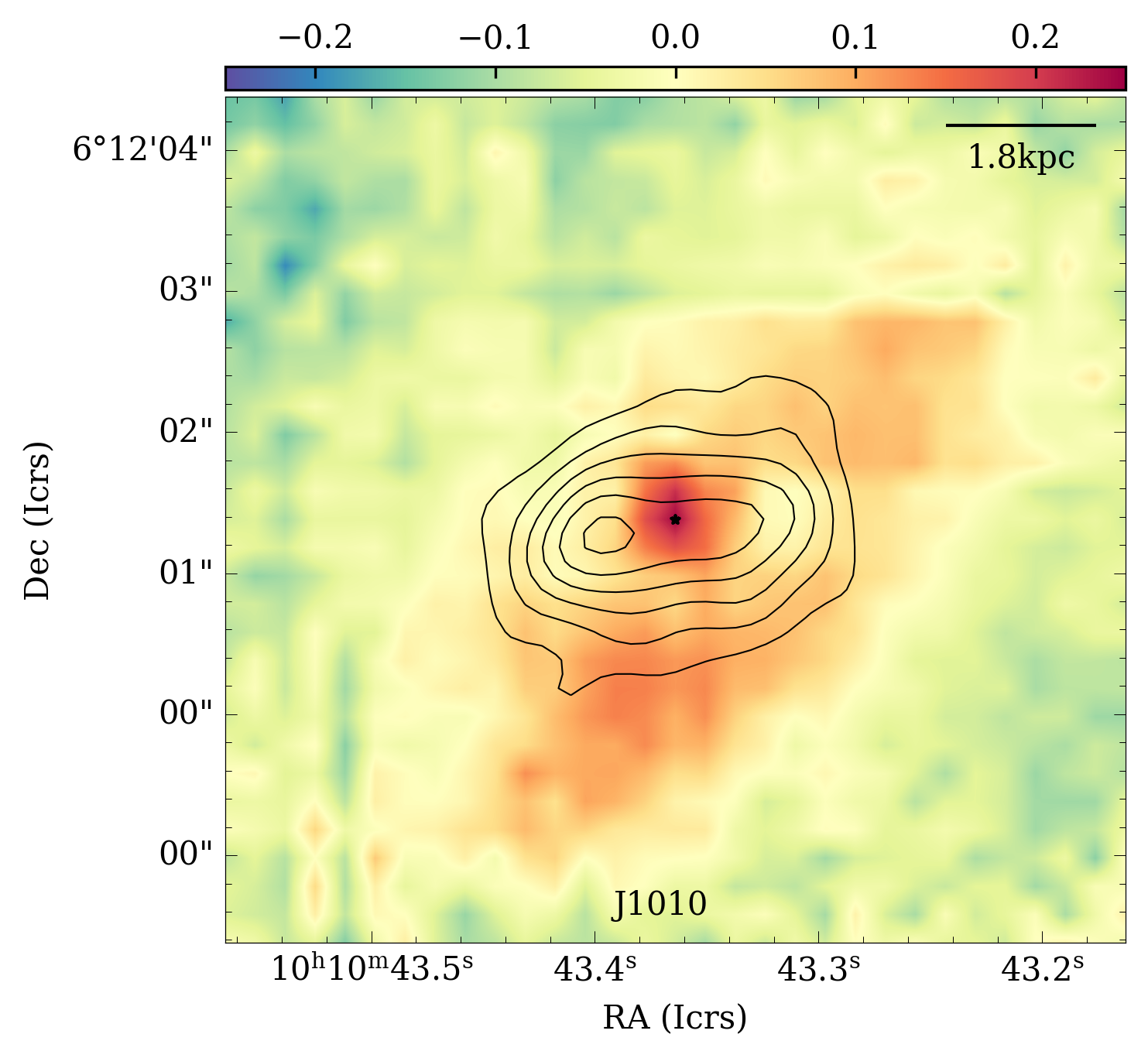}
\caption{CO(2$-$1) contours (0.2, 0.35, 0.5, 0.65, 0.8, 0.95 times the peak of CO emission) of J1010 overlaid on the R-I color map, shown in color scale, derived from continuum images obtained from VLT/MUSE data (see Table \ref{tab4a}), and centered at 6200 and 8500 \AA. The star corresponds to the peak of the ALMA continuum image and the color map corresponds to the central 6\arcsec$\times$6\arcsec. North is up and east to the left.} 
\label{figA2}
\end{figure}

The QSO2 host galaxy is in clear interaction with a small companion at $\sim$7\arcsec~(13 kpc) SW (see Figure \ref{fig1}). A long tidal tail of several kpc in size linking them is also evident. As can be seen from Figure \ref{figA1}, the small companion and another faint source at $\sim$9.5\arcsec~(17 kpc) SW are detected in CO(2$-$1). The faintest companion lacks an optical counterpart in either the SDSS or PanSTARRS images. None of the CO-emitting companions show continuum emission at 1.3 mm. Figure \ref{figA2} corresponds to the R-I color map constructed from the VLT/MUSE datacube (see Table \ref{tab4a}), with the CO(2$-$1) contours overlaid. The optical images have been shifted to match the ALMA astrometry. Redder colors are seen SW from the nucleus, produced by the dust lane detected in the two optical images.

\section{SDSS J135646.10+102609.0 (J1356)}
\label{appendixd}

J1356 is the only QSO2 in our sample with published spatially resolved CO and adjacent continuum observations to date. \citet{2014ApJ...790..160S} observed the CO(1$-$0) and CO(3$-$2) emission lines with ALMA in Cycle 0 (beam size 1.9\arcsec$\times$1.3\arcsec) and Cycle 1 (0.35\arcsec$\times$0.29\arcsec), respectively. The peak of the 3 mm continuum as measured from the Cycle 0 data coincides with the position of the QSO2 (the N nucleus). This N nucleus is detected in CO(1$-$0) and CO(3$-$2), as well as the ``western arm'' (W arm) and the S nucleus. 

\begin{figure*}
\centering
    {\par\includegraphics[width=1.035\columnwidth]{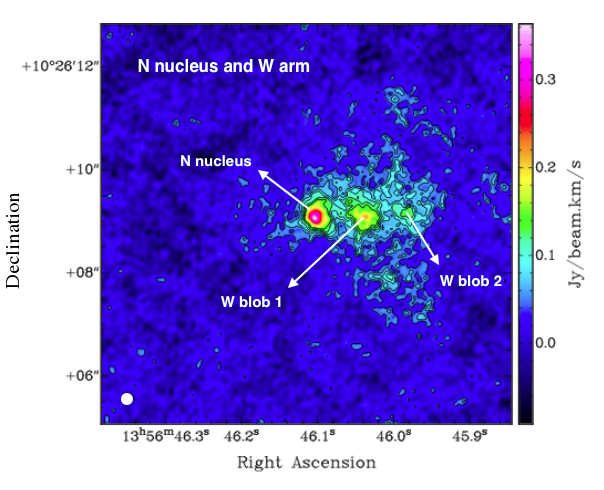}
    \includegraphics[width=1.00\columnwidth]{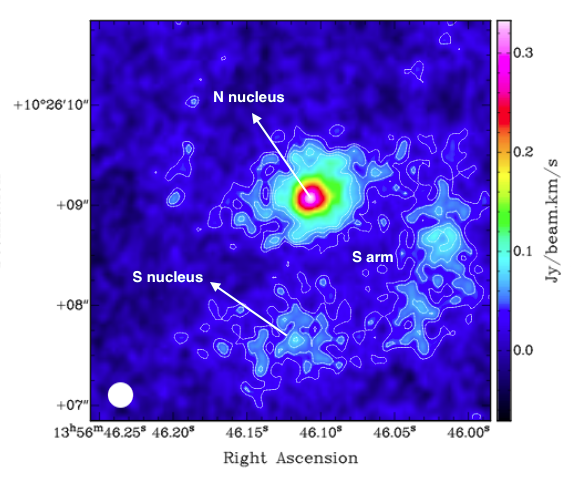}\par}
    \caption{CO(2$-$1) moment 0 maps of J1356 integrating over the emission detected in the ``W arm,'' westward of the N nucleus (left panel; black
    contours go from 2 to 7$\sigma$ with $\sigma$=0.0196 mJy/beam), and in the S nucleus, $\sim$1.1\arcsec~south of the N nucleus (right panel; white contours go from 2 to 5$\sigma$ with $\sigma$=0.0156 mJy/beam). 
    The N and S nuclei are connected by a CO-emitting bridge or spiral arm. 
    The beamsize is indicated in the bottom left corner of each panel.}
    \label{fig5bb}
\end{figure*}

When we integrate over the broad CO(2$-$1) line (see Figure \ref{fig3bis}) we only detect the N nucleus and the W arm. In Figure \ref{fig5aa} the latter appears as blueshifted emission at $\sim$1\arcsec~westward of the N nucleus, extending to a maximum distance of 2.3\arcsec~(5 kpc) from the N nucleus (see left panel of Figure \ref{fig5bb}). It coincides with the location of the base of an extended stellar feature in the HST continuum images \citep{2014ApJ...790..160S}. If we integrate over the narrower and blueshifted CO emission detected in this region we see three blobs, including the N nucleus, along PA=-90\degr, all of them surrounded by diffuse emission (see left panel of Figure \ref{fig5bb}). In Table \ref{tabB} we report the integrated flux, FWHM, and velocity of CO(2$-$1) extracted from the 4$\sigma$ contours at the positions of the two western blobs shown in Figure \ref{fig5bb}. The molecular gas at these positions is blueshifted by 200 km~s$^{-1}$, with a FWHM of 150--260 km~s$^{-1}$, in agreement with the CO(1$-$0) measurements reported by \citet{2014ApJ...790..160S}. These authors suggested that the W arm could be a tidal feature resulting from the merger, slightly offset from its stellar counterpart. Offsets between the gas and the stars have been observed in tidal features (e.g., \citealt{2000AJ....119.1130H}).

\begin{table}[!ht]
\caption{CO(2$-$1) line properties of J1356.}
\centering
\begin{tabular}{lcccc}
\hline
\hline
Region & rms$_{\rm10\ km s^{-1}}$ & S$\Delta$v$_{\rm CO}$ & FWHM          & V$_c$ \\
     & (mJy beam$^{-1}$)        &  (Jy~km~s$^{-1}$)       & (km~s$^{-1}$) &  (km~s$^{-1}$) \\
\hline
W blob1  & 0.020 & 1.64$\pm$0.12 & 148 & -226$\pm$7 \\
W blob2  & 0.020 & 0.56$\pm$0.07 & 148$\pm$22 & -202$\pm$8  \\
S nuc & 0.016 & 0.58$\pm$0.12 & 125$\pm$33 & -45$\pm$13  \\
S arm  & 0.016 & 1.28$\pm$0.14 & 125 & -45$\pm$8  \\
\hline	   					 			    								      
\end{tabular}
\tablefoot{CO(2$-$1) line properties measured in the two blobs detected in the W arm and in the S nucleus and arm (see Figure \ref{fig5bb}). 
The FWHMs of W blob 1 and S arm were fixed to match the value of W blob 2 and S nucleus, respectively.}
\label{tabB}
\end{table}

We also detect CO(2$-$1) emission at $\ge$3$\sigma$ at the position of the S nucleus when we integrate over the narrower and slightly blueshifted CO profile there detected at $\ge$3$\sigma$ (see right panel of Figure \ref{fig5bb}). Our data also reveal a bridge/arm connecting the S and N nuclei. Corresponding CO line measurements for the S nucleus and S arm are reported in Table \ref{tabB}. We measure a blueshift of -45 km~s$^{-1}$, in good agreement with the blueshift estimated from the 2$\sigma$ CO(1$-$0) detection at the S nucleus \citep{2014ApJ...790..160S}. No CO(3--2) emission was found at this position. 

\citet{2014ApJ...790..160S} measured a total molecular gas mass in the merging system of 4.9$\times$10$^9$ M$_{\sun}$ if we assume an $\alpha_{\rm CO}$=4.35 M$_{\sun}(\rm K~km~s^{-1}~pc^2)^{-1}$. 
This mass is smaller than our value of 11.5$\times 10^9 M_{\rm \sun}$, but consistent within the large uncertainties. According to \citet{2014ApJ...790..160S}, one third of this molecular gas would be distributed in a compact rotating disk at the position of the QSO2, with a radius of 300 pc, and half of the molecular gas in the W arm. The S nucleus would contain $\sim$10\% of the molecular gas mass in the system. From our CO(2$-$1) data, we find that the N nucleus represents $\sim$55\% of the total mass (6.4$\times 10^9 M_{\rm \sun}$), the W arm $\sim$32\% (after including the diffuse emission at $\geq$3$\sigma$), and the S nucleus and S arm $\sim$13\%. 

\begin{figure}
\centering
\includegraphics[width=0.95\columnwidth]{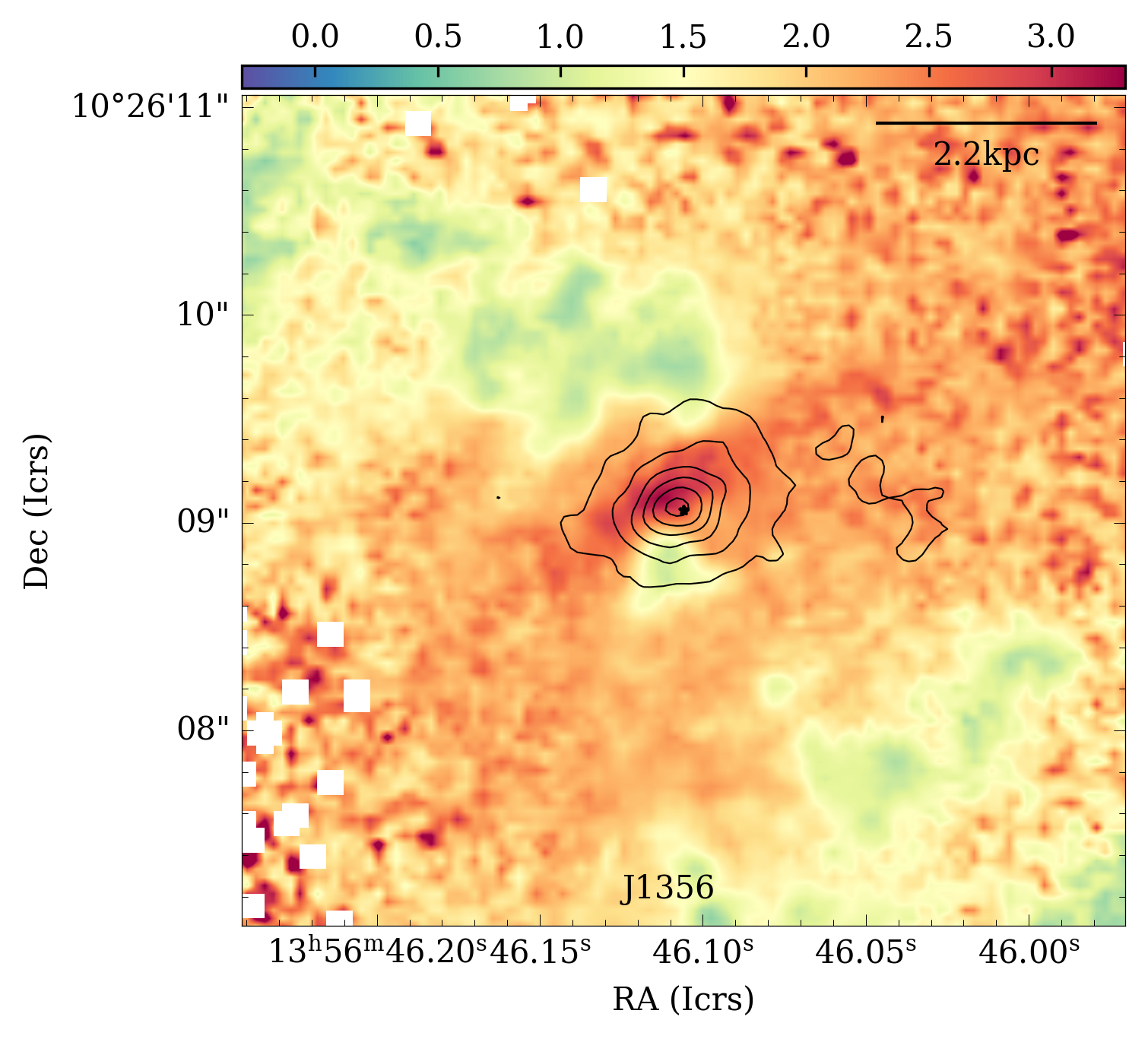}
\caption{Same as in Figure \ref{figA2} but with the CO(2$-$1) contours of J1356 overlaid on the B-I color map derived from HST/WFC3 F438W and F814W images. The color map corresponds to the central 4\arcsec$\times$4\arcsec.} 
\label{figD2}
\end{figure}

Figure \ref{figD2} shows the B-I color map constructed from the HST/WFC3 F438W and F814W images of J1356 described in \citet{2015ApJ...806..219C}, with the CO(2$-$1) contours overlaid. The HST images have been shifted to match the ALMA astrometry. A dust lane crosses the N nucleus, and slightly redder colors are observed toward the N of the CO disk.

\section{SDSS J143029.88+133912.0 (J1430)}
\label{appendixe}

\begin{figure}
\centering
\includegraphics[width=0.95\columnwidth]{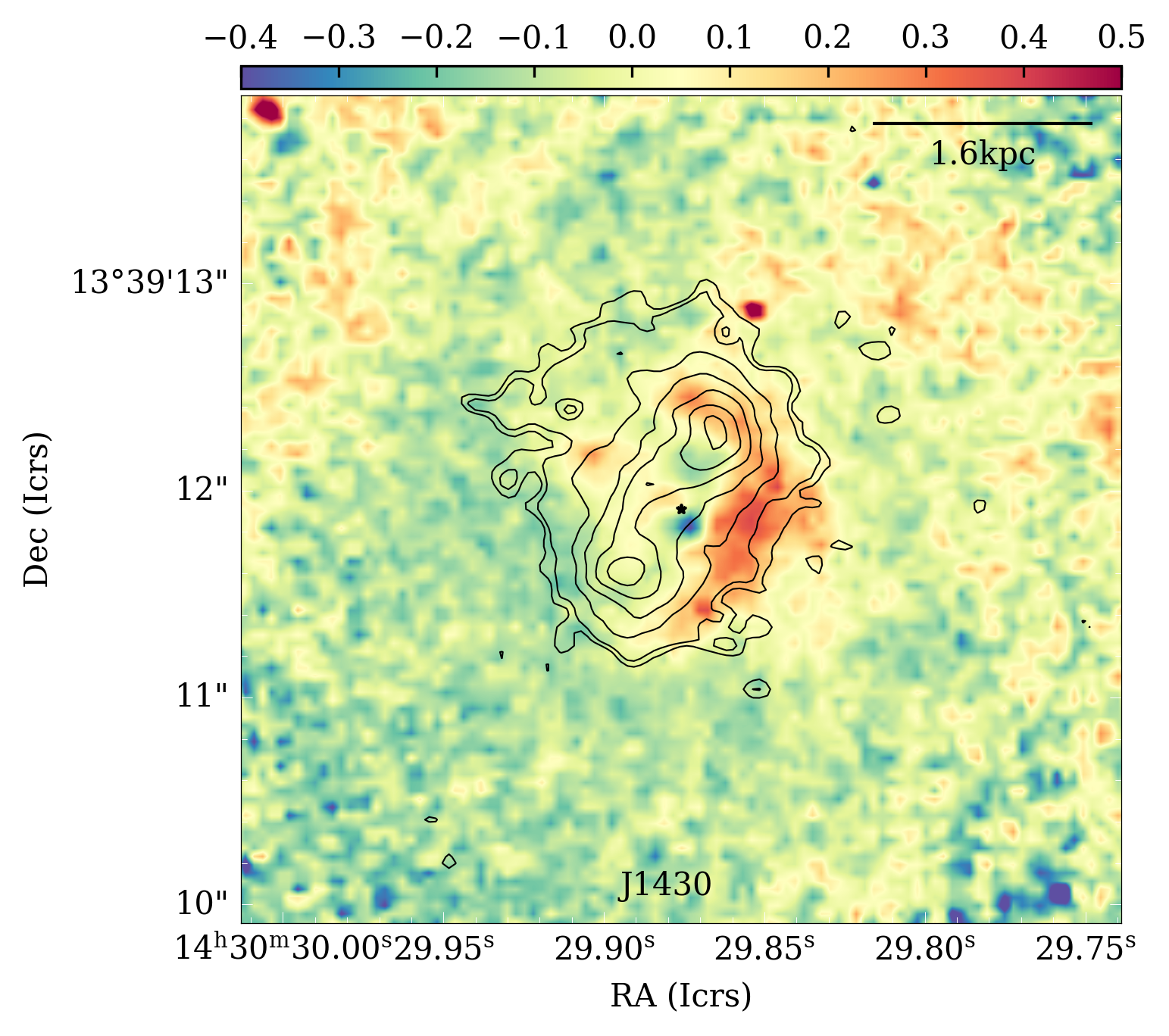}
\caption{Same as in Figure \ref{figA2} but with the CO(2$-$1) contours of J1430 (0.15, 0.2, 0.35, 0.5, 0.65, 0.8, and 0.95 times the peak of CO emission), overlaid on the r-i color map derived from HST/WFC3 F621M and F763M images. The color map corresponds to the central 4\arcsec$\times$4\arcsec.} 
\label{figE1}
\end{figure}

In Figure \ref{figE1} we show the r-i color map constructed from the HST/WFC3 F621M and F763M images of J1430 described in \citet{2015AJ....149..155K}, with the CO(2$-$1) contours overlaid. The HST images have been shifted to match the ALMA astrometry. An intricate system of shells can be clearly seen in the two images westward of the nucleus, where red colors are observed.

\end{appendix}

%
%

\end{document}